\documentclass[pdftex,twocolumn,3]{jour3}          
\usepackage{tabularx}
\usepackage{array}
\usepackage{cancel}
\usepackage{mathtools}
\usepackage[utf8]{inputenc}
\usepackage{epsf}
\usepackage{latexsym,amssymb,euscript}
\usepackage{widetext}
\usepackage[dvips]{graphicx}
\usepackage[numbers,sort&compress]{natbib}
\usepackage{comment}
\usepackage{amsmath}
\usepackage{nicefrac}
\usepackage{slashed}
\usepackage{booktabs}
\usepackage{hyperref}
\usepackage{braket}
\usepackage{chngcntr}
\usepackage{bm}
\usepackage{bbold}
\usepackage{graphics}
\usepackage{graphicx}
\usepackage{mciteplus}
\usepackage{bbold}
\usepackage{pdfpages}
\usepackage[titletoc]{appendix}
\usepackage{etoolbox}
\usepackage{titlesec}

\usepackage{ulem}
\usepackage{microtype} 

\graphicspath{{./figures/}}
\hypersetup{
 linktocpage = true,
 urlcolor = urlblue,
 colorlinks = true,
 linkcolor = urlblue,
 anchorcolor = citegreen,
 citecolor = citegreen,
 pdfstartview = {XYZ null null 1.25} 
           }
\usepackage[left=2cm, right=2cm]{geometry}
\usepackage{pstricks}
\usepackage{color}
\usepackage{xcolor}
\definecolor{dgreen}{rgb}{0,0.325,0}
\definecolor{urlblue}{rgb}{0.2,0.4,0.7}
\definecolor{citegreen}{rgb}{0,0.4,0.2}
\definecolor{linkred}{rgb}{0.9,0.2,0.1}
\usepackage{float}
\usepackage{academicons}
\definecolor{orcidlogocol}{HTML}{A6CE39}
\usepackage{fancyhdr}
\newcommand{\drv}{{\rm d}}

\newcommand{\as}{\alpha_s}

\newcommand{\MSb}{\overline{\rm MS}}

\newcommand{\LL}{{\rm LL/LO}}
\newcommand{\NLL}{{\rm NLL/NLO}}
\newcommand{\NLLp}{{\rm NLL/NLO^+}}
\newcommand{\NLLpp}{{\rm NLL/NLO^{(+)}}}
\newcommand{\HENLOp}{{\rm HE}\mbox{-}{\rm NLO^+}}

\newcommand{\CnLL}{{\cal C}_n^\LL}

\newcommand{\CnNLLp}{{\cal C}_n^\NLLp}

\newcommand{\CnHENLOp}{{\cal C}_n^{{\rm HE}\text{-}{\rm NLO}^+}}
\newcommand{\DY}{\Delta Y}

\newcommand{\E}{{\cal E}}

\newcommand{\HQ}{{\cal H}_Q}
\newcommand{\Hc}{{\cal H}_c}
\newcommand{\Hb}{{\cal H}_b}
\newcommand{\Hcb}{{\cal H}_{c,b}}
\newcommand{\etQ}{\eta_Q}
\newcommand{\etc}{\eta_c}
\newcommand{\etb}{\eta_b}
\newcommand{\etcb}{\eta_{c,b}}
\newcommand{\Jpsi}{J/\psi}
\newcommand{\Yps}{\Upsilon}
\newcommand{\BCs}{B_c(^1S_0)}
\newcommand{\Bss}{B_c(^3S_1)}

\newcommand{\Q}{\cal Q}

\newcommand{\TQc}{T_{4c}}

\newcommand{{\HFNRevo}}{\tt HF-NRevo}

\newcommand{{\Jethad}}{\tt JETHAD}
\newcommand{{\symJethad}}{\tt symJETHAD}
\newcommand{{\Hell}}{\tt HELL}
\newcommand{{\RadISH}}{\tt RadISH}
\newcommand{{\Pegasus}}{\tt QCD-PEGASUS}
\newcommand{{\HOPPET}}{\tt HOPPET}
\newcommand{{\QCDNUM}}{\tt QCDNUM}
\newcommand{{\APFEL}}{\tt APFEL}
\newcommand{{\APFELpp}}{\tt APFEL++}
\newcommand{{\APFELppp}}{\tt APFEL(++)}
\newcommand{{\EKO}}{\tt EKO}
\newcommand{{\FeynCalc}}{\tt FeynCalc}

\newcommand{\LHAPDF}{{\tt LHAPDF}}
\newcommand{{\HCFF}}{{\tt HCFF1.0}}
\newcommand{{\NRFF}}{{\tt NRFF1.0}}

\newcommand{\tarr}{
\begin{array}}
\newcommand{\earr}{\end{array}}

\newcommand{\orcidFGC}{\href{https://orcid.org/0000-0003-3299-2203}{\includegraphics[scale=0.1]{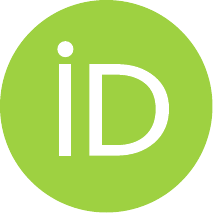}}}

\newcommand{\orcidFL}{\href{https://orcid.org/0009-0003-2052-4549}{\includegraphics[scale=0.1]{logo-orcid.pdf}}}

\smartqed  

\journalname{}

\begin{document}

\normalem

\title{Pseudoscalar heavy quarkonium hadroproduction \\[0.25cm] from nonrelativistic fragmentation at NLL/NLO$^+$
}

\subtitle{}

\author{
Francesco Giovanni Celiberto
\and
\thanksref{e1,addr1} \orcidFGC
\; 
Francesca Lonigro
\thanksref{e2,addr1} \orcidFL
}

\thankstext{e1}{{\it e-mail}:
\href{mailto:francesco.celiberto@uah.es}{francesco.celiberto@uah.es} (corresponding author)}
\thankstext{e2}{{\it e-mail}:
\href{mailto:francesco.celiberto@uah.es}{francesca.lonigro@uah.es}}

\institute{Universidad de Alcal\'a (UAH), Departamento de F\'isica y Matem\'aticas, Alcal\'a de Henares, E-28805, Madrid, Spain \label{addr1}
}

\date{\today}

\maketitle

\newcounter{appcnt}


\section*{Abstract}
We investigate the inclusive hadroproduction of pseudoscalar heavy quarkonia, $\eta_c$ and $\eta_b$ mesons, in high-energy proton collisions.
Our framework is based on the single-parton collinear fragmentation within a variable-flavor number scheme, tailored to describe the moderate to large transverse momentum regime.
To this end, we construct a new set of collinear fragmentation functions, denoted as {\tt NRFF1.0}, which evolve via standard DGLAP equations with a consistent treatment of flavor thresholds.
Initial conditions for all parton-induced channels are computed using next-to-leading-order nonrelativistic QCD.
We perform our analysis within the NLL/NLO$^+$ hybrid factorization framework, employing the {\tt JETHAD} numerical interface together with the {\tt symJETHAD} symbolic engine.
These tools allow us to deliver predictions for high-energy observables sensitive to quarkonium  final states at the 13~TeV LHC.
To the best of our knowledge, the {\tt NRFF1.0} sets represent the first-ever release of collinear fragmentation functions for heavy quarkonia that consistently include all partonic channels within collinear factorization.
\vspace{0.30cm} \hrule
\vspace{0.20cm}
{
 \setlength{\parindent}{0pt}
 \textsc{Keywords}: 
 Pseudoscalar quarkonium, Heavy flavor, {\HFNRevo}, {\NRFF}, Fragmentation, High-energy resummation, Natural stability
}
\vspace{-0.55cm}

\setcounter{tocdepth}{3}
\renewcommand{\baselinestretch}{1.0}\normalsize
\tableofcontents
\renewcommand{\baselinestretch}{1.0}\normalsize

\clearpage


\section{Introduction}
\label{sec:intro}

Hadrons with heavy quarks play a key role in the search for New Physics. 
As natural probes of potential Beyond the Standard Model (BSM) interactions, they are sensitive to rare processes and symmetry violations, making them valuable tools in precision experiments and theoretical studies.

Concurrently, studying their production mechanisms in high-energy collisions has been equally instrumental in advancing our understanding of Quantum Chromodynamics (QCD). 
In particular, these systems offer a unique window into the transition between perturbative and nonperturbative regimes, shedding light on color confinement and the hadronization process.

A key category of heavy-flavored hadrons consists of mesons whose dominant Fock component is a heavy quark--antiquark pair. These are known as heavy quarkonia. 
The field of quarkonium physics emerged in 1974 during the so-called ``November Revolution,'' marked by the discovery of a new vector meson, the $\Jpsi$, with mass around 3.1 GeV and quantum numbers matching those of the photon. 
This particle was independently observed at SLAC~\cite{SLAC-SP-017:1974ind} and BNL~\cite{E598:1974sol}, and soon confirmed at Frascati~\cite{Bacci:1974za}.

The hadronic nature of $\Jpsi$ was established by analyzing of the hadron-to-muon ratio in electron-positron annihilation at resonance, which implies strong hadronic decays. 
This discovery offered concrete proof of the existence of the charm quark (previously hypothesized by Bjorken and Glashow~\cite{Bjorken:1964bo}) and a clear indication that quarks are physical constituents, not theoretical constructs. 
The charm quark also played a critical role in the Glashow-Iliopoulos-Maiani (GIM) mechanism~\cite{Glashow:1970gi}, which explained the suppression of flavor-changing neutral currents (FCNCs) in weak interactions.

Quarkonium systems were among the earliest tools to reveal key features of QCD, most notably asymptotic freedom. 
The discovery of $\psi(2S)$, the first radial excitation of the $\Jpsi$, demonstrated that the strong force weakens at short distances, leading to a binding potential with a Coulomb-like behavior at small separations and a confining term at large ones. 
The $\Jpsi$ itself was the first observed charmonium state ($|c\bar{c}\rangle$), soon followed by other charmonia, such as the $P$-wave $\chi_c$ and the pseudoscalar $\etc$, as well as singly charmed $D$ mesons~\cite{Wiss:1976gd}.

A few years later, in 1977, the first bottomonium state ($|b\bar{b}\rangle$) was discovered: $\Upsilon$, a vector meson analogous to $\Jpsi$, but composed of bottom quarks~\cite{Herb:1977ek}. 
This milestone opened the way to the observation of excited bottomonium states like the $\Upsilon(2S)$ and singly bottomed $B$ mesons~\cite{CLEO:1980oyr}.

Despite decades of intense theoretical efforts, the production of heavy quarkonia remains one of the most intricate aspects of QCD. 
While their experimental signatures---especially for vector mesons---are remarkably clean, the underlying mechanisms involve nontrivial combinations of perturbative and nonperturbative dynamics.

Several approaches have been developed to describe the hadronization of a heavy [$Q\bar{Q}$] pair into a physical bound state, each relying on different assumptions about color flow and spin correlations. 
Among them, the Nonrelativistic Quantum Chromodynamics (NRQCD) effective field theory offers a systematic framework to organize both color-singlet and color-octet contributions in a double expansion in $\alpha_s$ and the relative velocity $v_{\cal Q}$~\cite{Caswell:1985ui,Thacker:1990bm,Bodwin:1994jh,Cho:1995vh,Cho:1995ce,Leibovich:1996pa,Bodwin:2005hm}.

NRQCD allows for a clear separation between the short-distance dynamics, computable in perturbation theory, and the long-distance hadronization step, which must be modeled or extracted from data.
Nonetheless, the large number of free parameters introduced by the theory, and the delicate interplay between direct and feed-down components, still pose open challenges for precision predictions and global fits~\cite{Brambilla:2010cs,Lansberg:2019adr}.

Another complication arises from the so-called ``nonprompt'' production: quarkonia like $\Jpsi$ can originate from $B$ meson decays~\cite{Halzen:1984rq}, leading to delayed vertices. 
Experiments can isolate prompt components by separating these vertices. 
Within prompt production, further subdivision exists: indirect contributions come from higher quarkonia decays, such as [$\chi_c \rightarrow \Jpsi$, while the direct component (produced in the hard process itself) can be inferred by subtraction.

At low transverse momentum $|\vec q_T|$, quarkonium production is dominated by \emph{short-distance} generation mechanisms, where a [$Q\bar{Q}$] pair is produced in the hard scattering and subsequently hadronizes into the final state.
Because the quark and antiquark are created with a relative transverse separation of order $1/|\vec q_T|$, this channel becomes suppressed as $|\vec q_T|$ increases.

At high $|\vec q_T|$, another production mechanism becomes relevant: the \emph{fragmentation} of a single high-energy parton into a quarkonium plus other partons. 
This process is described by collinear fragmentation functions (FFs), whose energy evolution is controlled by the well-known Dokshitzer-Gribov-Lipatov-Altarelli-Parisi (DGLAP) equations~\cite{Gribov:1972ri,Gribov:1972rt,Lipatov:1974qm,Altarelli:1977zs,Dokshitzer:1977sg}.
Initial-scale input for these quarkonium FFs contains perturbative ingredients that can be computed in NRQCD~\cite{Braaten:1993rw}. 
The [$Q \bar Q$ pair is formed with a typical separation of order $1/m_Q$.
Although this mechanism appears at higher perturbative order, it is enhanced by a factor $(Q/m_Q)^2$, making it dominant when $Q \gg m_Q$.

Phenomenological leading-order (LO) studies of the transition region between short-distance and fragmentation mechanisms appeared in Refs.~\cite{Doncheski:1993xm,Braaten:1994xb,Cacciari:1994dr,Cacciari:1995yt,Cacciari:1995fs}. 
Studies on $S$- and $P$-wave channels were conducted in Refs.~\cite{Ma:2013yla,Ma:2014eja}, showing that single-parton fragmentation becomes increasingly important at higher transverse momenta.

The gluon FFs to $S$-wave vector and pseudoscalar quarkonia were calculated in NRQCD at LO in Ref.~\cite{Braaten:1993rw}, and at next-to-leading order (NLO) in Refs.~\cite{Braaten:1993rw,Artoisenet:2014lpa,Zhang:2018mlo}. 
Correspondingly, the quark FFs were obtained at LO in Ref.~\cite{Braaten:1993mp} and at NLO in Ref.~\cite{Zheng:2019dfk,Zheng:2021mqr,Zheng:2021ylc}.
Building on these fixed-order calculations as input at the initial energy scale, the first determinations of DGLAP-evolved FFs for vector quarkonia within the variable flavor number scheme (VFNS)~\cite{Mele:1990cw,Cacciari:1993mq} were presented in Refs.~\cite{Celiberto:2022dyf,Celiberto:2023fzz}, under the name {\tt ZCW19$^+$}.
An extension to charmed $B$ mesons, known as {\tt ZCFW22}, was subsequently developed in Refs.~\cite{Celiberto:2022keu,Celiberto:2024omj}.

While the {\tt ZCW19$^+$} sets faithfully implement the NLO NRQCD FFs for constituent heavy quarks and gluons, as well as a numerically exact DGLAP evolution, they lack a consistent treatment of partonic flavor thresholds across all species. 
To address this limitation, the \emph{heavy-flavor nonrelativistic evolution} ({\HFNRevo}) scheme was developed~\cite{Celiberto:2024mex,Celiberto:2024bxu,Celiberto:2024rxa,Celiberto:2025xvy}. 
This framework is specifically tailored to evolve heavy-hadron FFs from nonrelativistic initial conditions, incorporating heavy quark threshold effects in a fully consistent way.

Building on this foundation, a broader effort was launched to derive a family of nonrelativistic-based, DGLAP-evolved FFs for quarkonium states, collectively named \emph{NonRelativistic Fragmentation Functions} ({\NRFF}). 
These functions are intended to supersede the {\tt ZCW19$^+$} sets for $\Jpsi$ and $\Yps$ vectors, and to provide the first-ever collinear FFs for $\etc$ and $\etb$ pseudoscalars. 
Preliminary results for charm and bottom quark fragmentation into these states have been presented in Refs.~\cite{Celiberto:2024mex,Celiberto:2024bxu}, while early developments toward quarkonium-in-jet fragmentation~\cite{Ernstrom:1996am,Baumgart:2014upa,Kang:2016ehg,Kang:2017yde,Bain:2016clc,Bain:2017wvk,Makris:2018npl,Makris:2017sfs,Cooke:2023ukz,Gambhir:2025afb} within the {\HFNRevo} approach were reported in Ref.~\cite{Celiberto:2024rxa}.

In this work, we focus on the pseudoscalar quarkonium sector.
To this end, we construct and publicly release the first version of the {\NRFF} {\HFNRevo} functions describing the collinear fragmentation of $\etc$ and $\etb$ mesons.
Studying pseudoscalar quarkonia offers theoretical and phenomenological advantages that make this sector a compelling testing ground for fragmentation dynamics. 

\begin{figure*}[!t]
\centering
\includegraphics[width=0.32\textwidth]{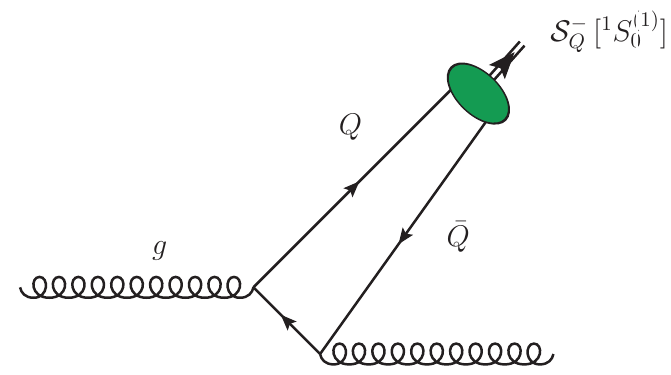}
\hspace{0.00cm}
\includegraphics[width=0.32\textwidth]{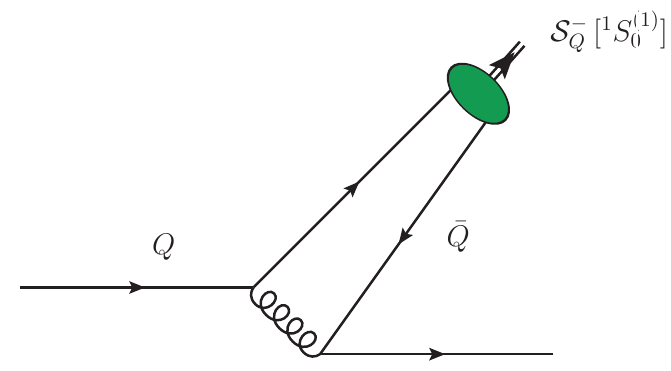}
\hspace{0.00cm}
\includegraphics[width=0.32\textwidth]{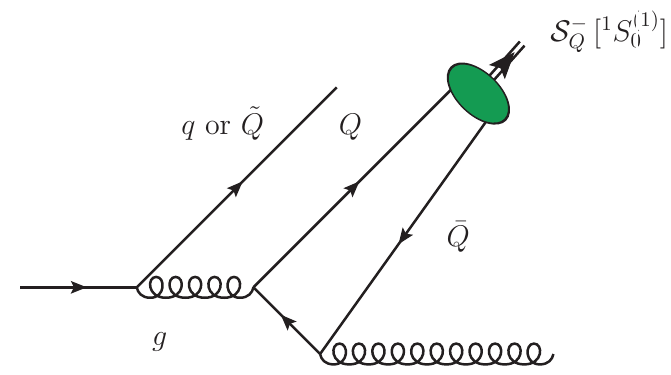}

\caption{Left panel: A leading diagram for the fragmentation of a gluon to ${\cal S}^-_Q \, [^1S_0^{(1)}]$ at ${\cal O}(\alpha_s^2)$.
Central panel: One of the leading diagrams for the fragmentation of a constituent heavy quark ($Q$) to a $S$-wave color-singlet pseudoscalar quarkonium ${\cal S}^-_Q \, [^1S_0^{(1)}]$ at ${\cal O}(\alpha_s^2)$.
Right panel: A leading diagram for the fragmentation of a nonconstituent light ($\tilde{q}$) or heavy ($\tilde{Q}$) quark to ${\cal S}^-_Q \, [^1S_0^{(1)}]$ at ${\cal O}(\alpha_s^3)$.
Green blobs refer to the ${\cal S}^-_Q \, [^1S_0^{(1)}]$ nonperturbative NRQCD LDME.}
\label{fig:PSQ_FF_diagrams}
\end{figure*}

\begin{figure*}[!t]
\centering
\includegraphics[width=0.32\textwidth]{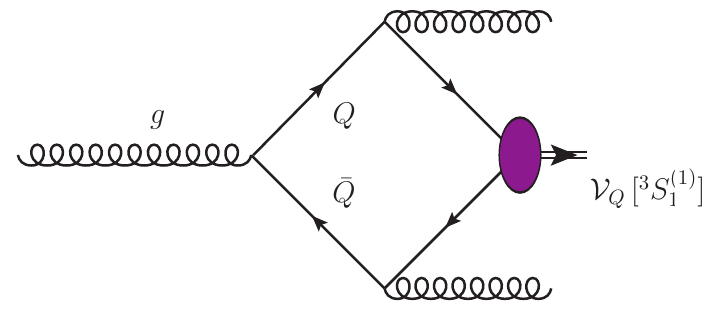}
\hspace{0.00cm}
\includegraphics[width=0.32\textwidth]{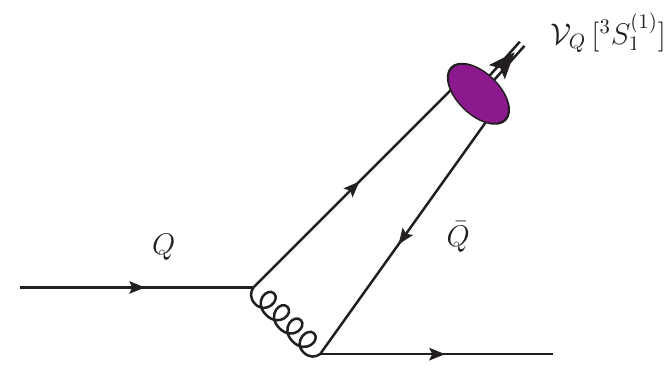}
\includegraphics[width=0.32\textwidth]{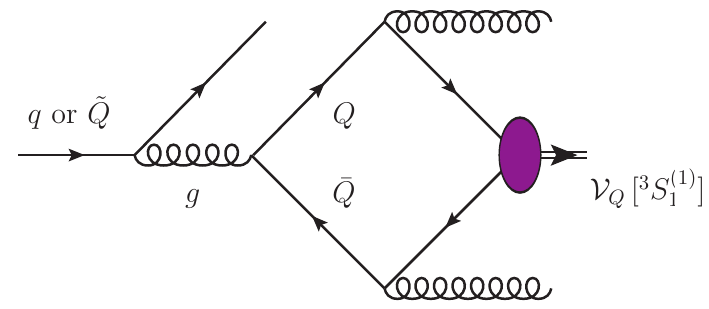}

\caption{Left panel: A leading diagram for the fragmentation of a gluon to ${\cal V}_Q \, [^3S_1^{(1)}]$ at ${\cal O}(\alpha_s^3)$.
Central panel: One of the leading diagrams for the fragmentation of a constituent heavy quark ($Q$) to a $S$-wave color-singlet vector quarkonium ${\cal V}_Q \, [^3S_1^{(1)}]$ at ${\cal O}(\alpha_s^2)$.
Right panel: A leading diagram for the fragmentation of a nonconstituent light ($\tilde{q}$) or heavy ($\tilde{Q}$) quark to ${\cal V}_Q \, [^3S_1^{(1)}]$ at ${\cal O}(\alpha_s^4)$.
Violet blobs refer to the ${\cal V}_Q \, [^3S_1^{(1)}]$ nonperturbative NRQCD LDME.}
\label{fig:VQ_FF_diagrams}
\end{figure*}

First, as illustrated in Fig.~\ref{fig:PSQ_FF_diagrams} and summarized in the first row of Table~\ref{tab:FF_order}, the NRQCD initial-scale inputs are known at NLO for all partonic channels: gluon~\cite{Artoisenet:2014lpa,Zhang:2018mlo}, constituent heavy quark~\cite{Zheng:2021ylc}, and nonconstituent quarks~\cite{Zheng:2021mqr}. 
In contrast, for vector quarkonia, the nonconstituent quark channels vanish within NLO accuracy (see Fig.~\ref{fig:VQ_FF_diagrams} and lower row of Table~\ref{tab:FF_order}). 
This makes pseudoscalar states well suited to testing the relative weight of all NRQCD fragmentation channels simultaneously, as well as the combined impact of DGLAP evolution across flavor thresholds. 
In this respect, the pseudoscalar sector stands out as an ideal laboratory for validating the {\HFNRevo} methodology.

Another key theoretical feature is the predicted suppression of color-octet mechanisms in pseudoscalar quarkonium production. 
As discussed in prior studies of exotic hadrons, color-octet transitions play a critical role in vector quarkonium phenomenology, particularly for the hadroproduction of $\Jpsi$ at moderate and large transverse momentum. 
In that case, singlet-only approaches fail to reproduce observed rates, necessitating the inclusion of gluon and heavy quark fragmentation into color-octet $|c\bar{c}\rangle$ states---most notably the ${}^3S_1^{(8)}$, ${}^1S_0^{(8)}$, and ${}^3P_J^{(8)}$ configurations~\cite{Braaten:1993rw,Cho:1995vh,Cho:1995ce,Beneke:1996tk, Butenschoen:2010rq,Chao:2012iv, Gong:2012ug}. 
However, for pseudoscalar states such as $\etc$ and $\etb$, NRQCD predicts a dominant ${}^1S_0^{(1)}$ singlet contribution, with octet channels suppressed by spin and parity selection rules~\cite{Braaten:1993rw,Han:2014jya}. 
This theoretical expectation is supported by the available LHCb data~\cite{LHCb:2014oii,LHCb:2019zaj}, which can be well described without invoking octet terms. 

A similar suppression holds in the $\Upsilon(nS)$ family~\cite{LHCb:2022byt}, where larger masses and reduced sensitivity to soft-gluon effects minimize octet contributions~\cite{Artoisenet:2008fc}, though minor corrections may still be relevant at large transverse momentum~\cite{Gong:2010bk}. 
As a result, it is theoretically well justified and methodologically advantageous to restrict our study to the color-singlet channel in this pioneering work.

Then, from a phenomenological perspective, future colliders will offer increasingly favorable conditions for accessing the pseudoscalar sector. 
Precise measurements of $\etc$ and $\etb$ production will be within reach at the LHC and its high-luminosity upgrade (HL-LHC)~\cite{Chapon:2020heu,LHCspin:2025lvj}, as well as at next-generation machines such as the Electron-Ion Collider (EIC)~\cite{Boer:2024ylx,AbdulKhalek:2021gbh,Khalek:2022bzd,Hentschinski:2022xnd,Amoroso:2022eow,Abir:2023fpo,Allaire:2023fgp}, NICA-SPD~\cite{Arbuzov:2020cqg,Abazov:2021hku}, the International Muon Collider (IMC)~\cite{Accettura:2023ked,InternationalMuonCollider:2024jyv,MuCoL:2024oxj,MuCoL:2025quu,Black:2022cth,InternationalMuonCollider:2025sys}, and the Future Circular Collider (FCC)~\cite{FCC:2025lpp,FCC:2025uan,FCC:2025jtd}. 
These facilities will open the way to detailed studies of semi-inclusive processes and quarkonium-in-jet observables in kinematic regimes where DGLAP-based predictions become testable.

Finally, we stress that, to date, no public collinear FFs exist for pseudoscalar quarkonia. 
The public release of the {\NRFF} sets for $\etc$ and $\etb$ thus fills a significant gap in the available phenomenological toolkit, enabling both precision theoretical predictions and dedicated comparisons with upcoming experimental data.

To validate our fragmentation framework, we investigate the semi-inclusive production of a pseudoscalar quarkonium state $\etQ$ accompanied by either a singly heavy-flavored hadron $\HQ$ or a jet, with both final-state objects well separated in rapidity. 
The process is studied within the $\NLLp$ hybrid factorization (HyF) formalism, which combines NLO collinear factorization with the resummation of high-energy logarithms beyond next-to-leading logarithmic (NLL) accuracy.

Our predictions are obtained using the {\Jethad} numerical interface in synergy with the {\symJethad} symbolic engine~\cite{Celiberto:2020wpk,Celiberto:2022rfj,Celiberto:2023fzz,Celiberto:2024mrq,Celiberto:2024swu}. 
These tools enable precise and efficient computations of high-energy observables sensitive to heavy quarkonium final states in the forward regime of the 13~TeV LHC.
For completeness, we refer the reader to Refs.~\cite{Baranov:2002cf,Kniehl:2006sk,Kniehl:2016sap,Saleev:2012hi,Nefedov:2013qya,Boussarie:2017oae,Cisek:2017gno,Cisek:2017ikn,Maciula:2018bex,Prokhorov:2020owf,Babiarz:2019mag,Baranov:2023ckv,Chernyshev:2023qea,Nefedov:2024swu,Bautista:2016xnp,Peredo:2023oym,Hentschinski:2025ovo,Kang:2013hta,Ma:2014mri,Ma:2015sia,Ma:2018qvc,Ducloue:2015gfa,Ducloue:2016pqr,Lappi:2020ufv,Stebel:2021bbn,Mantysaari:2021ryb,Mantysaari:2022kdm,Cheung:2024qvw,Gimeno-Estivill:2024gbu,Penttala:2024hvp,Siddikov:2025tzh} for other studies of quarkonium production at high energies.

This work is structured as follows. 
Section~\ref{sec:fragmentation} outlines the technical aspects underlying the construction of the new {\NRFF} collinear FFs for $\etQ$ mesons and provides details on our treatment of the collinear fragmentation of $\HQ$ hadrons.
Section~\ref{sec:HE_resummation} offers a technical overview of the HyF framework employed with $\NLLp$ accuracy.
Section~\ref{sec:results} presents a phenomenological study of the semi-inclusive associated production of $\etQ$ plus jet systems, including predictions for rapidity distributions, angular multiplicities, and transverse momentum spectra.
Finally, Section~\ref{sec:conclusions} offers concluding remarks and discusses perspectives for future developments.

\section{Theoretical setup: Heavy-flavor fragmentation}
\label{sec:fragmentation}

 \begin{table*}
 \begin{center}
 \begin{tabular}[c]{|c||c|c|c|c|}
 \hline
   Quarkonium [${\cal Q}$] 
 & $g \to {\cal Q}$ 
 & $Q \to {\cal Q}$ 
 & $\tilde{q} \mbox{ or } \tilde{Q} \to {\cal Q}$ 
 \\
 \hline
   $\hspace{-0.10cm} {\cal Q} \equiv \textcolor{dgreen}{\boldsymbol{{\cal S}_Q^- \, [^1S_0^{(1)}]}} \equiv \etc, \etb$ 
 & $\begin{array}{ll}
    {\cal O}(\alpha_s^2) \equiv \mbox{LO} 
   \vspace{0.10cm} \\
    {\cal O}(\alpha_s^3) \equiv \mbox{NLO}
   \vspace{0.10cm} \\
    \textcolor{gray}{{\cal O}(\alpha_s^4) \equiv \mbox{NNLO}}
   \vspace{-0.10cm} \\
    \hspace{1.00cm} \textcolor{gray}{\vdots}
   \end{array}$
 & $\begin{array}{ll}
    {\cal O}(\alpha_s^2) \equiv \mbox{LO} 
   \vspace{0.10cm} \\
    {\cal O}(\alpha_s^3) \equiv \mbox{NLO}
   \vspace{0.10cm} \\
    \textcolor{gray}{{\cal O}(\alpha_s^4) \equiv \mbox{NNLO}}
   \vspace{-0.10cm} \\
    \hspace{1.00cm} \textcolor{gray}{\vdots}
   \end{array}$
 & $\begin{array}{ll}
    - 
   \vspace{0.10cm} \\
    {\cal O}(\alpha_s^3) \equiv \mbox{NLO}
   \vspace{0.10cm} \\
    \textcolor{gray}{{\cal O}(\alpha_s^4) \equiv \mbox{NNLO}}
   \vspace{-0.10cm} \\
    \hspace{1.00cm} \textcolor{gray}{\vdots}
   \end{array}$
 \\
 \hline
 $\hspace{0.10cm} {\cal Q} \equiv \textcolor{violet}{\boldsymbol{{\cal V}_Q \, [^3S_1^{(1)}]}} \equiv \Jpsi, \Yps$
 & $\begin{array}{ll}
    - 
   \vspace{0.10cm} \\
    {\cal O}(\alpha_s^3) \equiv \mbox{NLO}
   \vspace{0.10cm} \\
    \textcolor{gray}{{\cal O}(\alpha_s^4) \equiv \mbox{NNLO}}
   \vspace{-0.10cm} \\
    \hspace{1.00cm} \textcolor{gray}{\vdots}
   \end{array}$
 & $\begin{array}{ll}
    {\cal O}(\alpha_s^2) \equiv \mbox{LO} 
   \vspace{0.10cm} \\
    {\cal O}(\alpha_s^3) \equiv \mbox{NLO}
   \vspace{0.10cm} \\
    \textcolor{gray}{{\cal O}(\alpha_s^4) \equiv \mbox{NNLO}}
   \vspace{-0.10cm} \\
    \hspace{1.00cm} \textcolor{gray}{\vdots}
   \end{array}$
 & $\begin{array}{ll}
    - 
   \vspace{0.10cm} \\
    -
   \vspace{0.10cm} \\
    \textcolor{gray}{{\cal O}(\alpha_s^4) \equiv \mbox{NNLO}}
   \vspace{-0.10cm} \\
    \hspace{1.00cm} \textcolor{gray}{\vdots}
   \end{array}$
 \\
 \hline
  \end{tabular}
 \caption{Perturbative QCD expansion of NRQCD initial-scale inputs for parton to color-singlet pseudoscalar (green, first line) and vector (violet, second line) quarkonium fragmentation. All channels, if nonzero, have been calculated within ${\cal O}(\alpha_s^3) \equiv \mbox{NLO}$ accuracy. They are unknown (gray) starting from  ${\cal O}(\alpha_s^4) \equiv \mbox{NNLO}$.}
 \label{tab:FF_order}
 \end{center}
 \end{table*}

The fragmentation of heavy hadrons presents a richer and more nuanced structure compared to that of their light counterparts.
This added complexity arises from the fact that heavy quark masses fall within the perturbative regime of QCD when considering their lowest Fock components.
As a result, unlike light-hadron FFs---which are fully nonperturbative at the starting scale---FFs for heavy hadrons inherently involve both perturbative inputs and nonperturbative modeling.

For singly heavy systems such as $D$ and $B$ mesons or ${\rm \Lambda}_Q$ baryons, the fragmentation mechanism is typically separated into two well-defined stages~\cite{Cacciari:1996wr,Cacciari:1993mq,Jaffe:1993ie,Kniehl:2005mk,Helenius:2018uul,Helenius:2023wkn}.
In the first, a parton $a$ generated with large transverse momentum during the hard scattering evolves into a heavy quark $Q$.
Because the strong coupling at the heavy quark scale remains moderate, $\alpha_s(m_Q) < 1$, this transition can be addressed using perturbative QCD.

This perturbative contribution is described by short-distance coefficients (SDCs) for the [$a \to Q$] transition, with $a$ being the fragmenting parton.
SDCs capture the perturbative dynamics at timescales much shorter than hadronization.
NLO determinations of these coefficients were provided in Ref.~\cite{Mele:1990yq,Mele:1990cw}, and were later extended to higher accuracy in a series of works~\cite{Rijken:1996vr,Mitov:2006wy,Blumlein:2006rr,Melnikov:2004bm,Mitov:2004du,Biello:2024zti}.
Very recently, single- and double-parton fragmentation has been investigated using a soft-gluon factorization approach~\cite{Jia:2024cvv,Saleev:2025ryh}.

The second stage, occurring at longer timescales, is governed by nonperturbative QCD and accounts for the hadronization of the heavy quark into a physical hadron.
This phase is modeled either through empirical FFs inspired by fits to data~\cite{Kartvelishvili:1977pi,Bowler:1981sb,Peterson:1982ak,Andersson:1983jt,Collins:1984ms,Colangelo:1992kh}, or within the language of effective field theories, such as Heavy Quark Effective Theory (HQET)~\cite{Georgi:1990um,Eichten:1989zv,Grinstein:1992ss,Neubert:1993mb,Jaffe:1993ie}.

Mathematically, the fragmentation of a parton $a$ into a singly heavy-flavored hadron $\HQ$ at the initial scale $\mu_{F,0} \sim m_Q$, where $m_Q$ is the heavy quark mass, is described by~\cite{Cacciari:1996wr,Cacciari:1997du}
\begin{equation}
\label{FFs_HF_initial}
D_a^{\HQ} (z, \mu_{F,0}) =
\int_z^1 \frac{\drv \zeta}{\zeta} D_a^Q (\zeta, \mu_{F,0}) \, D_{\rm [np]}^{\HQ} \left( \frac{z}{\zeta} \right) \;.
\end{equation}
In this expression, $D_a^Q$ denotes the perturbative component of the initial-scale FF, namely the SDC associated with a massless parton fragmenting into a massive heavy quark $Q$ via a perturbative QCD cascade.
The function $D_{\rm [np]}^{\HQ}$, by contrast, represents the nonperturbative fragmentation into the physical hadron.
It is assumed to be universal, independent of the initiating parton flavor, and scale-independent with respect to $\mu_{F,0}$.

To obtain a fully evolved set of FFs within a VFNS, it is essential to incorporate energy-scale dependence.
Assuming the initial conditions are free from scaling violations, the FFs are evolved through the DGLAP timelike equations, at the appropriate perturbative order, to generate predictions across a wide kinematic range.

We now turn to quarkonia, mesons whose lowest Fock level is $|Q\bar{Q}\rangle$.
Charmed $B$ mesons, having as lowest Fock states $|c\bar{b}\rangle$ or $|\bar{c}b\rangle$, are sometimes classified as generalized quarkonium states.
The presence of two heavy quarks makes their formation more complex than that of heavy-light hadrons.
The earliest model proposed was the color evaporation model (CEM)~\cite{Fritzsch:1977ay,Halzen:1977rs}, which assumes that the color of the produced [$Q\bar{Q}$] pair is fully randomized before hadronization.
The CEM cannot describe polarization or relative production rates of distinct states, such as $\Jpsi$ versus $\chi_c$, limiting its predictive power~\cite{Beneke:1996yw,Beneke:1998re,Lansberg:2005aw,Lansberg:2006dh,Brambilla:2010cs}.

A more structured approach is the color-singlet mechanism~\cite{Berger:1980ni,Baier:1981uk}, where hadronization preserves both color and spin.
The $|Q\bar{Q}\rangle$ system must be produced directly in a color-singlet state with the same quantum numbers as the final quarkonium.
Assuming $m_{\cal Q} \approx 2m_Q$, one adopts a static approximation, treating the quarks as at rest in the meson frame.
For $S$-wave states, the only nonperturbative input is the wave function at the origin, ${\cal R}_{\cal Q}(0)$, while for $P$-wave states one uses its derivative, ${\cal R}_{\cal Q}^\prime(0)$.
These are typically extracted from leptonic decay widths.
However, infrared divergences arise at NLO in $P$-wave channels~\cite{Barbieri:1976fp,Bodwin:1992ye}, pointing to the limitations of the color-singlet mechanism.

Ref.~\cite{Bodwin:1992ye} showed that these divergences are canceled by matching singularities in color-octet matrix elements, demonstrating that octet contributions are essential for consistency.
This led to the formulation of NRQCD, an effective theory that incorporates both singlet and octet channels~\cite{Caswell:1985ui,Thacker:1990bm,Bodwin:1994jh,Cho:1995vh,Cho:1995ce,Leibovich:1996pa,Bodwin:2005hm}---see also Refs.~\cite{Grinstein:1998xb,Kramer:2001hh,QuarkoniumWorkingGroup:2004kpm,Pineda:2011dg} for reviews.
Within the NRQCD framework, the quarkonium is expressed as a linear combination of Fock states, ordered via a double expansion in $\alpha_s$ and the relative velocity $v_{\cal Q}$ of the quark pair.

NRQCD offers a systematic approach to disentangle short-distance and long-distance contributions in quarkonium production.
Treating heavy quark and antiquark fields as nonrelativistic in the effective Lagrangian allows a consistent factorization between perturbative SDCs, which describe the production of the $|Q\bar{Q}\rangle$ intermediate state, and long-distance matrix elements (LDMEs), which encode the hadronization process.
LDMEs are nonperturbative and must be extracted from experimental data, computed via potential models~\cite{Eichten:1994gt}, or obtained from lattice QCD~\cite{Lepage:1992tx,Davies:1994mp}.

NRQCD was originally designed under the assumption that quarkonium is formed through a hard, \emph{short-distance} production of a $|Q\bar{Q}\rangle$ state, followed by nonperturbative hadronization.
As mentioned in the Introduction (Section~\ref{sec:intro}), the pair is generated with a transverse separation of order $1/Q$, where $Q$ identifies a characteristic energy scale of the considered process~\cite{Mangano:1995yd}.
In the high-$|\vec q_T|$ regime, $Q \sim |\vec q_T|$, so the time and volume available for hadronization shrink as $1/|\vec q_T|$ and $1/|\vec q_T|^3$, respectively.
This reduces the amplitude for direct formation of the physical state~\cite{Mangano:1995yd,Braaten:1996pv,Artoisenet:2009zwa}.

This suppression is compensated by the emergence of the \emph{fragmentation} mechanism.
At large $|\vec q_T|$, a single high-energy parton produced in the hard subprocess can fragment into a quarkonium state plus additional hadronic radiation.
Though formally starting at higher order in $\alpha_s$, the fragmentation contribution scales like $(|\vec q_T|/m_Q)^2$ and dominates at high energies~\cite{Braaten:1993rw,Kuhn:1981jy,Kuhn:1981jn,Cacciari:1994dr,Braaten:1994xb}.
Adapting NRQCD to this context enabled LO calculations of gluon and charm FFs to $S$-wave color-singlet charmonia in Refs.~\cite{Braaten:1993rw,Braaten:1993mp}, later extended to $P$-wave states~\cite{Braaten:1994kd,Ma:1995ci,Yuan:1994hn}.

Early phenomenological analyses on the transition between short-distance and fragmentation regimes focused mainly on charmonium production~\cite{Cacciari:1994dr,Roy:1994ie,Cacciari:1995yt,Cacciari:1996dg,Lansberg:2019adr}.
These works showed that gluon fragmentation becomes dominant for transverse momenta above $10 \div 15$~GeV.
A similar threshold was identified for charmed $B$ mesons~\cite{Kolodziej:1995nv}, though later analyses~\cite{Artoisenet:2007xi} suggest the onset may occur at even higher $|\vec q_T|$ values.

Because fragmentation is rooted in collinear factorization, one must consistently connect NRQCD to the QCD fragmentation correlator formalism.
Recent developments leverage NRQCD as a robust model for setting initial conditions of FFs at the scale $\mu_{F,0}$~\cite{Kang:2011mg,Ma:2013yla,Ma:2014eja}, enabling a unified and predictive framework for heavy-flavor fragmentation to quarkonia.

This approach offers two main benefits.
On the one side, NRQCD enables the factorization of initial FF inputs into a convolution of perturbative SDCs and nonperturbative LDMEs, paralleling the treatment of singly heavy-flavored hadrons (see Eq.~\eqref{FFs_HF_initial}).
On the other side, it provides a robust framework for calculating SDCs and interpreting LDMEs physically.
From these inputs, VFNS quarkonium FFs can be constructed via DGLAP evolution.
Recent applications of VFNS fragmentation to $\psi(2S)$ production, including the impact of relativistic corrections, have further confirmed its phenomenological robustness~\cite{Bertone:2025jex}.

The {\tt ZCW19$^+$}~\cite{Celiberto:2022dyf,Celiberto:2023fzz} and {\tt ZCFW22}~\cite{Celiberto:2022keu,Celiberto:2024omj} sets represent the first attempt at determining VFNS collinear FFs for vector quarkonia ($\Jpsi$, $\Yps$) and charmed $B$ mesons ($\BCs$, $\Bss$), based on NLO NRQCD calculations for gluon and heavy quark channels~\cite{Braaten:1993rw,Chang:1992bb,Braaten:1993jn,Ma:1994zt,Zheng:2019gnb,Zheng:2021sdo,Feng:2021qjm,Feng:2018ulg}.
These studies provided supporting evidence that gluon FFs with smooth, nondecreasing $\mu_F$ dependence serve as effective stabilizers for high-energy resummed cross sections in semi-inclusive hadroproductions.
This striking feature, observed for both singly and multiply heavy-flavored bound states, is referred to as the \emph{natural stability} of high-energy resummation~\cite{Celiberto:2022grc}.

Beyond the phenomenology of the high-energy resummation sector, FFs for (generalized) quarkonium states serve as powerful tools for precision investigations of collinear dynamics and hadronization.
A notable example is provided by the recent analysis in Ref.~\cite{Celiberto:2024omj}, where rapidity and transverse momentum distributions of charmed $B$ mesons, described using the {\tt ZCFW22} FF set, were studied in detail.
That study confirmed the estimate by LHCb~\cite{LHCb:2014iah,LHCb:2016qpe} that the production rate hierarchy between $\BCs$ mesons and singly bottomed $B$ mesons remains below the 0.1\% level.

This served as a simultaneous benchmark for both the hybrid factorization scheme and the NRQCD-based fragmentation framework applied to charmed $B$ mesons at our reference transverse masses.
It reinforced the robustness of using leading-power NRQCD initial-scale inputs, which are subsequently evolved to higher energy scales via DGLAP equations.

From studies performed using {\tt ZCW19$^+$} and {\tt ZCFW22} functions, it became evident that a consistent framework is needed to integrate NRQCD-based inputs within collinear factorization, ensuring the correct implementation of DGLAP evolution thresholds across all partonic fragmentation channels.

To meet this requirement, the novel {\HFNRevo} scheme was developed~\cite{Celiberto:2024mex,Celiberto:2024bxu,Celiberto:2024rxa,Celiberto:2025xvy}.
This framework is specifically designed to describe the DGLAP evolution of heavy-hadron FFs starting from nonrelativistic inputs.

The {\HFNRevo} methodology relies on three essential components: \emph{interpretation}, \emph{evolution}, and \emph{uncertainties}.

The first component enables a reinterpretation of the short-distance mechanism---dominant at low transverse momentum---as a fixed-flavor number scheme (FFNS) two-parton fragmentation process that goes beyond the leading-power approximation (see Refs.~\cite{Alekhin:2009ni,Fleming:2012wy} for further details).
This picture is supported by the observation that introducing transverse momentum dependence reveals distinct singularity structures, particularly in the transition region between low-$|\vec q_T|$ shape functions~\cite{Echevarria:2019ynx} and large-$|\vec q_T|$ FFs~\cite{Boer:2023zit}.

The second component concerns the implementation of DGLAP evolution within {\HFNRevo}, which proceeds in two stages.
First, an \emph{expanded}, semianalytical, and \emph{decoupled} evolution module ({\tt EDevo}) handles threshold matching for all partonic channels.
Second, the standard \emph{all-order} evolution ({\tt AOevo}) is numerically activated, providing a seamless transition to the higher-scale regime.

The third component addresses the computation of perturbative and nonperturbative uncertainties. 
The former correspond to missing higher-order uncertainties (MHOUs), stemming from scale variations associated with threshold settings in the evolution process. 
The latter arise from the genuinely nonperturbative part of the fragmentation, namely the LDMEs in our case.

By leveraging a subset of the core features of the {\HFNRevo} scheme, pioneering determinations of VFNS collinear FFs have been achieved for a variety of exotic heavy-flavored systems.
These include the {\tt TQHL1.x} and {\tt TQ4Q1.x}\footnote{Recent studies suggest that NRQCD factorization may describe double $\Jpsi$ excitations~\cite{LHCb:2020bwg,ATLAS:2023bft,CMS:2023owd} as fully charmed tetraquarks~\cite{Zhang:2020hoh,Zhu:2020xni}.
Initial NRQCD inputs for $\TQc$ FFs were computed in Refs.~\cite{Feng:2020riv,Bai:2024ezn,Bai:2024flh}.} sets for doubly heavy and fully heavy tetraquarks~\cite{Celiberto:2023rzw,Celiberto:2024mrq,Celiberto:2024beg,Celiberto:2025dfe,Celiberto:2025ziy}, the {\tt PQ5Q1.0} set for $S$-wave pentacharm baryons~\cite{Celiberto:2025ipt}, and the {\tt OMG3Q1.0} set for rare triply charmed $\Omega$ baryons in the heavy-flavor Omega sector~\cite{Celiberto:2024mex}.
These results mark a foundational step toward a unified framework for the collinear fragmentation of exotic hadrons, built upon nonrelativistic inputs and evolved through threshold-matched DGLAP dynamics.

In the next section (Section~\ref{ssec:NRFF}), we present in detail our strategy for constructing the {\NRFF} functions for pseudoscalar quarkonia.
In the final part (Section~\ref{ssec:HCFF}), we discuss the main features of the {\HCFF} functions for singly charmed hadrons.

\subsection{\NRFF: NRQCD FFs for pseudoscalars}
\label{ssec:NRFF}

First and pioneering determinations of collinear FFs for doubly heavy-flavored mesons were obtained by comparing NRQCD-based versus collinear factorization-inspired formul{\ae} of cross sections featuring the emission of that meson~\cite{Chang:1992bb,Chen:1993ii}.
A correlatorlike definition of doubly heavy-flavored meson FFs in terms of nonlocal, gauge-invariant operators~\cite{Collins:1981uw} was provided a couple of years later~\cite{Ma:1994zt}. Making use of fragmentation correlator significantly eases the extension to higher perturbative orders.

According to NRQCD, the function describing the fragmentation process of a parton $a$ into a physical quarkonium state $\cal Q$ with momentum fraction $z$ takes a simple and intuitive form
\begin{equation}
 \label{FFs_NRQCD}
 D^{\cal Q}_a(z, \mu_{F,0}) = \sum_{[n]} {\cal D}^{\cal Q}_{a}(z, [n]) \langle {\cal O}^{\cal Q}([n]) \rangle \;.
\end{equation}
This expression can be expanded in powers of $\alpha_s$ and contains DGLAP-type logarithms to be subsequently resummed.
Then, $\langle {\cal O}^{\cal Q}([n]) \rangle$ stands for the nonperturbative NRQCD LDME, which embodies suppression factors in the relative velocity $v_{\cal Q}$ of the [$Q \bar Q$ system.
LDMEs have to be extracted through global fits or estimated via
lattice or potential-model QCD techniques.
Finally, $[n] \equiv \,^{2S+1}L_J^{(c)}$ denotes the set of quarkonium quantum numbers in the spectroscopic notation, with the $(c)$ superscript referring to the color state, singlet (1) or octet (8).
From Eq.~\eqref{FFs_NRQCD} the two founding assumptions of NRQCD emerge.
First, all possible Fock states contribute to the physical quarkonium by means of a linear superposition given by the sum over quantum numbers $[n]$.
Second, all terms contributing to the formation mechanism are organized as a double expansion in powers of $\alpha_s$ and $v_{\cal Q}$.

A close comparison between Eq.~\eqref{FFs_HF_initial} and Eq.~\eqref{FFs_NRQCD} highlights both structural analogies and fundamental differences in the treatment of heavy-flavored hadron and quarkonium fragmentation.
In the case of singly heavy-flavored hadrons $\HQ$, the [$a \to \HQ$] FF at the initial scale $\mu_{F,0}$ is expressed as a convolution between a perturbatively calculable FF, $D_a^Q(\zeta, \mu_{F,0})$, and a nonperturbative transition function, $D_{\rm [np]}^{\HQ}(z/\zeta)$, which models the long-distance hadronization of a heavy quark $Q$ into the hadron $\HQ$.
This convolution over the light cone momentum fraction $\zeta$ reflects the fact that the nonperturbative transition is $z$-dependent, encoding the probability for a heavy quark with momentum fraction $\zeta$ to hadronize into a final hadron carrying a fraction $z$ of the momentum of the fragmenting parton.

Conversely, in the NRQCD factorization approach for quarkonium, the $D_a^{\cal Q}(z, [n])$ FF is decomposed as a sum over Fock states $[n]$, where each term is a product of a SDC function, $\mathcal{D}_a^{Q \bar Q}(z, [n])$, and a corresponding LDME, $\langle \mathcal{O}^{\cal Q}([n]) \rangle$.
This structure arises from the fact that in NRQCD, the transition from the perturbative [$Q\bar Q$] pair to the bound quarkonium state is localized in momentum space, and governed by a finite number of hadronic matrix elements with no explicit $z$ dependence.
In other words, the LDMEs act as multiplicative constants that parametrize the nonperturbative probabilities of hadronization for each Fock state $[n]$ independently of the momentum fraction $z$.

This conceptual difference has a direct physical interpretation: in singly heavy-flavored hadron production, the heavy quark combines with soft, light constituents from the vacuum, and the energy-momentum sharing is dynamically distributed, hence the need for a convolution.
In contrast, quarkonia are composed entirely of heavy constituents produced in the short-distance process---the long-distance transition only involves spin and color rearrangements, not momentum reshuffling.
Therefore, the nonperturbative component of the fragmentation probability into a quarkonium depends only on the internal quantum numbers and not on the longitudinal momentum fraction, resulting in LDMEs being pure numbers.

Figures~\ref{fig:PSQ_FF_diagrams} and~\ref{fig:VQ_FF_diagrams}, together with Table~\ref{tab:FF_order}, provide a visual and analytic summary of the leading representative Feynman diagrams and the perturbative QCD expansion of the initial-scale FFS for color-singlet quarkonium states in the NRQCD formalism. 
All partonic channels are divided into three main categories: gluon [$g \to {\cal Q}$], constituent heavy quark [$Q \to {\cal Q}$], and nonconstituent (light or heavy, [$\tilde{q} \mbox{ or } \tilde{Q} \to {\cal Q}$]) quark fragmentation.
The figures show representative LO diagrams for each channel, while the table summarizes the perturbative order at which each channel contributes.

For comparison, we also show the case of vector quarkonia ${\cal V}_Q \, [^3S_1^{(1)}$] such as $\Jpsi$ or $\Yps$ (Fig.~\ref{fig:VQ_FF_diagrams} and the second row of Table~\ref{tab:FF_order}), even though in this work we focus exclusively on pseudoscalar quarkonia ${\cal S}^-_Q$ such as $\etc$ or $\etb$ (Fig.~\ref{fig:PSQ_FF_diagrams} and the first row of Table~\ref{tab:FF_order}).
As a general convention, we classify as LO the lowest nonvanishing term in the perturbative expansion for each channel, namely, ${\cal O}(\alpha_s^2)$ for the gluon and heavy quark (constituent) initiated channels in the case of pseudoscalar quarkonia ${\cal S}^-_Q \, [^1S_0^{(1)}$], and for the heavy quark initiated channel only in the vector case ${\cal V}_Q \, [^3S_1^{(1)}$]. 

The key difference lies in the perturbative accessibility of the various channels.
For pseudoscalars, all three channels are active within NLO: [$g \to \mathcal{Q}$] and [$Q \to \mathcal{Q}$] at $\mathcal{O}(\alpha_s^2)$, and [$\tilde{q} \mbox{ or }  \tilde{Q} \to \mathcal{Q}$] at $\mathcal{O}(\alpha_s^3)$. 
In contrast, for vector quarkonia, the latter channel starts at next-to-NLO (NNLO), $\mathcal{O}(\alpha_s^4)$, beyond our working accuracy.
This feature makes pseudoscalar quarkonia especially suitable for testing the {\HFNRevo} framework in its full structure, since all relevant channels can be consistently included at the initial scale. 
In particular, all FF channels receive initial-scale inputs from NRQCD, rather than being generated only through DGLAP evolution.

Finally, we note that the higher-order activation of the gluon nonconstituent quark channels for vector quarkonia is a consequence of the Landau-Yang selection rule~\cite{Landau:1948kw,Yang:1950rg}. 
This rule forbids the coupling of a vector particle to two on-shell massless vector bosons, implying that an additional gluon emission is required to make the process nonvanishing. 
As a result, the leading diagrams appear at higher order in $\alpha_s$ compared to the pseudoscalar case.

In this section we will make use of NRQCD calculations for gluon (Section~\ref{sssec:NRFF_g}), constituent heavy quark (Section~\ref{sssec:NRFF_Q}), and nonconstituent quark (Section~\ref{sssec:NRFF_q}) channels as \emph{initial-scale} inputs to build the novel NLO DGLAP-evolving FFs for pseudoscalar quarkonium mesons.
In Section~\ref{sssec:NRFF_ns} we will discuss the DGLAP evolution of these functions from {\HFNRevo}.

From this point onward in the text, we fix the values of the heavy‑quark masses to 
$m_c = 1.27\mbox{ GeV}$ and $m_b = 4.18\mbox{ GeV}$, 
as commonly adopted in modern $\MSb$ determinations~\cite{ParticleDataGroup:2024cfk}.

\subsubsection{LDMEs}
\label{sssec:LDMEs}

The nonperturbative LDMEs entering the NRQCD factorization formalism play a key role in determining the hadronization probability of a heavy quark--antiquark pair into a physical quarkonium state. 
Although LDMEs are typically unknown quantities that must be extracted from experimental data or computed using nonperturbative methods such as lattice QCD or potential models, certain assumptions can simplify their structure.

One such simplification arises from the so-called \emph{vacuum saturation approximation} (VSA)~\cite{Shifman:1978bx,Gilman:1979bc,Bodwin:1994jh}, which assumes that contributions from intermediate states other than the vacuum are suppressed by powers of the relative velocity $v_{\cal Q}$ of the heavy quarks. 
Under this approximation, the matrix element for the transition into a quarkonium ${\cal Q}$ reduces to the squared matrix element between the vacuum and the lowest-order Fock state, $|Q\bar Q\rangle$. 
Symbolically, the VSA reads
\begin{equation}
\label{VSA}
 \hspace{-0.15cm}
 \langle 0 | \chi^\dagger \kappa_n \psi \, \mathcal{P}_{\cal Q} \, \psi^\dagger \kappa_n^\prime \chi | 0 \rangle 
 \simeq 
 \langle 0 | \chi^\dagger \kappa_n \psi | {\cal Q} \rangle \langle {\cal Q} | \psi^\dagger \kappa_n^\prime \chi | 0 \rangle 
\end{equation}
where $\kappa_n$ and $\kappa_n^\prime$ are spin-color matrices projecting onto the relevant NRQCD state, and $\mathcal{P}_{\cal Q}$ is the projector onto the physical quarkonium ${\cal Q}$.
Here, $\chi$ and $\psi$ are the nonrelativistic Pauli spinor fields that annihilate a heavy antiquark and create a heavy quark, respectively, in the NRQCD effective theory.

For pseudoscalar $S$-wave quarkonium states such as $\etc$ and $\etb$ mesons, the relevant matrix element corresponds to a [$^1S_0^{(1)}$] configuration. 
In this case, the VSA leads to a direct relation between the LDME and the radial wave function at the origin, ${\cal R}_{{\cal S}_Q^-}(0)$, namely
\begin{equation}
\label{LDMEs_radialWF}
\langle \mathcal{O}^{{\cal S}_Q^-} [^1S_0^{(1)}] \rangle = \frac{N_c}{2\pi} \, |{\cal R}_{{\cal S}_Q^-}(0)|^2 \;,
\end{equation}
with $N_c = 3$ representing the number of colors in QCD.
This matching coefficient arises from the color and spin projection of the singlet state and from the normalization conventions used in NRQCD.

The factor $3$ stems from the dimension of the color space (number of color polarizations for a singlet), while the $2\pi$ arises from phase-space integration and amplitude normalization. This compact formula encapsulates the nonperturbative content of the pseudoscalar quarkonium NRQCD fragmentation, and allows one to express the full LDME in terms of a well-defined, physically meaningful wave function parameter.

While color-octet transitions play a crucial role in the production of vector quarkonia, particularly in the high-$|\vec q_T|$ regime of $\Jpsi$ hadroproduction, their impact in the case of pseudoscalar quarkonia is predicted to be strongly suppressed.
This suppression originates from spin and parity selection rules in NRQCD, which disfavor contributions from channels such as ${}^3S_1^{(8)}$ and ${}^3P_J^{(8)}$ for ${\cal S}_Q^- = \etc, \etb$ states.
As a result, the dominant hadronization mechanism for pseudoscalar quarkonia proceeds via the color-singlet configuration ${}^1S_0^{(1)}$, as captured by Eq.~\eqref{LDMEs_radialWF}.

This theoretical expectation is supported by recent LHCb measurements~\cite{LHCb:2014oii,LHCb:2019zaj}, which are well described by singlet-only calculations, and has motivated our choice to focus exclusively on the color-singlet channel in this study.
A more comprehensive discussion of the so-called quarkonium production puzzle, including its phenomenological implications and relevance for vector versus pseudoscalar states, is provided in Appendix~\hyperlink{app:onium_puzzle}{A}.

A few remarks are in order regarding the scale dependence of LDMEs and their interplay with DGLAP evolution.
In the NRQCD framework, the LDMEs are defined at an intermediate scale $\mu_\Lambda$, characteristic of the binding dynamics of the heavy quark--antiquark pair, and typically of the order of the relative momentum $m_Q v_{\cal Q}$ inside the quarkonium. 
These matrix elements are assumed to encapsulate all soft contributions from lower energy scales, and their scale dependence is formally governed by NRQCD-specific renormalization group equations~\cite{Bodwin:1994jh,Brambilla:2010cs}.

However, in most phenomenological applications, LDMEs are treated as static, process-independent parameters, either extracted from data or estimated via potential models, implicitly assuming that the evolution from infrared scales up to $\mu_\Lambda$ has already been resummed into their fitted value. 
This practical treatment simplifies the use of NRQCD while maintaining consistency with factorization.

Such separation of scales ensures that the DGLAP evolution of FFs performed in our formalism above $\mu_\Lambda$ remains theoretically sound. 
The collinear evolution acts on the FF $D_{a \to {\cal Q}}(x,\mu_F)$ as a whole, while the LDMEs enter as fixed nonperturbative coefficients in the NRQCD matching performed at the scale $\mu_\Lambda$. 

As a result, our use of DGLAP equations to evolve the FFs starting at scales above $\mu_\Lambda$ toward the final factorization scale $\mu_F$ is fully compatible with the NRQCD approach and does not entail any implicit double evolution of the LDMEs themselves. 
This distinction, while sometimes overlooked, is crucial to correctly interpret the role and scale dependence of nonperturbative inputs in quarkonium fragmentation~\cite{Bodwin:1994jh}.

In our numerical analysis, we employ specific values for the color-singlet LDMEs entering Eq.~\eqref{LDMEs_radialWF}, corresponding to the pseudoscalar quarkonium states ${\cal S}_Q^-$ considered in this work.
These have been estimated using a combination of potential NRQCD models and inputs from previous phenomenological studies.

For the charmonium state $\etc$, we adopt the central value
\begin{equation}
\label{R0_etc}
 |{\cal R}_{\etc}(0)|^2 = (0.810 \pm 0.052)\mbox{ GeV}^3 \;,
\end{equation}
derived from potential-model calculations consistent with the Buchm{\"u}ller-Tye~\cite{Buchmuller:1980su} and Eichten-Quigg~\cite{Eichten:1995ch} frameworks, and cross-checked against values used in global NRQCD fits, such as the one by Butenschoen and Kniehl~\cite{Butenschoen:2011yh}.
The quoted uncertainty, approximately $6.45\%$, reflects the spread among different models and is propagated linearly to the LDME via Eq.~\eqref{LDMEs_radialWF}.

For the bottomonium state $\etb$, we use the analogous potential-model estimate
\begin{equation}
\label{R0_etb}
 |{\cal R}_{\etb}(0)|^2 = (6.477 \pm 0.836)\mbox{ GeV}^3 \;,
\end{equation}
where the relative uncertainty has been conservatively set to twice that of the $\etc$ case, in light of the absence of direct fits for bottomonium pseudoscalar LDMEs.
This choice ensures a realistic estimate of the theoretical systematics affecting the normalization of the FFs, and has been validated by cross-checks against predictions based on the Cornell potential~\cite{Eichten:1995ch}.

We note that the extraction of the wave function at the origin from leptonic decay widths is highly sensitive to higher-order QCD corrections, with recent studies showing large variations between NLO and NNLO estimates~\cite{ColpaniSerri:2021bla}. 
Although the analysis in Ref.~\cite{ColpaniSerri:2021bla} concerns vector quarkonia, the underlying issue extends to pseudoscalar states as well, since potential-model approaches commonly assume identical values of the wave function for spin-singlet and spin-triplet $S$-wave quarkonia.
Notably, the extracted value can nearly double at each successive perturbative order, reflecting substantial scale uncertainties and renormalization scheme dependence. 

Our analysis, however, is consistently performed at NLO in both the NRQCD matching and the DGLAP evolution. 
Therefore, we accordingly employ input values of $|{\cal R}_{\etQ}(0)|^2$ consistent with this level of accuracy, and adopt conservative uncertainties to account for model dependence, thus protecting ourselves from precision issues associated with higher-order extractions.

\subsubsection{Gluon channel}
\label{sssec:NRFF_g}

The [$g \to \etQ$] SDC in the $^1S_1^{(0)}$ quantum configuration (see Fig.~\ref{fig:PSQ_FF_diagrams}, left diagram) reads
\begin{equation}
\begin{split}
\label{SDC_SQ_g}
 {\cal D}^{\cal Q}_{g}(z, ^{1\!\!}S^{(0)}) \,&=\, 
 \as^2 \, \drv_{g}^{\rm [LO]} (z, ^{1\!\!}S^{(0)}) 
 \\[0.10cm]
 \,&+\,
 \as^3 \, \drv_{g}^{\rm [NLO]} (z, ^{1\!\!}S_1^{(0)})
 \;,
\end{split}
\end{equation}
where the QCD running coupling $\alpha_s(\mu_R \equiv \mu_{R,0})$ is evaluated at $\mu_{R,0} = 2 m_Q$, corresponding to the minimal invariant mass of the gluon in the LO fragmentation process~\cite{Braaten:1993rw,Artoisenet:2014lpa,Zhang:2018mlo}.
The LO SDC, first derived in Ref.~\cite{Braaten:1993rw}, is
\begin{equation}
\begin{split}
\label{SDC_SQ_g_LO}
 \drv_{g}^{\rm [LO]} (z, ^{1\!\!}S_0^{(1)}) =
 \frac{3z - 2z^2 + 2(1 - z)\ln(l - z)}{36 m_Q^3}
 \;.
\end{split}
\end{equation}

The NLO correction has been computed in Refs.~\cite{Artoisenet:2014lpa,Zhang:2018mlo}.
Reference~\cite{Artoisenet:2014lpa} offers a comprehensive discussion of the subtraction and matching procedures.
In turn, Ref.~\cite{Zhang:2018mlo} presents a fully equivalent semianalytical result, incorporating both real and virtual corrections.
For practical purposes, in the {\symJethad} implementation we have adopted the semi analytical expression of Ref.~\cite{Zhang:2018mlo}.
This choice enables a numerically stable and efficient evaluation of the NLO correction across the full $z$ range, avoiding spurious singularities near the end points.

We write
\begin{equation}
\begin{split}
\label{SDC_SQ_g_NLO}
 \drv_{g}^{\rm [NLO]} (z, ^{1\!\!}S_0^{(1)}) =
 \frac{1}{6\pi\, m_Q^3} \left[\right. 
 &\drv_{g, \mathrm{reg}}^{\rm [NLO]} (z, ^{1\!\!}S_0^{(1)}) 
 \\[0.10cm]
 +\; 
 &\drv_{g, \mathrm{dist}}^{\rm [NLO]} (z, ^{1\!\!}S_0^{(1)}) 
 \left.\right]
 \;.
\end{split}
\end{equation}

The NLO correction splits into a regular part, $\drv_{g, \mathrm{reg}}^{\rm [NLO]}$, which captures the resolved emission dynamics away from the [$z \to 1$] end point, and a distributional part, $\drv_{g, \mathrm{dist}}^{\rm [NLO]}$, containing plus-distributions and $\delta(1 - z)$ terms. 
The latter ensure the cancellation of infrared divergences and maintain consistency with DGLAP evolution.
As presented in Ref.~\cite{Zhang:2018mlo}, both components are expressed in a fully regularized form, suitable for direct numerical evaluation.
For completeness, we report their expressions in Appendix~\hyperlink{app:nlo_gluon_SDC}{B}.

The full expression for our initial-scale gluon FF is readily obtained by combining Eqs.~\eqref{FFs_NRQCD}, \eqref{SDC_SQ_g_LO}, and~\eqref{SDC_SQ_g_NLO}.

To benchmark the numerical implementation of our gluon SDC and facilitate a direct comparison with the results of Refs.~\cite{Artoisenet:2014lpa,Zhang:2018mlo}, we adopt the same strategy as employed in those studies.
Specifically, we apply a simplified, diagonal, timelike DGLAP evolution at NLO, using only the LO Altarelli-Parisi kernel $P_{gg}(z)$ for the gluon FF.

\begin{figure*}[!t]
\centering

 \includegraphics[scale=0.640,clip]{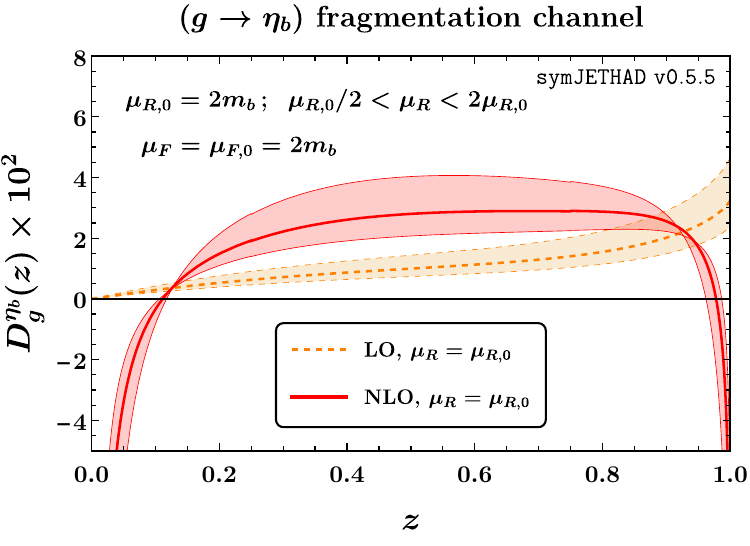}
 \hspace{0.50cm}
 \includegraphics[scale=0.640,clip]{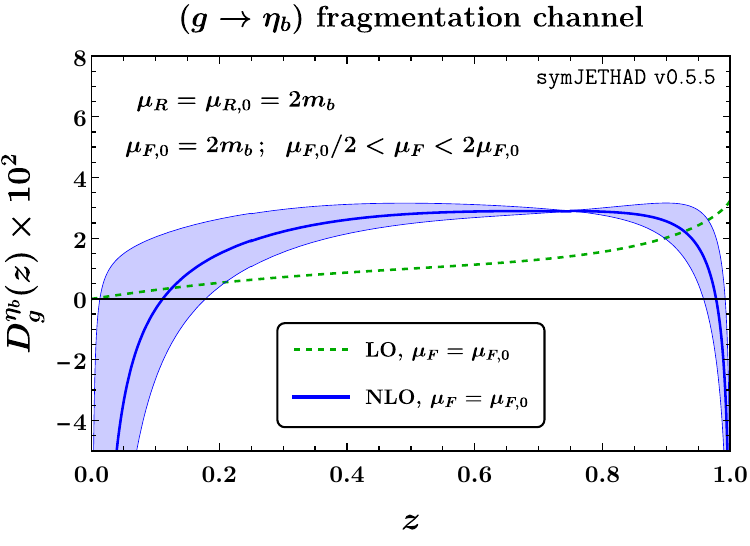}

\caption{NRQCD [$g \to \etb$] fragmentation channel at LO and NLO, for $\mu_{R,0}=\mu_{F,0}=2m_b$. 
Left plot: $\mu_R$ runs from $\mu_{R,0}/2$ to $2\mu_{R,0}$, while $\mu_F$ is fixed at $\mu_{F,0}$.
Right plot: $\mu_R$ is fixed at $\mu_{R,0}$, while $\mu_F$ runs from $\mu_{F,0}/2$ to $2\mu_{F,0}$.}
\label{fig:FF_g-to-etb_muRF-var}
\end{figure*}

For illustration purposes, Fig.~\ref{fig:FF_g-to-etb_muRF-var} shows the [$g \to \etb$] FF at LO and NLO, for different choices of factorization and renormalization scales. 
The left plot corresponds to $\mu_F = \mu_{F,0}$ with $\mu_R$ varied in the range $\mu_{R,0}/2$ to $2\mu_{R,0}$, while in the right plot $\mu_R$ is kept fixed at $\mu_{R,0}$ and $\mu_F$ is varied.

As mentioned earlier, we set the central renormalization and factorization scales to $\mu_{R,0} = \mu_{F,0} = 2m_Q$, corresponding to the minimal invariant mass of the gluon and the final heavy quark pair, respectively, in the LO fragmentation process~\cite{Braaten:1993rw,Artoisenet:2014lpa,Zhang:2018mlo}.

Here, for simplicity, the wave function at the origin is fixed to its central value, and uncertainty variations are not considered (see Section~\ref{sssec:LDMEs}).

We find that the behavior illustrated in Fig.~\ref{fig:FF_g-to-etb_muRF-var} is consistent with the results reported in Figs.~3 and~4 of Ref.~\cite{Artoisenet:2014lpa}, and in Figs.~8 and~10 of Ref.~\cite{Zhang:2018mlo}.

The plots in Fig.~\ref{fig:FF_g-to-etb_muRF-var} highlight several characteristic features of the gluon FFs into pseudoscalar quarkonia. 
At LO, the FF steadily increases with $z$ and reaches its maximum near the end point, [$z \to 1$]. 
This behavior is consistent with the original observation of Ref.~\cite{Braaten:1993rw}, where the LO color-singlet [$g \to \eta_{Q}$] FF exhibits a monotonic rise toward $z=1$ without vanishing. 
At NLO, the situation changes dramatically: the FF becomes negative in large regions of $z$, and develops logarithmic divergences both at [$z \to 0$] and [$z \to 1$].

These features are well known and have been discussed in Refs.~\cite{Artoisenet:2014lpa,Zhang:2018mlo}, where the divergence at small $z$ is shown to be harmless, as physical cross sections are obtained through convolution with partonic hard parts that strongly suppress the small-$z$ region. 
The divergence at [$z \to 1$], instead, might reflect a breakdown of fixed-order perturbation theory, and points to the need for threshold resummation techniques to restore predictive power.

We remark that the end point behavior observed here for $\etQ$ production is not unique. 
A similar phenomenon has been encountered in the study of color-singlet fragmentation into scalar or tensor fully heavy tetraquarks, where the NRQCD input for the initial-scale gluon FF, $D_g^{\TQc}(0^{++},2^{++})(z,\mu_{F,0})$, remains finite and nonvanishing as [$z \to 1$]~\cite{Feng:2020riv,Celiberto:2024mab,Celiberto:2025ziy}. 
In that context, the lack of suppression at the end point challenges the leading-twist interpretation of collinear factorization, where only one parton fragments into the hadron, and the probability of transferring all its momentum should ideally vanish.

Some authors argue that the divergent NLO evolution induced by this nonzero boundary condition may still yield physically meaningful results when convoluted with partonic cross sections~\cite{Artoisenet:2014lpa}, while others interpret it as a signal of perturbative instability and advocate for an extension of the formalism, possibly involving resummation~\cite{Zhang:2018mlo}. 
In our view, a deeper understanding of the end point structure of NRQCD-based FFs remains an open and relevant direction for future work.

\subsubsection{Constituent heavy quark channel}
\label{sssec:NRFF_Q}

The [$Q \to \etQ$] SDC in the $^1S_1^{(0)}$ quantum configuration (see Fig.~\ref{fig:PSQ_FF_diagrams}, central diagram) reads
\begin{equation}
\begin{split}
\label{SDC_SQ_Q}
 {\cal D}^{\cal Q}_{Q}(z, ^{1\!\!}S^{(0)}) \,&=\, 
 \as^2 \, \drv_{Q}^{\rm [LO]} (z, ^{1\!\!}S^{(0)}) 
 \\[0.10cm]
 \,&+\,
 \as^3 \, \drv_{Q}^{\rm [NLO]} (z, ^{1\!\!}S_1^{(0)})
 \;,
\end{split}
\end{equation}
where the QCD running coupling $\alpha_s(\mu_R \equiv \mu_{R,0})$ is evaluated at $\mu_{R,0} = 2 m_Q$, corresponding to the minimal invariant mass of the gluon in the LO fragmentation process~\cite{Braaten:1993mp,Zheng:2021ylc}.
The LO SDC, first derived in Ref.~\cite{Braaten:1993mp}, is
\begin{equation}
\begin{split}
\label{SDC_SQ_Q_LO}
 \drv_{Q}^{\rm [LO]} (z, ^{1\!\!}S_0^{(1)}) 
 \,&=\,
 \frac{16 \, z(1 - z)^2}{81 N_c (2 - z)^6 \, m_Q^3} 
 \\[0.10cm]
 \,&\times\,
 \left( 3z^4 - 8z^3 + 8z^2 + 48 \right)
 \;.
\end{split}
\end{equation}

The NLO correction has been computed in Ref.~\cite{Zheng:2021ylc} and can be expressed in terms of the following polynomial functions
\begin{subequations}
\begin{align}
\label{SDC_SQ_Q_NLO_etc}
 \drv_{\etc}^{\rm [NLO]}& (z, ^{1\!\!}S_0^{(1)})
  =
  \frac{2\pi}{N_c \, m_c^3} 
  \notag \\[0.10cm]
  \,&\times\,
  \left(\right.
  0.60361 z^8 - 3.48697 z^7 
  \notag \\[0.10cm]
  \,&+\,
  8.69232 z^6 - 11.43404 z^5 + 8.39029 z^4 
  \notag \\[0.10cm]
  \,&-\,
  3.73948 z^3 + 1.14322 z^2 - 0.16352 z 
  \notag \\[0.10cm]
  \,&-\,
  0.00453 - 0.001282/z 
 \left.\right)
 \;,
\\[0.35cm]
\label{SDC_SQ_Q_NLO_etb}
 \drv_{\etb}^{\rm [NLO]}& (z, ^{1\!\!}S_0^{(1)})
  =
  \frac{2\pi}{N_c \, m_b^3} 
  \notag \\[0.10cm]
  \,&\times\,
  \left(\right.
  1.16869 z^8 - 6.27381 z^7 
  \notag \\[0.10cm]
  \,&+\,
  14.08792 z^6 - 16.85613 z^5 + 11.51566 z^4 
  \notag \\[0.10cm]
  \,&-\,
  4.78788 z^3 + 1.33076 z^2 - 0.18076 z 
  \notag \\[0.10cm]
  \,&-\,
  0.00359 - 0.001282/z
 \left.\right)
 \;.
\end{align}
\end{subequations}
The difference between Eqs.~\eqref{SDC_SQ_Q_NLO_etc} and~\eqref{SDC_SQ_Q_NLO_etb} stems from the number of active heavy flavors considered in the gluon vacuum polarization loop.
In the charm case, only the charm quark contributes to the loop, while in the bottom case both charm and bottom quarks are included, resulting in distinct $\drv_{\etQ}^{\rm [NLO]}$ profiles.

The full expression for our initial-scale constituent heavy quark FF is obtained by combining Eqs.~\eqref{FFs_NRQCD}, \eqref{SDC_SQ_Q_LO},~\eqref{SDC_SQ_Q_NLO_etc}, and~\eqref{SDC_SQ_Q_NLO_etb}.
Due to the symmetry of the final-state quarkonium, FFs from the constituent quark and antiquark of the same flavor are identical.

To validate the numerical implementation of our constituent heavy quark SDC and facilitate a direct comparison with the results of Ref.~\cite{Zheng:2021ylc}, we follow the same strategy adopted in that work.
In particular, we employ a simplified timelike DGLAP evolution at NLO, retaining only the LO Altarelli-Parisi splitting kernels, namely $P_{gg}(z)$ and $P_{gq}(z)$, corresponding to gluon and quark contributions to the gluon FF evolution, respectively.
As initial-scale inputs, we use the NRQCD result for the quark-induced channel derived in this section, together with the gluon-induced FF discussed in the previous one.

\begin{figure*}[!t]
\centering

 \includegraphics[scale=0.640,clip]{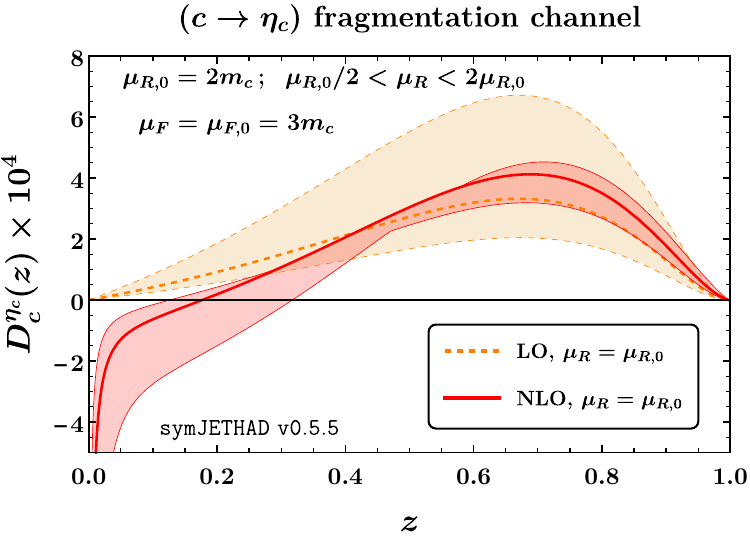}
 \hspace{0.50cm}
 \includegraphics[scale=0.640,clip]{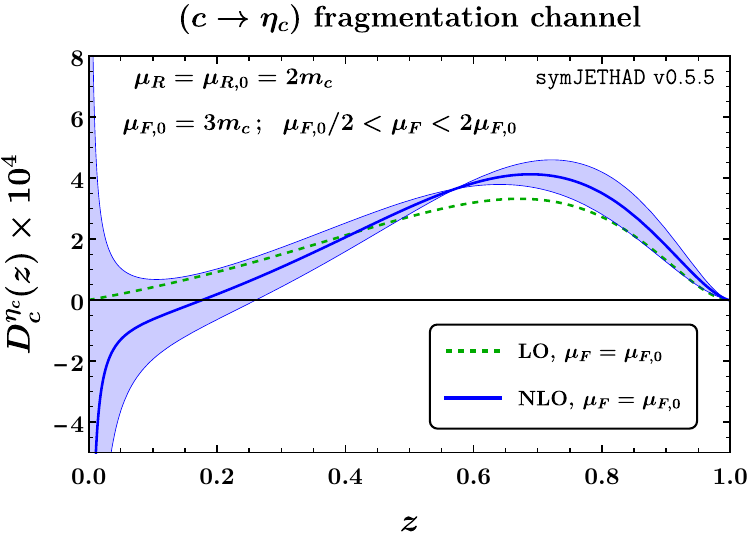}

 \vspace{0.35cm}

 \includegraphics[scale=0.640,clip]{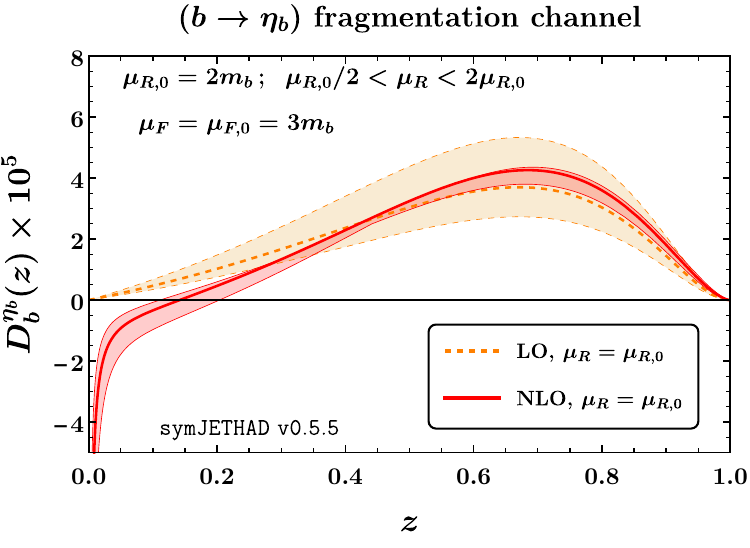}
 \hspace{0.50cm}
 \includegraphics[scale=0.640,clip]{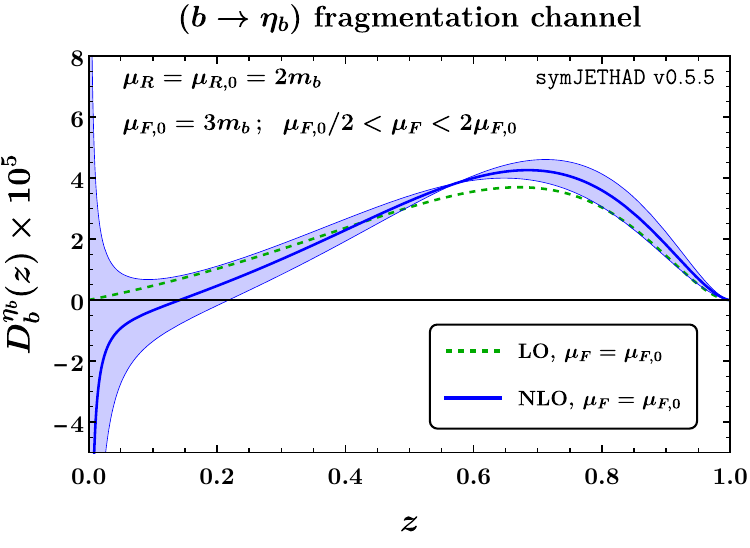}

\caption{NRQCD [$c \to \etc$, upper] and [$b \to \etb$, lower] fragmentation channels at LO and NLO, for $\mu_{R,0}=2m_Q$ and $\mu_{F,0}=3m_Q$. 
Left plot: $\mu_R$ runs from $\mu_{R,0}/2$ to $2\mu_{R,0}$, while $\mu_F$ is fixed at $\mu_{F,0}$.
Right plot: $\mu_R$ is fixed at $\mu_{R,0}$, while $\mu_F$ runs from $\mu_{F,0}/2$ to $2\mu_{F,0}$.}
\label{fig:FF_Q-to-etQ_muRF-var}
\end{figure*}

For illustration purposes, Fig.~\ref{fig:FF_Q-to-etQ_muRF-var} displays the FFs for [$c \to \etc$] (upper panels) and [$b \to \etb$] (lower panels) at LO and NLO, under different choices of factorization and renormalization scales.
The left-hand plots correspond to fixed $\mu_F = \mu_{F,0}$ with $\mu_R$ varied in the range $\mu_{R,0}/2$ to $2\mu_{R,0}$, while the right-hand plots keep $\mu_R = \mu_{R,0}$ fixed and explore variations in $\mu_F$.

We set the central renormalization and factorization scales to $\mu_{R,0} = 2m_Q$ and $\mu_{F,0} = 3m_Q$, respectively.
The choice $\mu_{R,0} = 2m_Q$ corresponds to the minimal invariant mass of the intermediate gluon that produces the [$Q\bar{Q}$] pair in the LO fragmentation process.
In contrast, $\mu_{F,0} = 3m_Q$ reflects the minimal invariant mass of the initial-state configuration in the quark-induced channel, where a heavy quark fragments via [$Q \to Q\bar{Q}Q$]. 
This differs from the gluon-induced case (see the previous section) and accounts for the distinct partonic content of the corresponding collinear subprocess.

As in the gluon case, for simplicity, the wave function at the origin is fixed to its central value, and uncertainty variations are not considered (see Section~\ref{sssec:LDMEs}).

Our results are in excellent agreement with those shown in Figs.~4 to~7 of Ref.~\cite{Zheng:2021ylc}, both in shape and magnitude across the entire $z$ range.
For both the charm and bottom fragmentation channels, we observe the same pattern of NLO corrections: moderate suppression in the intermediate $z$, an enhancement in the large $z$, and the onset of a $1/z$ divergence in the small-$z$ region, absent at LO. 
These features reflect the well-known behavior of the constituent quark FFs, driven by real emission contributions at NLO and consistent with the structure of light quark--initiated fragmentation as discussed in~\cite{Zheng:2021ylc}.

The sensitivity of the FFs to scale variations is also found to match that of Ref.~\cite{Zheng:2021ylc}. 
In particular, variations in the renormalization scale $\mu_R$ affect predominantly the low-$z$ region, where the NLO corrections are largest and the LO contribution vanishes. Conversely, changes in the factorization scale $\mu_F$ impact the entire $z$ range, but only at NLO, confirming that DGLAP evolution effects enter beyond LO in this channel. 
Notably, both [$c\to \etc$] and [$b \to \etb$] FFs exhibit good perturbative stability: the NLO bands lie close to the LO ones, with reduced uncertainty and no sign of pathological behavior. 
This agreement reinforces the reliability of our implementation and supports the applicability of our constituent heavy quark NRQCD inputs in phenomenological studies.

\subsubsection{Nonconstituent quark channel}
\label{sssec:NRFF_q}

\begin{figure*}[t]
\centering
 
 \vspace{0.35cm}
 
 \includegraphics[scale=0.640,clip]{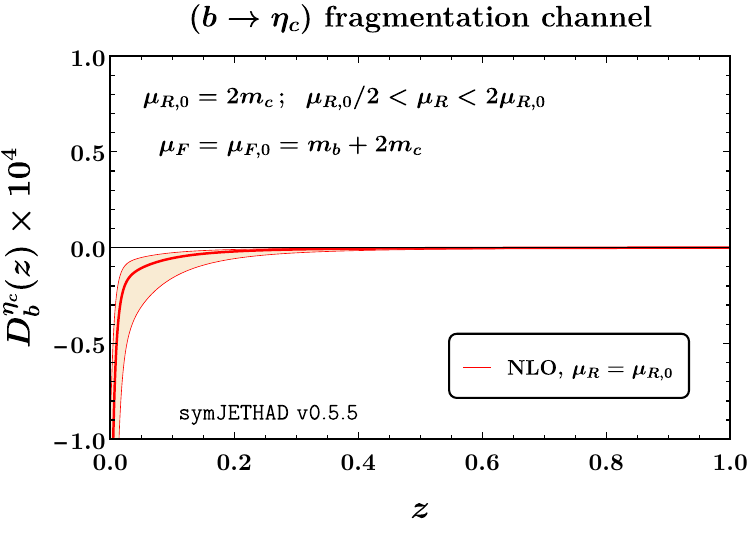}
 \hspace{0.50cm}
 \includegraphics[scale=0.640,clip]{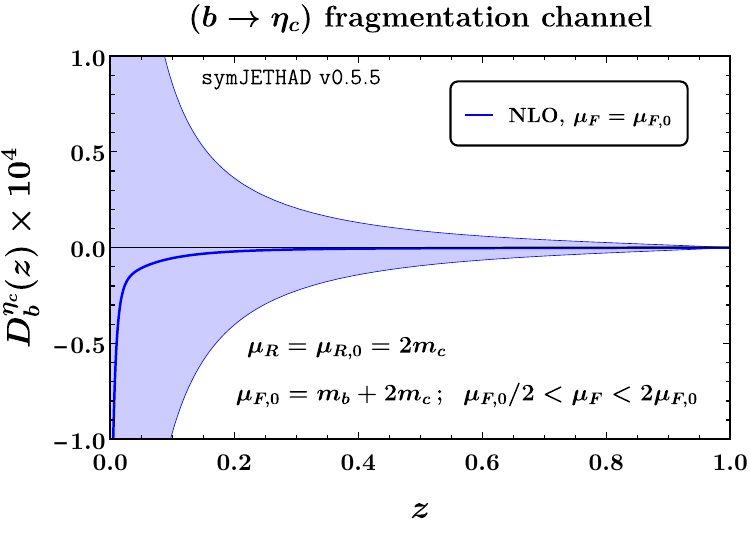}
 
 \vspace{0.35cm}

\caption{NRQCD [$b \to \etc$, lower] fragmentation channel at NLO for $\mu_{R,0}=2m_c$ and $\mu_{F,0}=m_b+2m_c$. 
Left plot: $\mu_R$ runs from $\mu_{R,0}/2$ to $2\mu_{R,0}$, while $\mu_F$ is fixed at $\mu_{F,0}$.
Right plot: $\mu_R$ is fixed at $\mu_{R,0}$, while $\mu_F$ runs from $\mu_{F,0}/2$ to $2\mu_{F,0}$.}
\label{fig:FF_b-to-etc_muRF-var}
\end{figure*}

The [$\tilde{q} \mbox{ or } \tilde{Q} \to \etQ$] SDC in the $^1S_1^{(0)}$ quantum configuration (see Fig.~\ref{fig:PSQ_FF_diagrams}, right diagram) was first derived in Ref.~\cite{Zheng:2021mqr}.
As mentioned earlier, this channel is absent at leading order and starts contributing only at NLO [namely at $\mathcal{O}(\alpha_s^3)$], as summarized in Table~\ref{tab:FF_order}.
This structural feature marks a key distinction from gluon and constituent quark channels, which both appear already at $\mathcal{O}(\alpha_s^2)$.
Due to the symmetry of the final-state quarkonium, SDCs from nonconstituent quark and antiquark of the same flavor are identical.
One has
\begin{equation}
\begin{split}
\label{SDC_SQ_ncq}
 {\cal D}^{\cal Q}&_{\tilde{q},\tilde{Q}}(z, ^{1\!\!}S^{(0)}) \,\equiv\,
 \as^3 \, \drv_{\tilde{q},\tilde{Q}}^{\rm [NLO]} (z, ^{1\!\!}S_1^{(0)})
 \\[0.05cm]
 \,&=\,
 \as^3 \, 
 \int_z^1 \! \drv \upsilon \int_{\frac{4\upsilon\,m_Q^2}{z}}^{\infty} \! \drv \omega\, \hat\drv_{\tilde{q},\tilde{Q}}^{\rm [NLO]} (\omega,\upsilon,z, ^{1\!\!}S_1^{(0)})
 \;,
\end{split}
\end{equation}
where the QCD running coupling $\alpha_s(\mu_R \equiv \mu_{R,0})$ is evaluated at $\mu_{R,0} = m_{\tilde{q},\tilde{Q}} + 2 m_Q$, corresponding to the minimal invariant mass of the nonconstituent heavy quark in the fragmentation process~\cite{Zheng:2021mqr}.
Furthermore, one writes
\begin{equation}
\begin{split}
\label{SDC_SQ_ncq_itgd}
\hat\drv&_{\tilde{q},\tilde{Q}}^{\rm [NLO]} (\omega,\upsilon,z, ^{1\!\!}S_1^{(0)})
\,=\,
\frac{2 \, \hat f_{\tilde{q},\tilde{Q}}}{9\pi \, N_c \, 
m_Q}
\\[0.10cm]
\,& \times\, 
\Bigg\{ (\upsilon-1)\Big[\omega^3 \big( \upsilon^4-2\upsilon^3(z+1)
\\[0.10cm]
\,&+\,
2\,\upsilon^2(z^2+6z+1) 
\\[0.10cm]
\,& -\, 
12\,\upsilon z(z+1)+12z^2\big)
\\[0.10cm]
\,&+\,
\omega^2 \big(2 \upsilon^4 m_{\tilde{q},\tilde{Q}}^2-4\upsilon^3(2m_Q^2(z+4)+z\, m_{\tilde{q},\tilde{Q}}^2)
\\[0.10cm]
\,&+\,
4 \upsilon^2(4 m_Q^2(3z+2)+z^2 m_{\tilde{q},\tilde{Q}}^2) 
- 48\,\upsilon z \, m_Q^2 \big)
\\[0.10cm]
\,& +\, 
16 \, \omega \upsilon^2 \, m_Q^2 \big(\upsilon^2 m_Q^2-\upsilon \, (2m_Q^2+z\,m_{\tilde{q},\tilde{Q}}^2)
\\[0.10cm]
\,& +\,
2 m_Q^2\big)+32 \, \upsilon^4 \, m_Q^4 m_{\tilde{q},\tilde{Q}}^2 \Big]
\\[0.10cm]
\,& -\,
\left[(1-\upsilon)\omega+\upsilon^2m_{\tilde{q},\tilde{Q}}^2\right]
\\[0.10cm]
\,& \times\,
\left[z^2(\omega-4\,\frac{\upsilon}{z}\,m_Q^2)^2+(\upsilon-z)^2\omega^2 \right] \upsilon^2
\;,
\\[0.05cm]
\end{split}
\end{equation}
with
\begin{equation}
\label{SDC_SQ_ncq_itgd_f}
\begin{split}
 \hat f_{\tilde{q},\tilde{Q}} \,&=\, \Big\{ \upsilon^4 \omega^2 (\omega-4m_Q^2)^2
\\[0.10cm]
\,& \times\,
 \left[(1-\upsilon)\,\omega+\upsilon^2m_{\tilde{q},\tilde{Q}}^2\right] \Big\}^{-1} \;.
\end{split}
\end{equation}
The integration variables $\upsilon$ and $\omega$ in Eqs.~\eqref{SDC_SQ_ncq} and~\eqref{SDC_SQ_ncq_itgd}, respectively, parametrize the longitudinal momentum fraction carried by the fragmenting quark and the invariant mass squared of the recoiling system.
The integrand function $\hat\drv_{\tilde{q},\tilde{Q}}^{\rm [NLO]}$ exhibits an explicit dependence on the mass of the constituent heavy quark, $m_Q$, and that of the nonconstituent one, $m_{\tilde{q},\tilde{Q}}$.
Numerical benchmarks have confirmed that our {\symJethad} implementation of our NLO SDC is in excellent agreement with the results of Ref.~\cite{Zheng:2021mqr} over the entire $z$ range.

For illustration purposes, Fig.~\ref{fig:FF_b-to-etc_muRF-var} shows the [$b \to \etc$] FF at NLO, for different choices of factorization and renormalization scales. 
The left plot corresponds to $\mu_F = \mu_{F,0}$ with $\mu_R$ varied in the range $\mu_{R,0}/2$ to $2\mu_{R,0}$, while in the right plot $\mu_R$ is kept fixed at $\mu_{R,0}$ and $\mu_F$ is varied.

We adopt a simplified NLO timelike DGLAP evolution, retaining only the LO $P_{gq}(z)$ splitting kernel. The initial inputs are given by the NRQCD-derived nonconstituent quark FF discussed above and the gluon-induced FF introduced earlier.

We set the central renormalization and factorization scales to $\mu_{R,0} = 2m_Q$ and $\mu_{F,0} = m_{\tilde{q},\tilde{Q}} + 2m_Q$, respectively.
The choice $\mu_{R,0} = 2m_Q$ corresponds to the minimal invariant mass of the intermediate gluon that produces the [$Q\bar{Q}$] pair in the fragmentation process, while $\mu_{F,0} = m_{\tilde{q},\tilde{Q}} + 2m_Q$ reflects the minimal invariant mass of the initial-state configuration in the quark-induced channel, where a nonconstituent quark fragments into itself, a [$Q\bar{Q}$ pair, and a gluon.

As in the previous cases, for simplicity, the wave function at the origin is fixed to its central value, and uncertainty variations are not considered (see Section~\ref{sssec:LDMEs}).

For $\mu_R = \mu_{R,0}$ and $\mu_F = \mu_{F,0}$, the FF is negative across the entire $z$ range.
This behavior is consistent with the logarithmic structure of the NLO SDC.
As $z$ increases, the FF smoothly approaches zero from below, and the scale variation bands become progressively narrower, indicating reduced sensitivity in the large-$z$ region.

In the left panel, the renormalization scale dependence mainly affects the normalization of the FF in the small-$z$ domain, as expected from the $\alpha_s^3(\mu_R)$ term in Eq.~\eqref{SDC_SQ_ncq}.
Conversely, the right panel shows that variations in $\mu_F$ lead to a distortion of the FF shape at low $z$, with the function becoming shallower and eventually positive as $\mu_F$ increases.
Despite the relatively small numerical size of this FF compared to the gluon and constituent heavy quark channels, the nonconstituent fragmentation components remain essential for a consistent, flavor-complete initial condition within the {\HFNRevo} scheme.

\subsubsection{DGLAP evolution from {\HFNRevo}}
\label{sssec:NRFF_ns}

The final step in the construction of our {\NRFF} FFs for pseudoscalar quarkonia $\eta_Q$ is the DGLAP evolution of the initial-scale functions introduced in the previous sections. 

\begin{table*}
 \begin{center}
 \begin{tabular}[c]{|c||c|c|c|c||c|}
 \hline
 ${\cal S}_Q^- \, [^1S_0^{(1)}]$ quarkonium & $\mu_{F,0}(g \to {\cal S}_Q^-)$ & $\mu_{F,0}(d,u,s \to {\cal S}_Q^-)$ & $\mu_{F,0}(Q \to {\cal S}_Q^-)$ & $\mu_{F,0}(\tilde{Q} \to {\cal S}_Q^-)$ & $Q_0 \equiv \max \left( \{ \mu_{F,0} \} \right)$ \\
 \hline
 $\etc$ & $2m_c$ & $m_{d,u,s}+2m_c$ & $3m_c$ & \textcolor{red}{$\boldsymbol{m_b+2m_c}$} & \textcolor{red}{$\boldsymbol{m_b+2m_c}$} \\
 \hline
 $\etb$ & $2m_b$ & $m_{d,u,s}+2m_b$ & \textcolor{red}{$\boldsymbol{3m_b}$} & $m_c+2m_b$ & \textcolor{red}{$\boldsymbol{3m_b}$} \\
 \hline
  \end{tabular}
 \caption{Central columns: initial factorization scale, $\mu_{F,0}$, for the fragmentation of a given parton species (gluon, light quark, constituent and nonconstituent heavy quark) to a color-singlet pseudoscalar quarkonium, ${\cal S}_Q^- \, [^1S_0^{(1)}] \equiv \etc, \etb$.
 Rightmost column: evolution-ready energy scale, $Q_0$, set to the maximum of $\mu_{F,0}$ values.
 For clarity, values of $Q_0$ are highlighted in bold red font.
 Due to the symmetry of the final-state quarkonium, FFs from quark and antiquark of the same flavor are identical.}
 \label{tab:muF0_Q0}
 \end{center}
\end{table*}

Unlike light-hadron fragmentation, where the gluon and light-quark FFs are defined down to very low scales, the $\eta_Q$ FFs relevant possess intrinsic thresholds arising from the kinematic structure of the perturbative splittings that generate the color-singlet [$Q\bar{Q}$] pair. 
These splittings are schematically represented via LO representative diagrams in Fig.~\ref{fig:PSQ_FF_diagrams} and reflect different underlying mechanisms depending on the parton flavor.

In the gluon-induced channel (left diagram), the minimal invariant mass corresponds to the threshold for producing an on-shell [$Q\bar{Q}$] pair.
Accordingly, we set [$\mu_{F,0}(g \to \eta_Q) = 2m_Q$] as the natural starting point for gluon fragmentation.

In the case of a constituent heavy quark $Q$ fragmenting into $\eta_Q$ (central diagram), LO kinematics require an additional spectator quark in the final state. 
This raises the minimal invariant mass to [$\mu_{F,0}(Q \to \eta_Q) = 3m_Q$], which we take as the starting scale for evolution in this channel.

For the nonconstituent channels (right diagram), where a light quark $\tilde{q} = u,d,s$ or a heavier nonconstituent quark $\tilde{Q}$ fragments via gluon emission into $\eta_Q$, the corresponding thresholds are [$\mu_{F,0}(\tilde{q} \to \eta_Q) = m_q + 2m_Q$] and [$\mu_{F,0}(\tilde{Q} \to \eta_Q) = m_{\tilde{Q}} + 2m_Q$], respectively.

All threshold scales $\mu_{F,0}$ and the scale $Q_0$, taken as the maximum among these values for a given $\eta_Q$, are summarized in Table~\ref{tab:muF0_Q0}.

To consistently incorporate scale evolution into our fragmentation framework, we adopt a dedicated strategy tailored to the pseudoscalar quarkonium case.
We rely on the {\HFNRevo} methodology~\cite{Celiberto:2024mex,Celiberto:2024bxu}, originally designed to describe the fragmentation of color-singlet quarkonium states from nonrelativistic inputs.
This approach enables a rigorous DGLAP evolution while consistently incorporating mass thresholds and the sequential activation of flavor channels.

As already mentioned, the framework is structured around three core components: a physical interpretation of the fragmentation process beyond the leading-power approximation, thus enabling a consistent matching between FFNS and VFNS schemes; a two-step evolution strategy that separates threshold matching from scaling; and a systematic assessment of perturbative uncertainties associated with renormalization and factorization scale choices.

In the present study, which investigates {\HFNRevo} evolution of pseudoscalar quarkonia at high transverse momentum, we defer the explicit implementation of matching techniques.
Our focus is instead placed on developing a consistent and accurate scheme for handling DGLAP evolution across all parton thresholds, and on a controlled propagation of LDME-related uncertainties, whose impact will be analyzed in detail.

According to the {\HFNRevo} scheme, the DGLAP evolution of FFs with multiple parton thresholds is structured into two sequential phases.
The first one, labeled as {\tt EDevo}, is a semianalytical and expanded evolution, valid from the lowest factorization threshold up to the highest one.
In this stage, flavor channels are activated dynamically and independently, allowing for a decoupled treatment of the partonic evolution across distinct threshold regimes.
The second phase, referred to as {\tt AOevo}, involves a standard all-order numerical evolution, performed from the highest initial scale $Q_0$ upward.
This two-step strategy ensures accurate control over threshold matching effects while maintaining perturbative stability throughout the evolution range.

\begin{figure*}[t]
\centering

 \includegraphics[scale=0.18,clip]{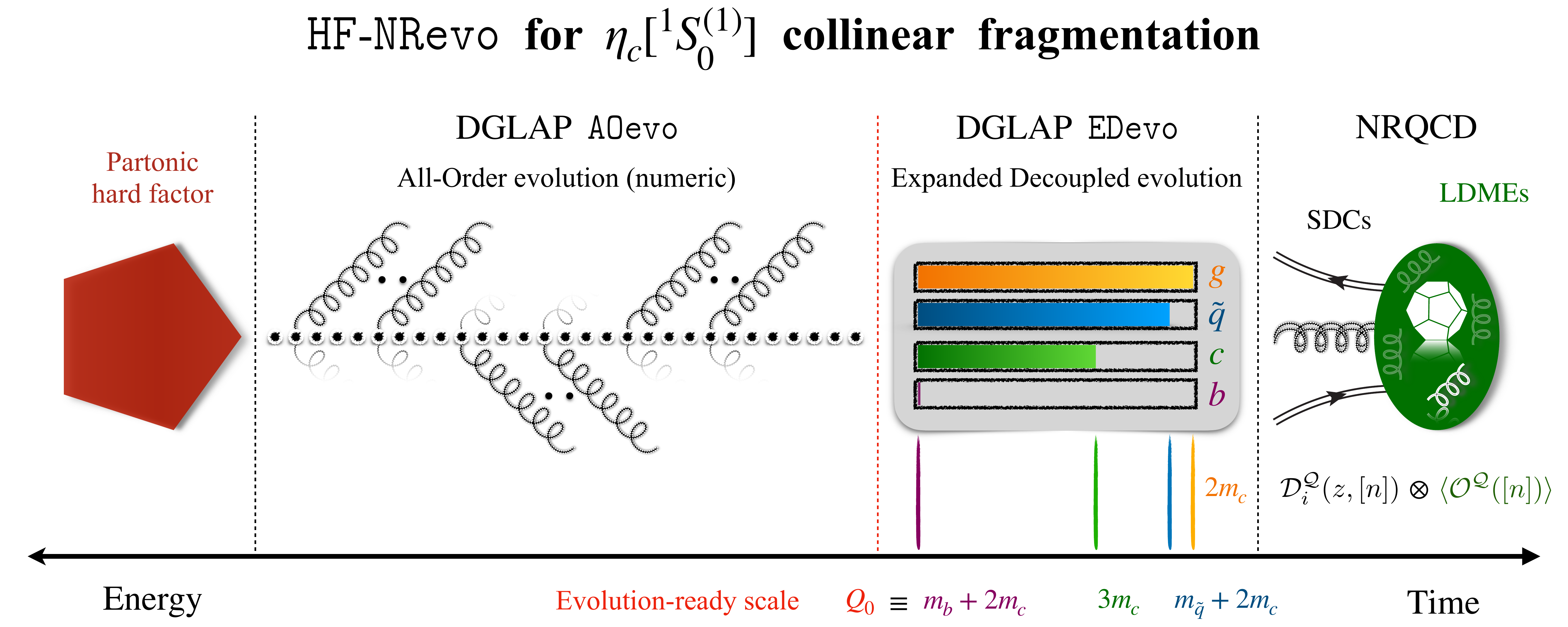}

\caption{Evolution workflow for $\etc[^1S_0^{(1)}]$ FFs within the {\HFNRevo} scheme.
The diagram illustrates the two-stage DGLAP evolution: a semianalytical {\tt EDevo} phase, starting from the lowest threshold and valid up to the evolution-ready scale $Q_0 = m_b + 2m_c$, where flavor channels are dynamically activated in a stepwise fashion, followed by a numerical {\tt AOevo} phase.}
\label{fig:etc_cs_HFNRevo_sketch}
\end{figure*}

\begin{figure*}[t]
\centering

 \includegraphics[scale=0.18,clip]{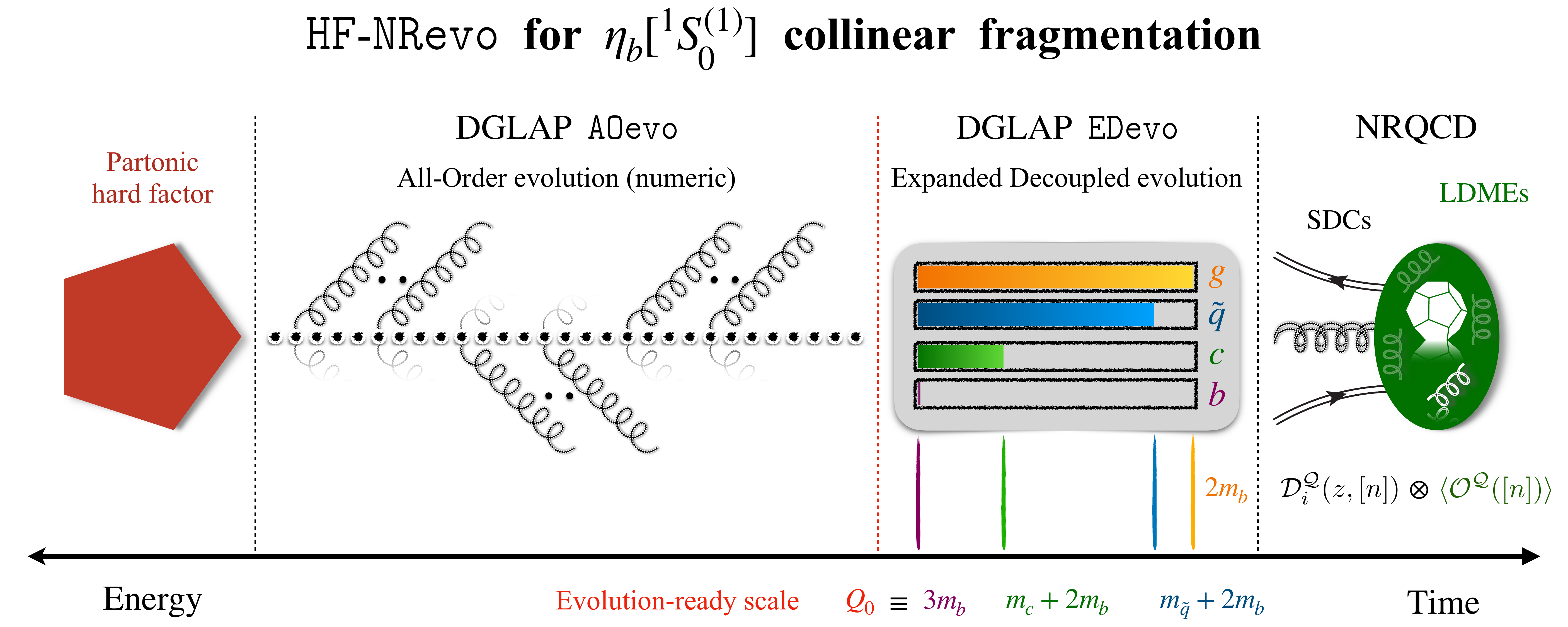}

\caption{Evolution workflow for $\etb[^1S_0^{(1)}]$ FFs within the {\HFNRevo} scheme.
The diagram illustrates the two-stage DGLAP evolution: a semianalytical {\tt EDevo} phase, starting from the lowest threshold and valid up to the evolution-ready scale $Q_0 = 3m_b$, where flavor channels are dynamically activated in a stepwise fashion, followed by a numerical {\tt AOevo} phase.}
\label{fig:etb_cs_HFNRevo_sketch}
\end{figure*}

To better visualize the {\HFNRevo} approach to quar\-ko\-nium fragmentation, we present in Figs.~\ref{fig:etc_cs_HFNRevo_sketch} and~\ref{fig:etb_cs_HFNRevo_sketch} a conceptual flowchart illustrating the two-stage evolution of the $\etc[^1S_0^{(1)}]$ and $\etb[^1S_0^{(1)}]$ color-singlet channels, respectively.
The horizontal axis encodes the flow of energy from infrared to ultraviolet scales (right to left), thus inverting the direction of physical time. 

If the diagram is read chronologically (from left to right), one observes a high-energy parton emitted from the hard-scattering kernel (leftmost block, depicted as a firebrick pentagon) radiating energy and fragmenting into the final-state quarkonium.
Conversely, reading the diagram in the direction of energy flow (right to left), one follows the construction workflow of the FFs: the process begins from NRQCD factorization inputs (rightmost block), built via the convolution of SDCs with LDMEs.

From there, the semianalytical {\tt EDevo} module (cen\-ter-right block) initiates DGLAP evolution at the lowest flavor threshold, $\mu_{F,0}(g \to \etc) = 2m_c$ (orange bar), and continues in a stepwise fashion by successively activating additional parton channels.
First, the nonconstituent light-quark threshold, $\mu_{F,0}(\tilde{q} \to \etc) = m_{\tilde{q}} + 2m_c$ (blue bar), is crossed---for simplicity, in our sketch we do not distinguish among $u$, $d$, and $s$ flavors, given their negligible mass differences.
Then, the charm-quark channel is activated at $\mu_{F,0}(c \to \etc) = 3m_c$ (green bar), followed by the bottom-quark threshold (purple bar).
This hierarchical sequence leads up to the highest scale in the system, $\mu_{F,0}(b \to \etc) = m_b + 2m_c$, which we identify as the \emph{evolution-ready} scale, $Q_0$ (marked in bold red in Table~\ref{tab:muF0_Q0}).

At this point, the numerical {\tt AOevo} module (center-left block) takes over and performs an all-order DGLAP evolution at NLO, acting on the fully active flavor set and evolving it to higher scales.
For this second step, we make use of the {\APFELpp} evolution library~\cite{Bertone:2013vaa,Carrazza:2014gfa,Bertone:2017gds}, while future developments will interface {\symJethad} with the {\EKO} framework~\cite{Candido:2022tld,Hekhorn:2023gul}.
The resulting evolved FFs are finally convoluted with the partonic hard-scattering cross section (left block) to yield physical predictions.

The overall structure of the evolution workflow for $\etb[^1S_0^{(1)}]$ FFs is analogous to the one discussed for $\etc$, and is sketched in Fig.~\ref{fig:etb_cs_HFNRevo_sketch}. 
The horizontal axis encodes the flow from infrared to ultraviolet energy scales (right to left) and, conversely, from early to late stages of the fragmentation process (left to right). 
The rightmost block corresponds to the NRQCD input, followed by the semianalytical {\tt EDevo} evolution (center-right), the all-order numerical {\tt AOevo} evolution (center-left), and the final convolution with the hard-scattering kernel (left block).
Colored bars indicate the activation of the $g$, $\tilde q$, $c$, and $b$ channels, using the same scheme as in the $\etc$ case.

In this case, the {\tt EDevo} module starts at the gluon threshold, $\mu_{F,0}(g \to \etb) = 2m_b$, and proceeds across all other channels in a flavor-ordered sequence. The light-quark nonconstituent channel is activated at $\mu_{F,0}(\tilde q \to \etb) = m_{\tilde q} + 2m_b$, followed by the charm threshold at $\mu_{F,0}(c \to \etb) = m_c + 2m_b$. 
Notably, this configuration differs from the $\etc$ case in the ordering and spacing of thresholds, leading to a distinct activation pattern. 
The final active channel is the bottom quark itself, with the evolution-ready scale identified as $Q_0 = 3m_b$, consistent with Table~\ref{tab:muF0_Q0}. 
The subsequent {\tt AOevo} stage, implemented via the {\APFELpp} library, performs the all-order NLO DGLAP evolution up to the hard-scattering scale.

Finally, among the two sources of FF uncertainty anticipated earlier---namely, the perturbative MHOUs from scale variations and the nonperturbative contributions from LDMEs---we focus in this first study on the latter. 
Specifically, we include only the LDME-induced uncertainties, as detailed in Section~\ref{sssec:LDMEs}. 
A systematic strategy for evaluating the perturbative MHOUs associated with the SDCs is currently under development and will be employed in future studies via advanced replicalike methods. 
Nevertheless, for completeness, in this work we also present a preliminary estimate of such uncertainties by means of a simple variation of the perturbative energy scales encoded in the FFs. 
We refer to these as fragmentation MHOUs (F-MHOUs), and we quantify their impact on the $z$-shape of the NLO FFs.

\begin{figure*}[!t]
\centering

 \hspace{-0.00cm}
 \includegraphics[scale=0.440,clip]{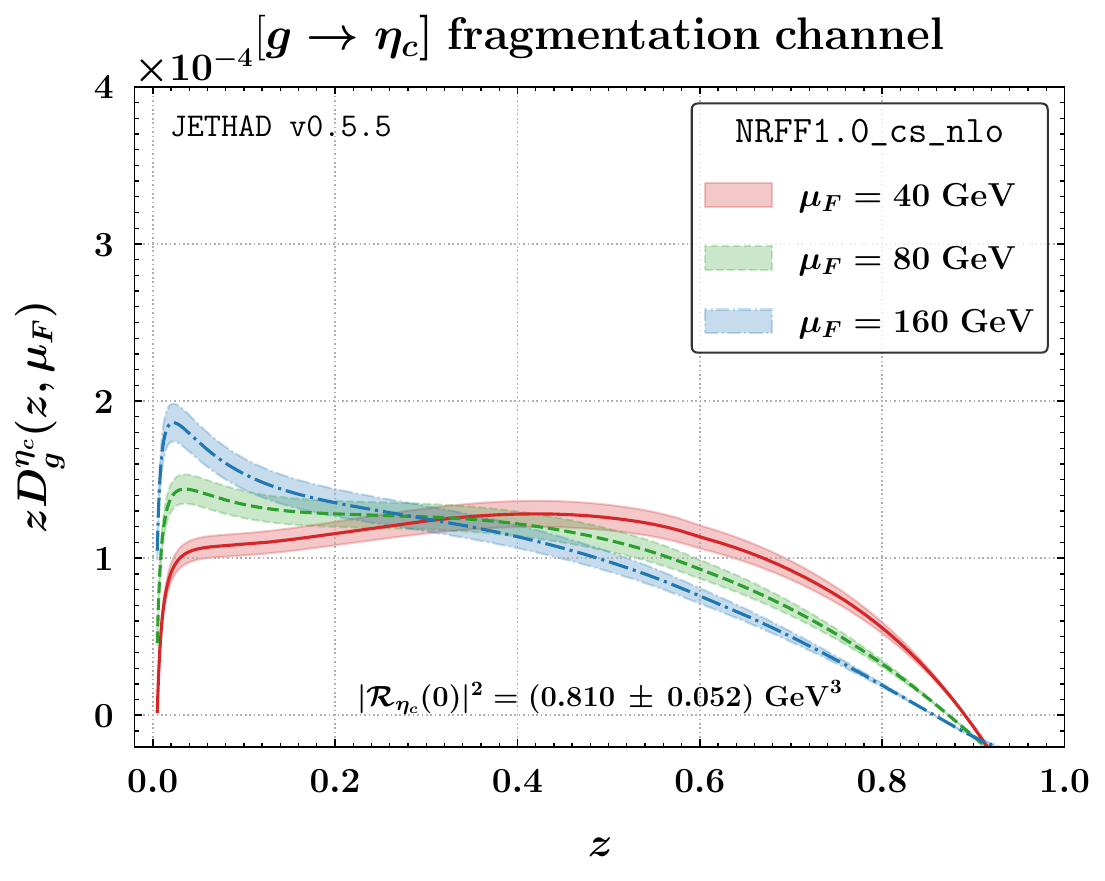}
 \hspace{0.00cm}
 \includegraphics[scale=0.440,clip]{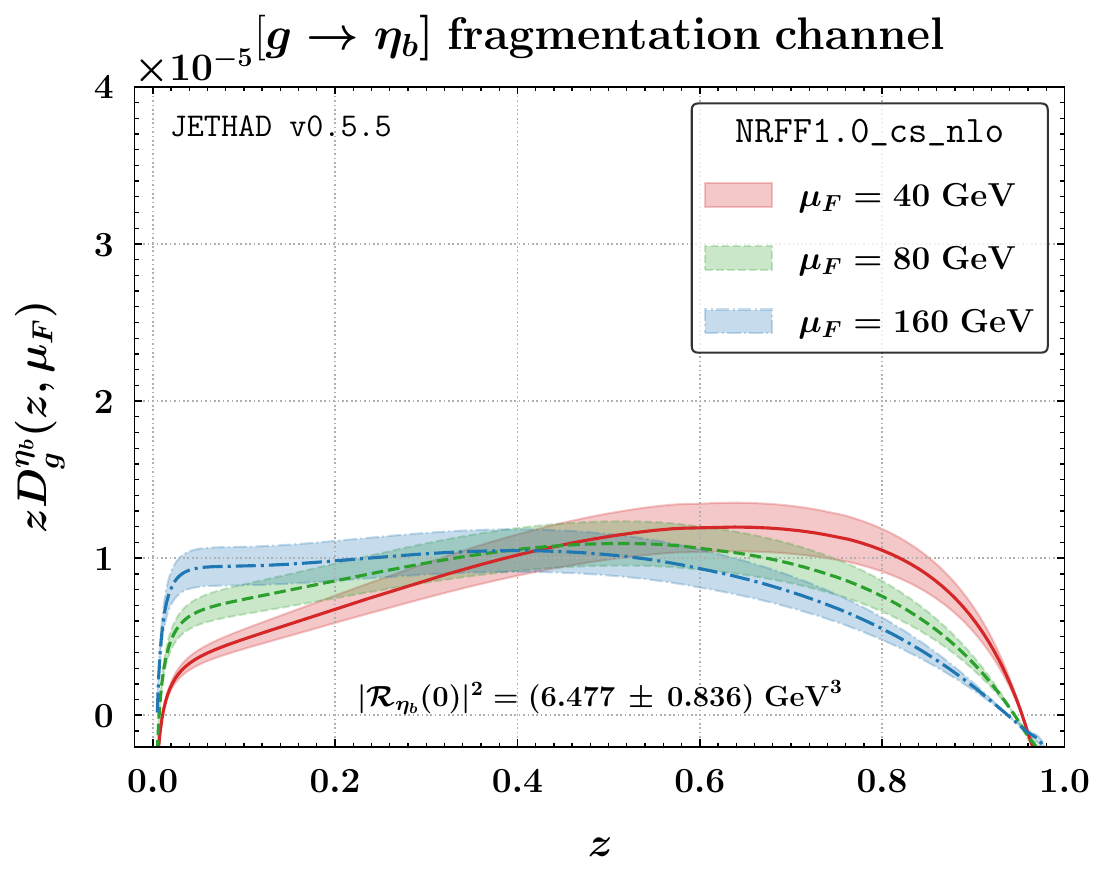}
 
 \vspace{0.35cm}
 
 \hspace{-0.10cm}
 \includegraphics[scale=0.440,clip]{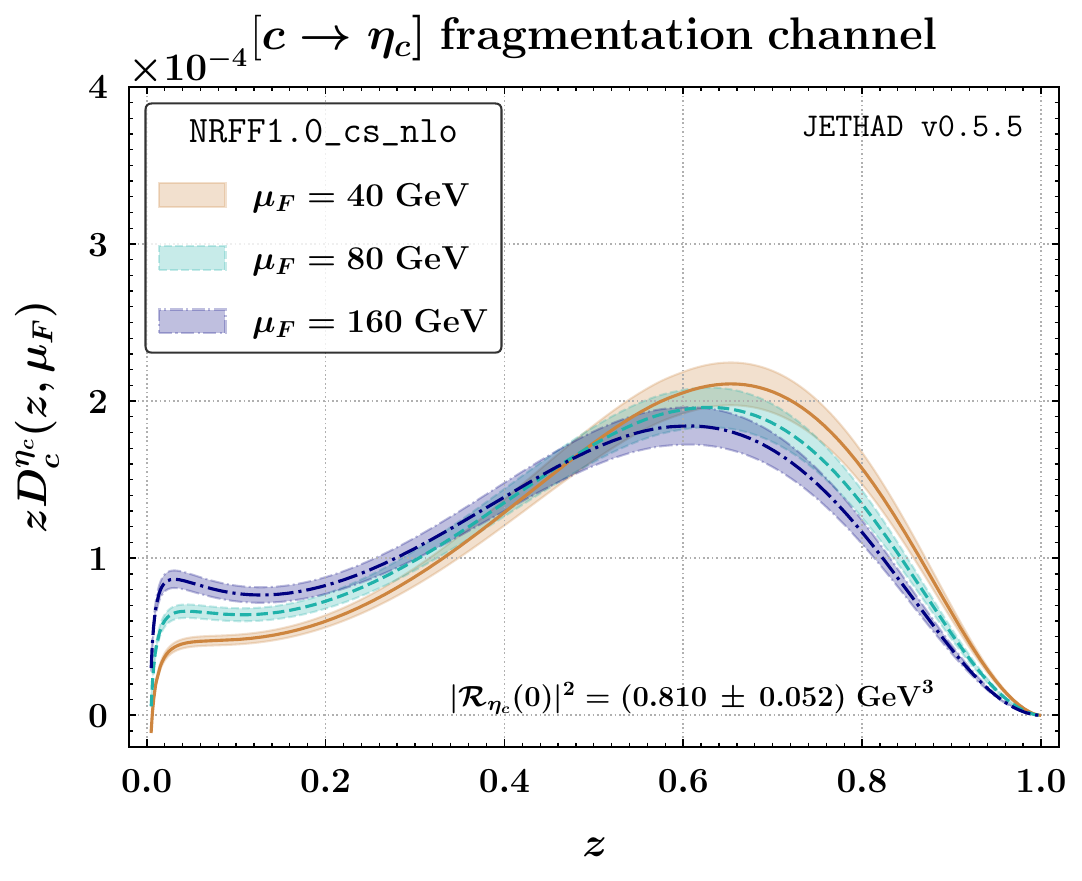}
 \hspace{0.20cm}
 \includegraphics[scale=0.440,clip]{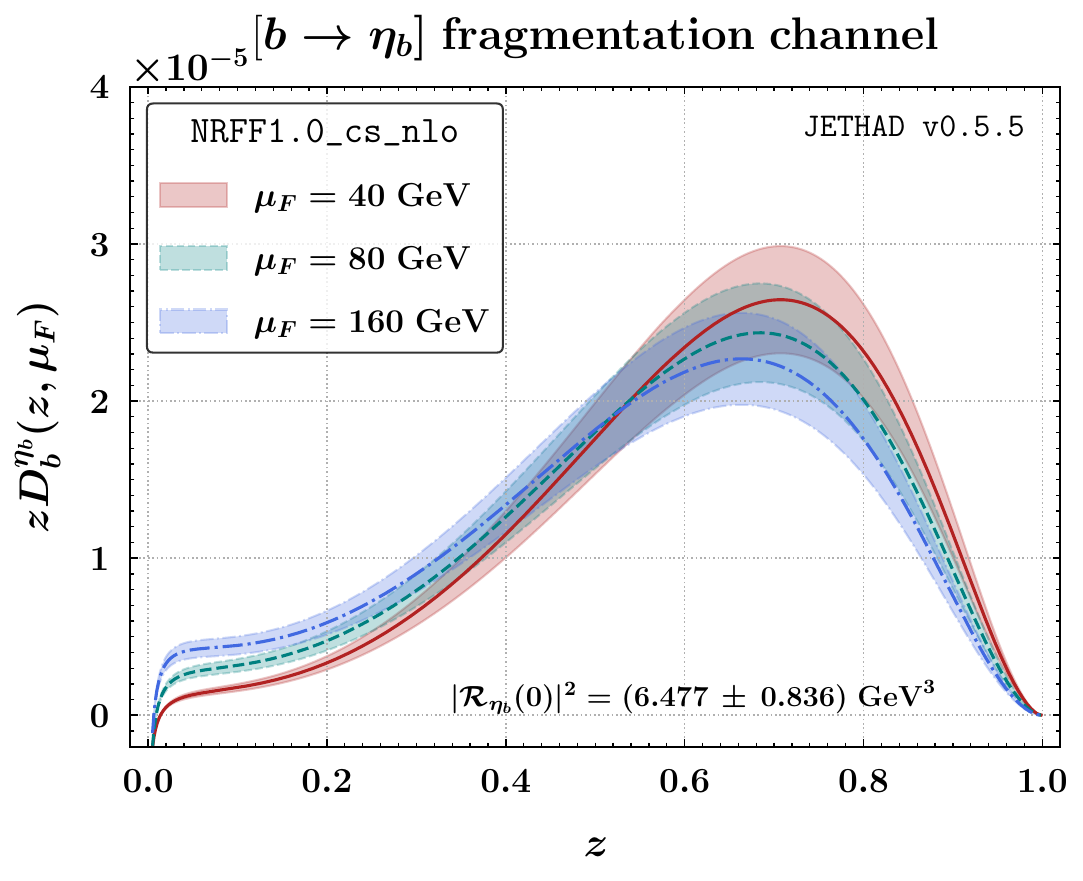}

\caption{$z$ dependence of the {\NRFF} NLO FFs~\cite{Celiberto:2025_NRFF10_cs_eQs} for $\etc[^1S_0^{(1)}]$ (left) and $\etb[^1S_0^{(1)}]$ (right) production at $\mu_F = 40$, 80, and 160~GeV.
The upper (lower) panels show the gluon (constituent heavy quark) fragmentation channel.
The shaded bands represent the uncertainty associated with the LDMEs.}
\label{fig:NRFF10_FFs-muF_g-Q}
\end{figure*}

\begin{figure*}[!t]
\centering

 \hspace{-0.00cm}
 \includegraphics[scale=0.435,clip]{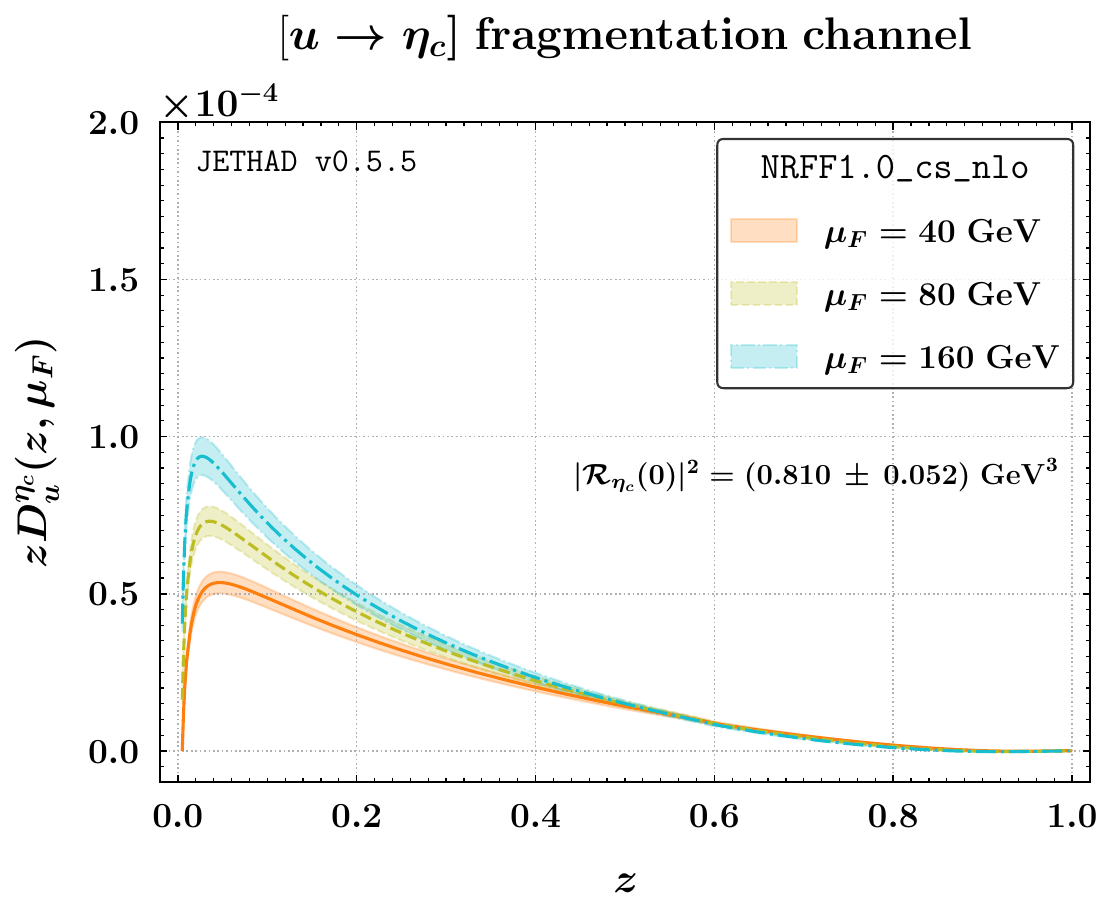}
 \hspace{0.00cm}
 \includegraphics[scale=0.435,clip]{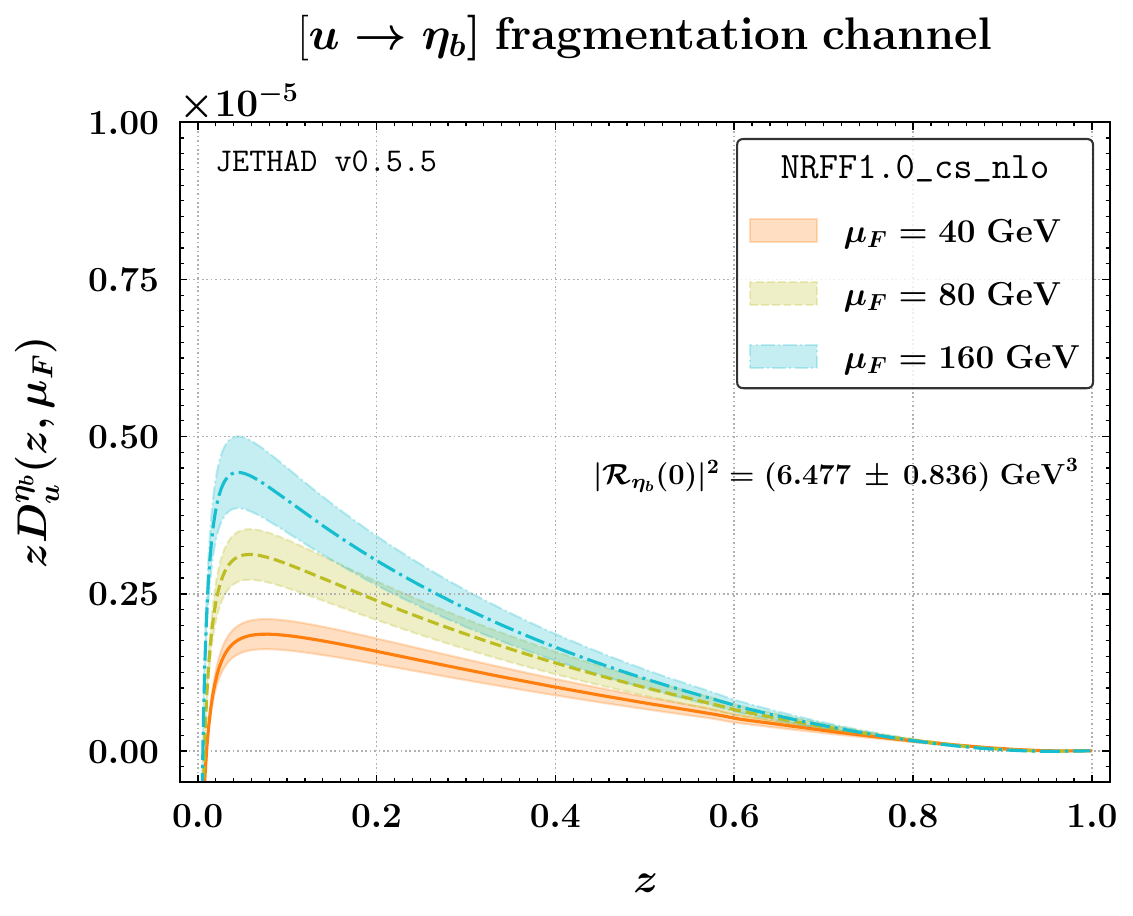}
 
 \vspace{0.35cm}
 
 \hspace{-0.00cm}
 \includegraphics[scale=0.435,clip]{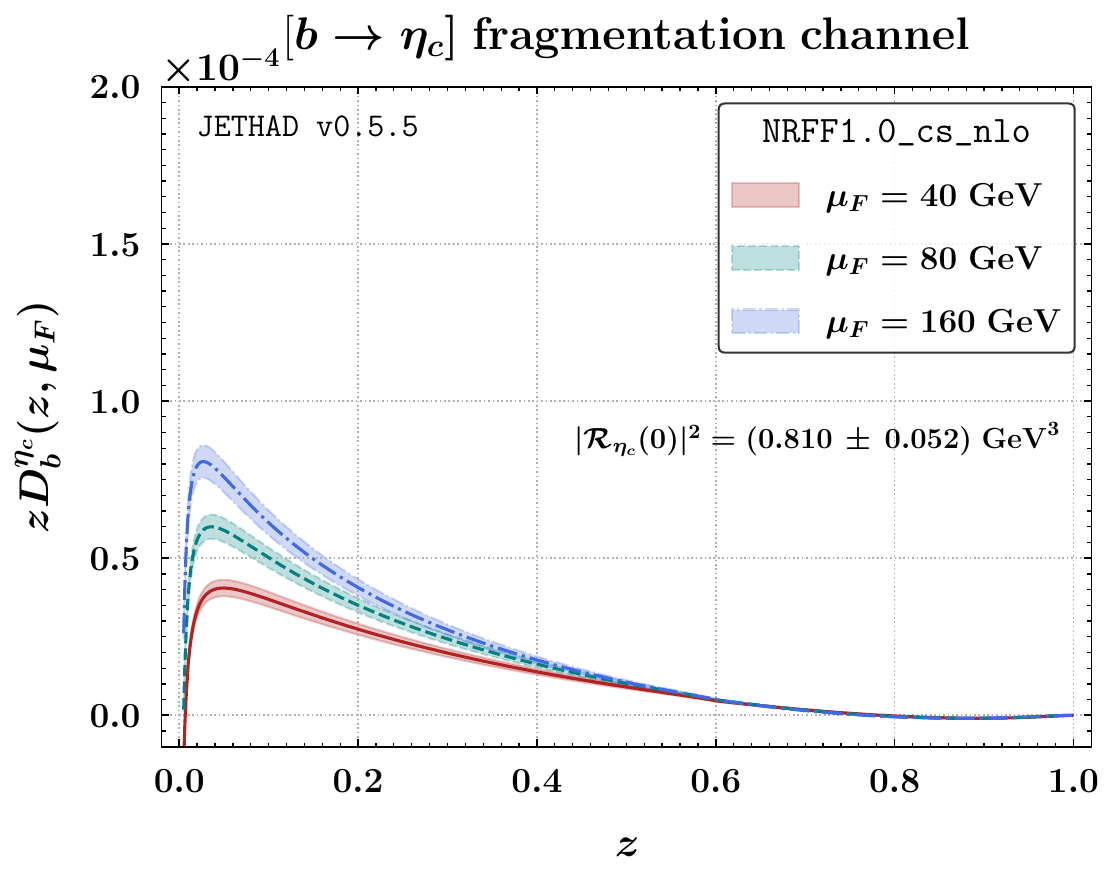}
 \hspace{0.00cm}
 \includegraphics[scale=0.435,clip]{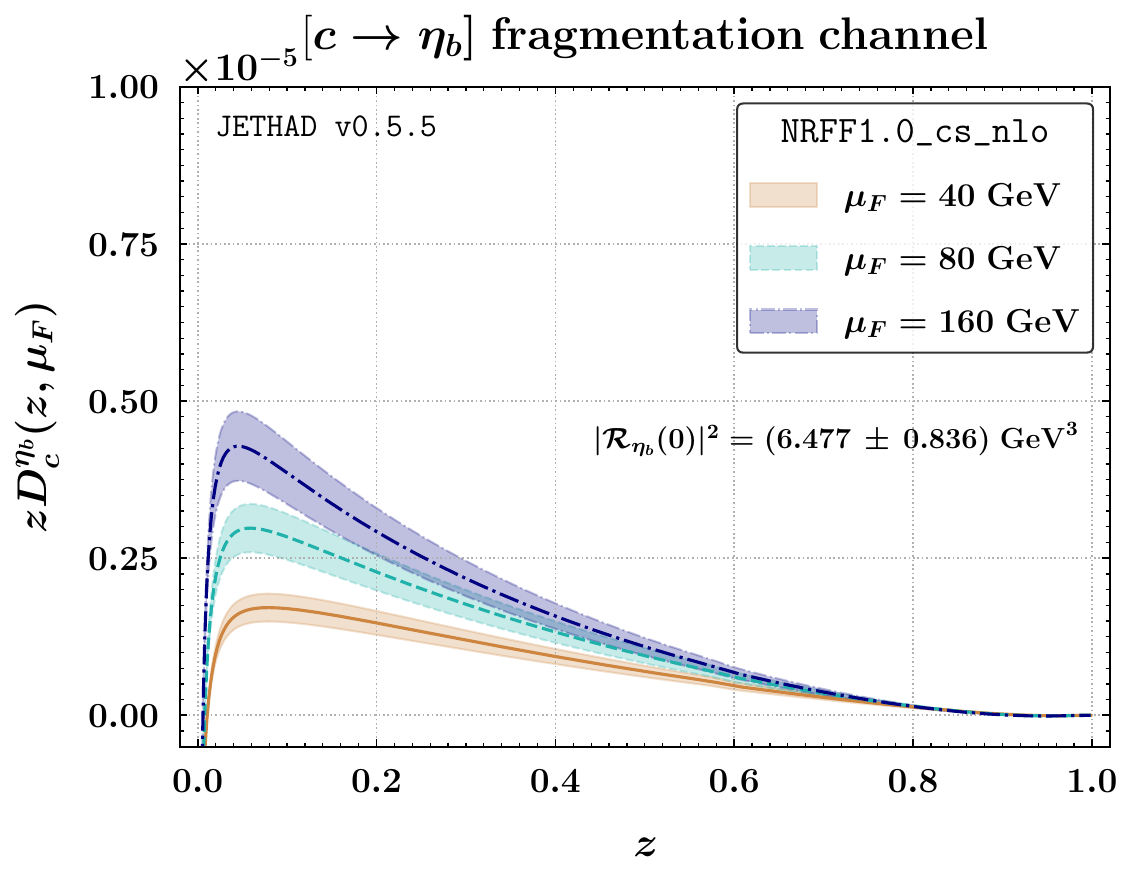}

\caption{$z$ dependence of the {\NRFF} NLO FFs~\cite{Celiberto:2025_NRFF10_cs_eQs} for $\etc[^1S_0^{(1)}]$ (left) and $\etb[^1S_0^{(1)}]$ (right) production at $\mu_F = 40$, 80, and 160~GeV.
The upper (lower) panels show the up-quark (nonconstituent heavy quark) fragmentation channel.
The shaded bands represent the uncertainty associated with the LDMEs.}
\label{fig:NRFF10_FFs-muF_ncq}
\end{figure*}

\vspace{1em}
\noindent
\textbf{Momentum dependence.}
We now examine the $z$ dependence of the {\NRFF} FFs at NLO, shown in Figs.~\ref{fig:NRFF10_FFs-muF_g-Q} and~\ref{fig:NRFF10_FFs-muF_ncq} for $\etc$ (left) and $\etb$ (right) pseudoscalar production. 
Each panel displays the evolved FFs at three representative factorization scales, $\mu_F = 40$, 80, and 160~GeV, with bands encoding LDME uncertainties.

\begin{figure*}[!t]
\centering

 \hspace{-0.00cm}
 \includegraphics[scale=0.440,clip]{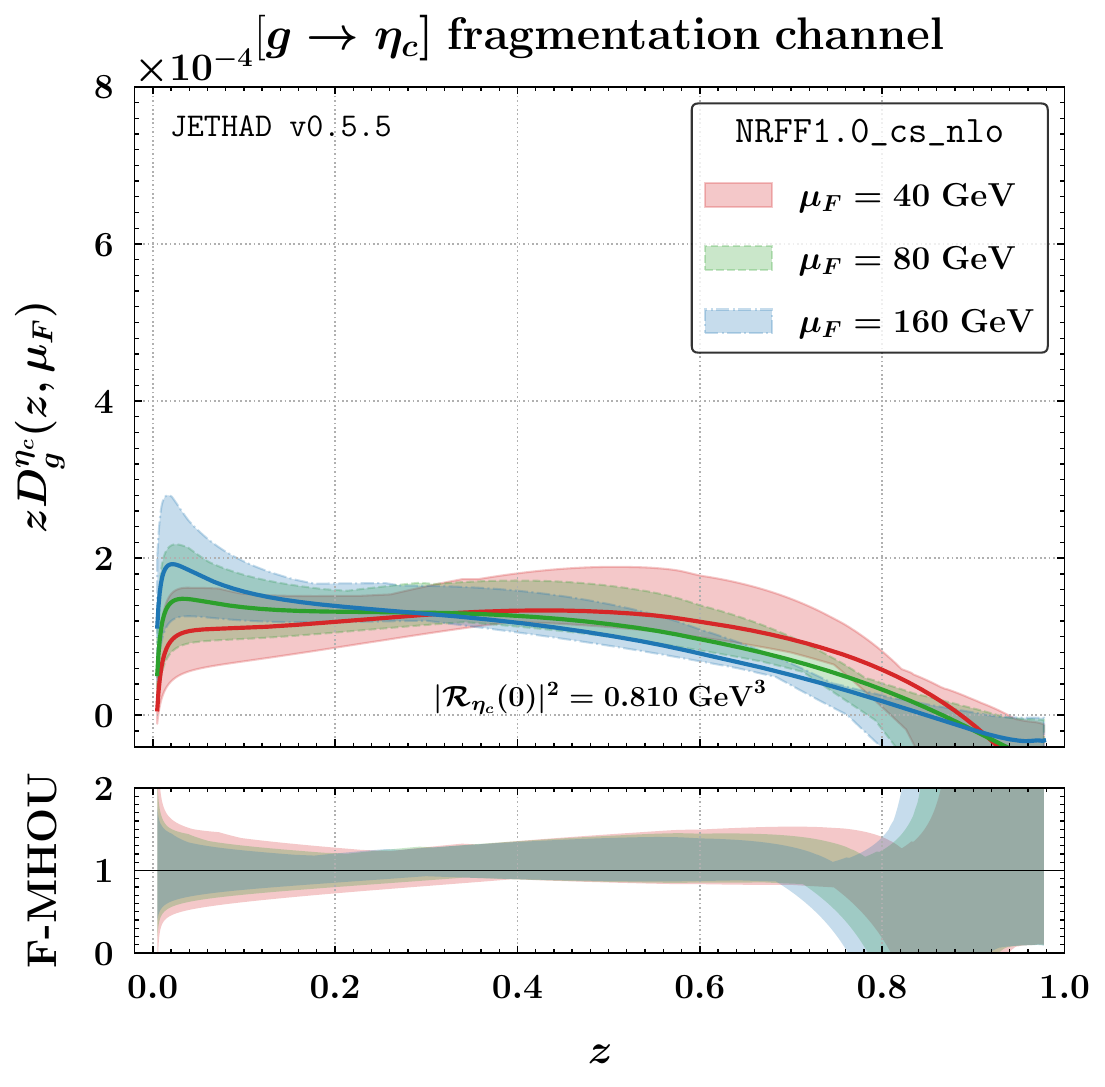}
 \hspace{0.00cm}
 \includegraphics[scale=0.440,clip]{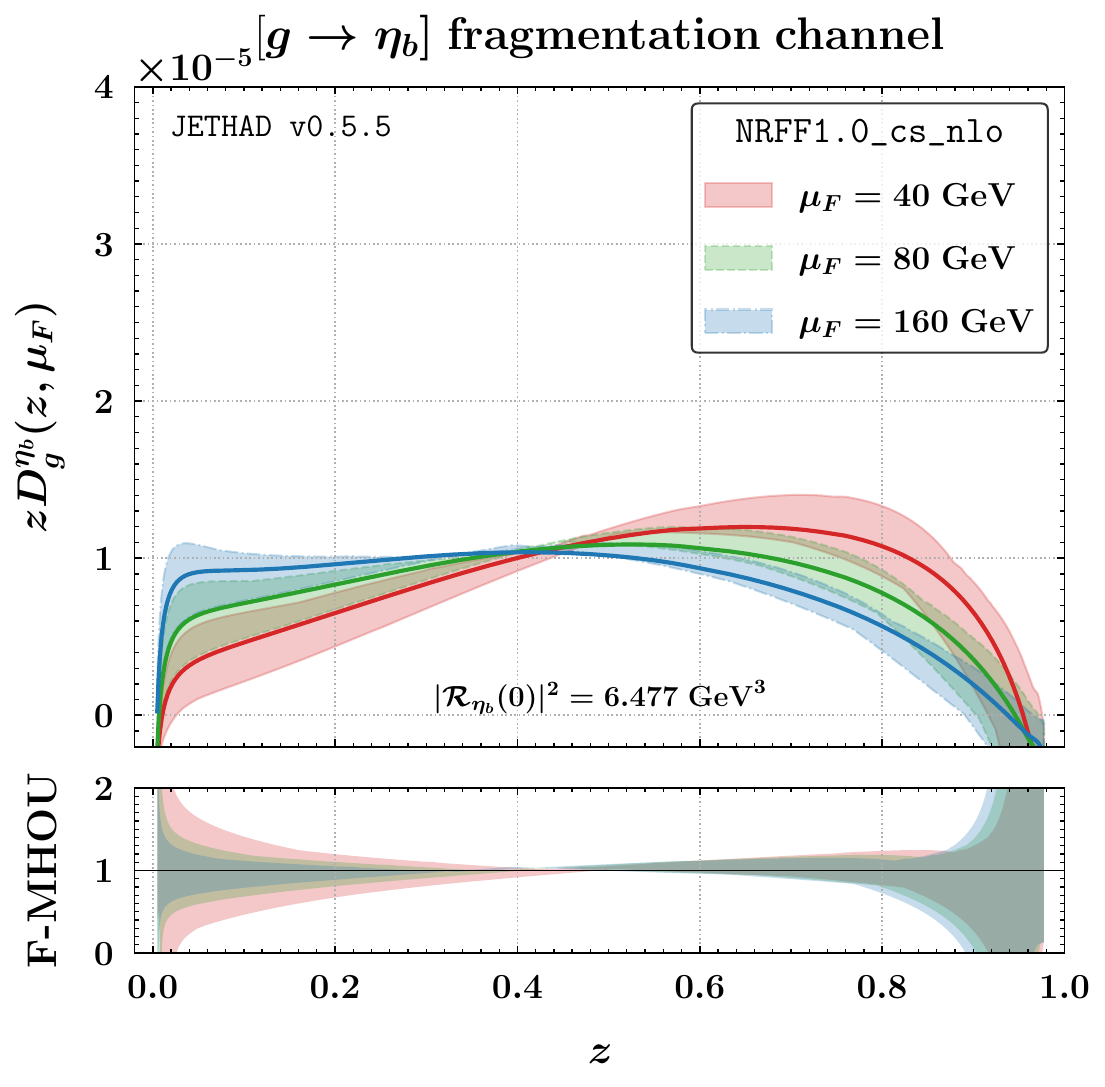}
 
 \vspace{0.35cm}
 
 \hspace{-0.10cm}
 \includegraphics[scale=0.440,clip]{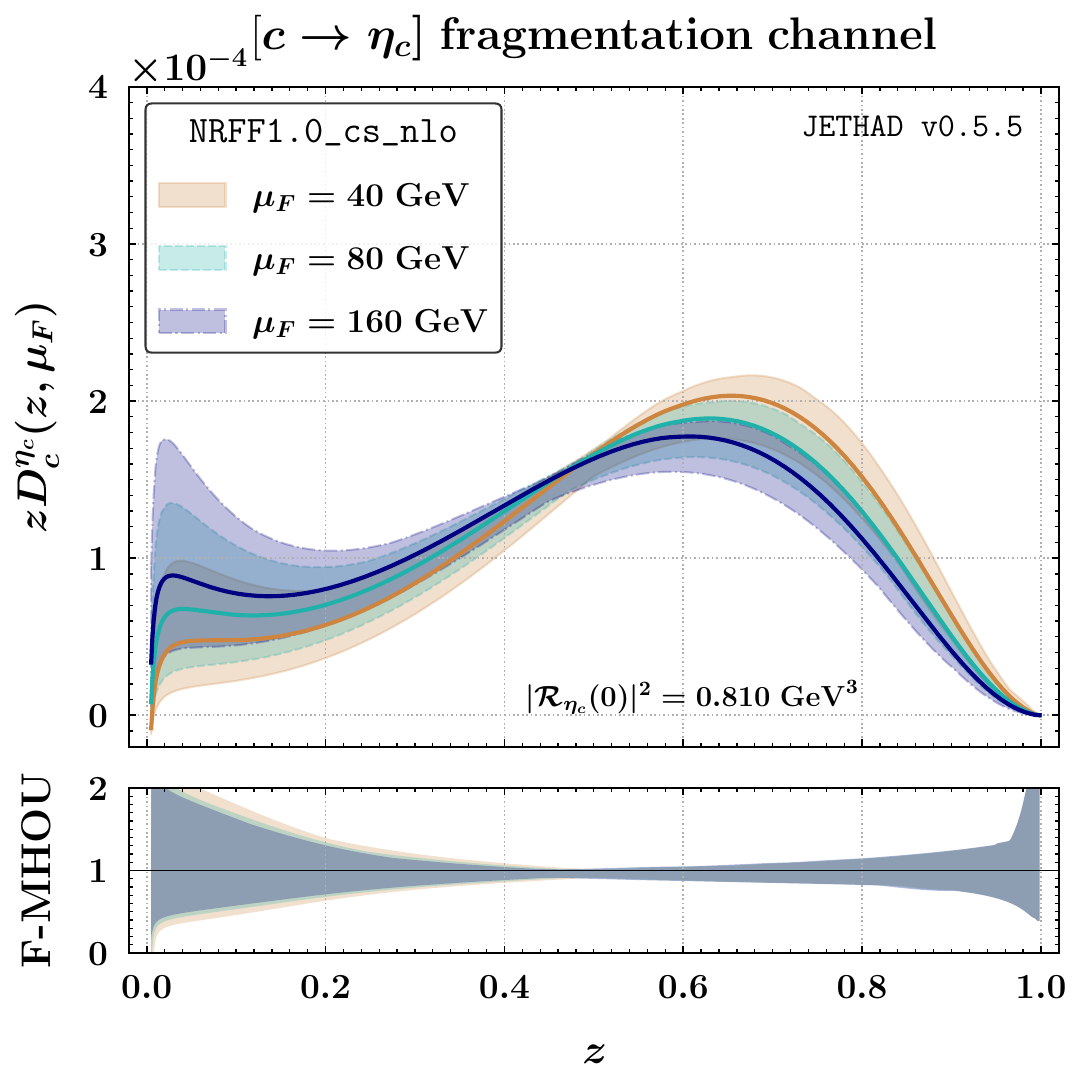}
 \hspace{0.20cm}
 \includegraphics[scale=0.440,clip]{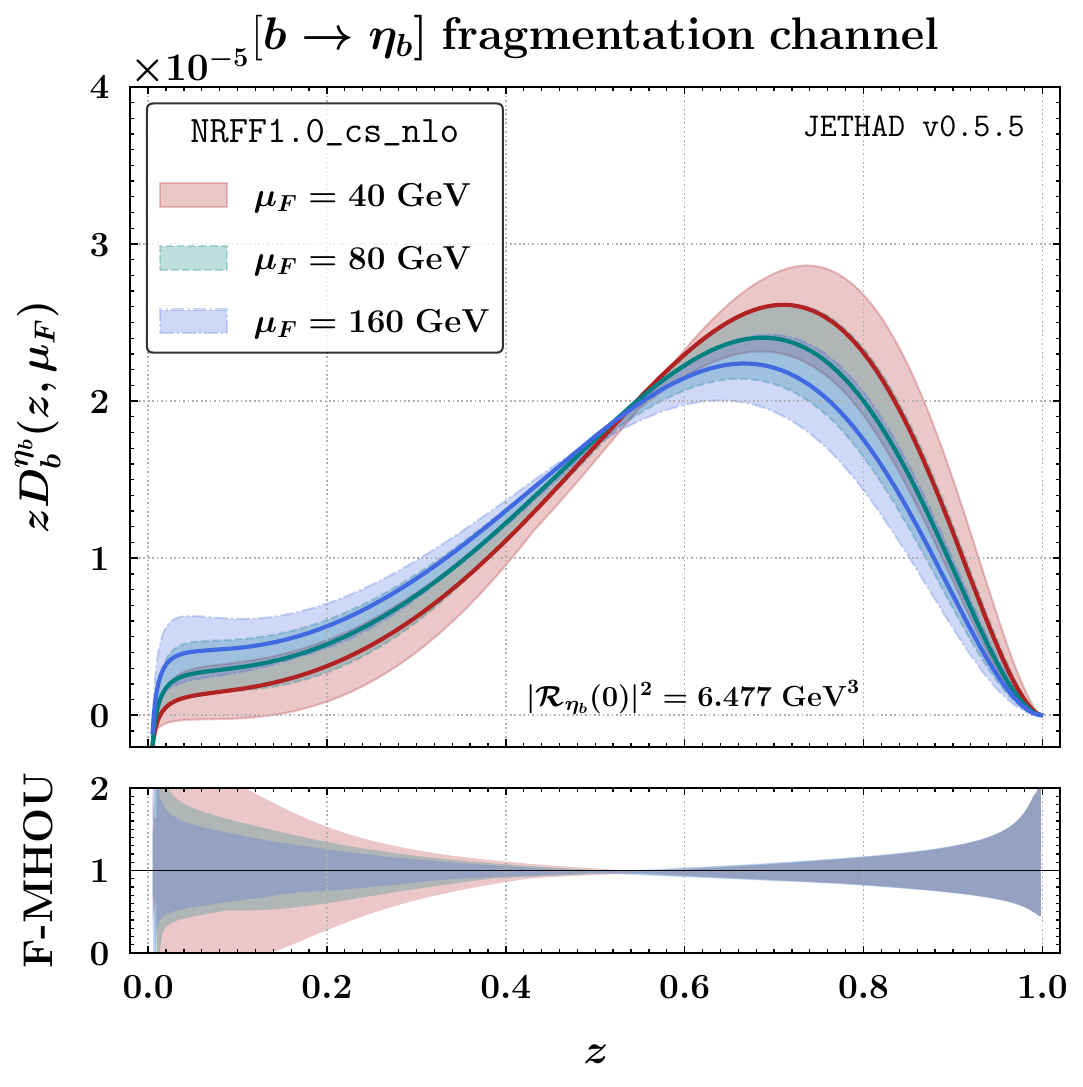}

\caption{$z$ dependence of the {\NRFF} NLO FFs~\cite{Celiberto:2025_NRFF10_cs_eQs} for $\etc[^1S_0^{(1)}]$ (left) and $\etb[^1S_0^{(1)}]$ (right) production at $\mu_F = 40$, 80, and 160~GeV.
The upper (lower) panels show the gluon (constituent heavy quark) fragmentation channel.
The shaded bands represent the uncertainty associated with the F-MHOUs, while LDMEs are kept at their central values as in Eqs.~\eqref{R0_etc} and~\eqref{R0_etb}.
The ancillary panels below each main plot display the ratio between the FFs evaluated at the central values of the evolution and renormalization scales, and the corresponding FFs obtained by varying each of these scales independently by a factor of 2 around its central value.
}
\label{fig:NRFF10_FFs-muF_g-Q_FMHOU}
\end{figure*}

The gluon and constituent quark channels are shown in the upper and lower panels of Fig.~\ref{fig:NRFF10_FFs-muF_g-Q}, respectively.
As expected, heavy quark FFs exhibit a pronounced peak in the moderate-$z$ region, around $z \simeq 0.6 \div 0.8$, reflecting the fact that a large fraction of the constituent quark's momentum is retained by the final-state quarkonium~\cite{Braaten:1993mp}. 
In contrast, gluon-induced FFs display a broader shape, with a maximum at lower $z$ and a gentle enhancement toward the low-$z$ region, due to the larger phase space available for collinear radiation~\cite{Braaten:1993rw}.

We observe a common and clear scale dependence: increasing $\mu_F$ leads to a moderate suppression in the bulk region and an enhancement in the low-$z$ tail, driven by logarithmic scaling violations encoded in the DGLAP evolution. 
This behavior is consistent with previous findings on heavy-flavor and quarkonium fragmentation~\cite{Zhang:2018mlo,Zheng:2021mqr,Zheng:2021ylc,Celiberto:2025dfe,Celiberto:2025ziy}.

Light-quark ($u$, shown as a proxy for $d$ and $s$, see upper panels of Fig.~\ref{fig:NRFF10_FFs-muF_ncq}) and nonconstituent heavy quark channels (lower panels) exhibit a characteristic enhancement at small $z$, followed by a smooth turnover and rapid suppression as [$z \to 1$].
This behavior reflects the expected kinematics of fragmentation from nonconstituent partons: only a limited fraction of the initial energy can be retained by the final-state quarkonium, with the rest dissipated through collinear emissions.

Although globally suppressed compared to constituent channels, these FFs show a pronounced sensitivity to scale evolution.
Increasing the factorization scale $\mu_F$ leads to a marked enhancement of the low-$z$ region, particularly visible at $\mu_F = 160$~GeV, consistent with the accumulation of logarithmic terms $\ln(1/z)$ in the perturbative expansion.

A direct comparison between the [$c \to \etb$] and [$b \to \etc$] channels reveals an intriguing cross-flavor asymmetry: while both are suppressed as they do not contribute to the leading Fock state of the final quarkonium, the different mass hierarchies introduce nontrivial variations in both the shape and normalization of the corresponding FFs.

At $z \simeq 0.05$, the fragmentation probability can increase by a factor of $3 \div 5$ when evolving from $\mu_F = 40$~GeV to 160~GeV, depending on the channel.
Despite their small absolute normalization, such growth at low $z$, combined with favorable parton luminosities at small $x$, may render these channels phenomenologically relevant in forward production environments.

As a general remark, when comparing charmonium and bottomonium channels in Figs.~\ref{fig:NRFF10_FFs-muF_g-Q} and~\ref{fig:NRFF10_FFs-muF_ncq}, we observe that the overall shape is preserved across quark flavors, while the normalization reflects the mass and LDME hierarchy, leading to systematically smaller values for $\etb$ FFs.
In addition, the uncertainty bands associated with [$b \to \etb$] and [$g \to \etb$] channels are visibly broader, as a direct consequence of the more conservative treatment of nonperturbative inputs in the bottom sector (see Sec.~\ref{sssec:LDMEs}).

For completeness, we also present in Fig.~\ref{fig:NRFF10_FFs-muF_g-Q_FMHOU} a first uncertainty study of the F-MHOUs, associated with the variation of perturbative energy scales entering the initial FFs. 
Specifically, we vary independently and simultaneously the scale $Q_0$ used in the DGLAP evolution and the renormalization scale $\mu_{R,0}$ appearing in the SDCs by a factor of two around their canonical values. 
For simplicity, we restrict this study to the gluon and constituent heavy quark fragmentation channels, in analogy with Fig.~\ref{fig:NRFF10_FFs-muF_g-Q}, and we keep the LDMEs fixed at their central values.

The bands shown in the lower insets of each panel represent the envelope of the ratio between the FF computed at central evolution and renormalization scales, and the functions obtained by varying each scale independently by a factor of two around its central value. 
For both $\eta_c$ and $\eta_b$ production, we observe that the F-MHOU bands are relatively narrow in the bulk region, especially for the gluon channel, and tend to grow significantly in the low-$z$ and high-$z$ end points. 
This behavior is consistent with the fact that DGLAP evolution amplifies the sensitivity to scale choices near the phase-space boundaries, where the FFs are enhanced or rapidly decreasing.

Compared to the uncertainty due to LDMEs, F-MHOUs often compete in size, and can even become the dominant source of uncertainty, particularly at small $z$ and in the charm sector. 
These observations support the need for a systematic treatment of F-MHOUs in future analyses, particularly in scenarios where the low-$z$ region is phenomenologically enhanced.

\begin{figure*}[t]
\centering

 \includegraphics[scale=0.42,clip]{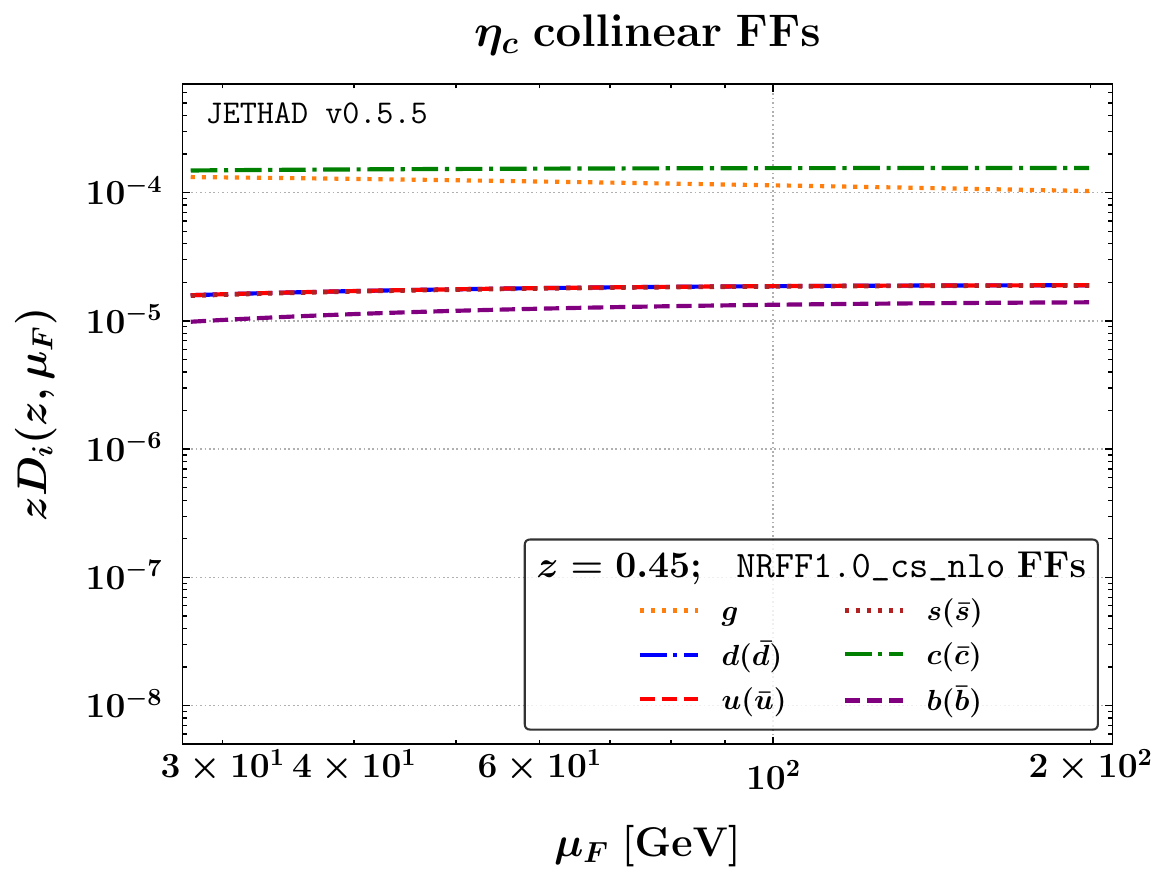}
 \includegraphics[scale=0.42,clip]{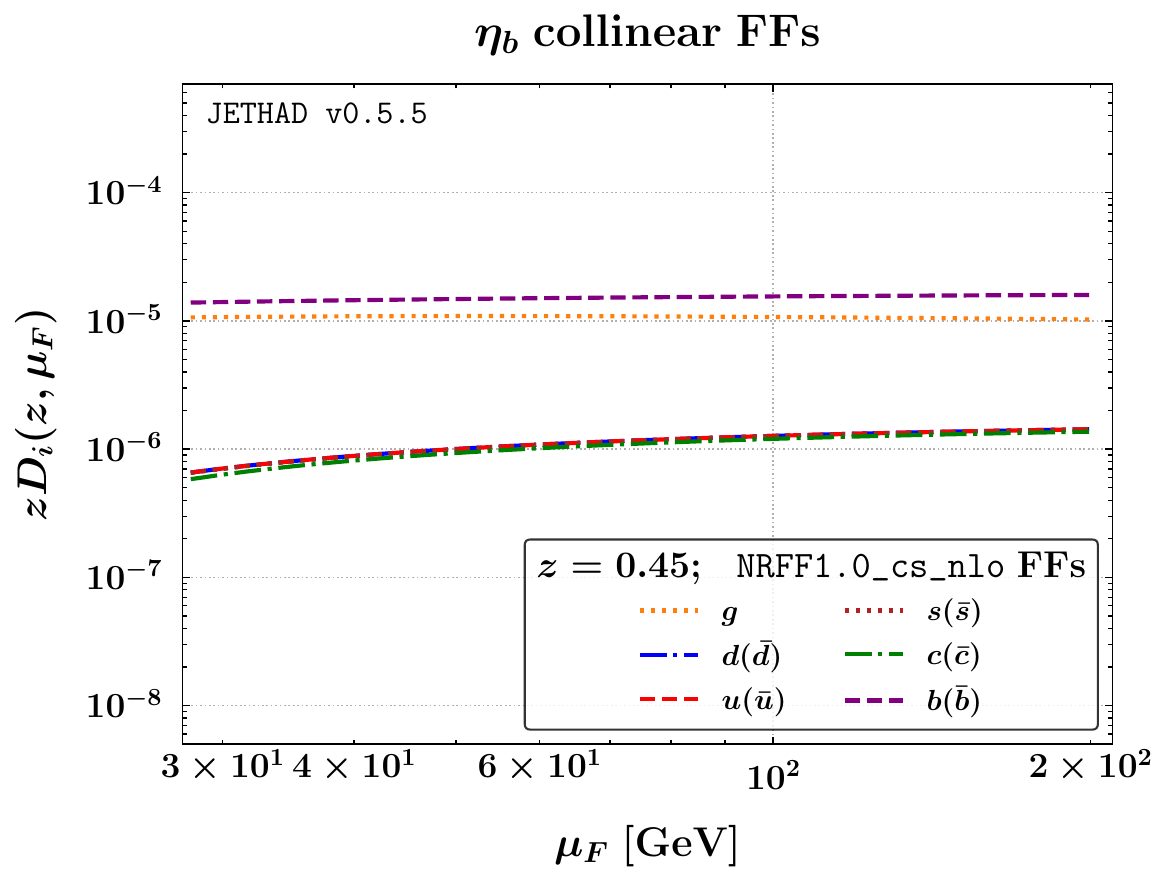}

\caption{Factorization scale dependence, at $z \equiv \langle z \rangle \simeq 4.5 \times 10^{-1}$, of {\NRFF} NLO FFs~\cite{Celiberto:2025_NRFF10_cs_eQs} depicting $\etc[^1S_0^{(1)}]$ (left) and $\etb[^1S_0^{(1)}]$ (right) pseudoscalar quarkonium production.
For simplicity, the analysis is performed using only the central value of the LDMEs.}
\label{fig:NRFF10_FFs-z}
\end{figure*}

\vspace{1em}
\noindent
\textbf{Energy dependence.}
For completeness, in Fig.~\ref{fig:NRFF10_FFs-z} we display the energy dependence of the {\NRFF} NLO FFs for $\etc$ (left panel) and $\etb$ (right panel), multiplied by $z$ and evaluated at a representative momentum fraction $z = \langle z \rangle \simeq 4.5 \times 10^{-1}$. 
This value is typical in phenomenological applications involving high-energy hadron collisions~\cite{Celiberto:2020wpk,Celiberto:2021dzy,Celiberto:2021fdp,Celiberto:2022dyf,Celiberto:2022keu,Celiberto:2024omj}.
For clarity, only the central value of the FFs is shown, with LDME-induced uncertainties not included in the present visualization.

The gluon FF shows an almost flat behavior with $\mu_F$, indicating a radiatively stable evolution across the full range. 
This regular pattern aligns with expectations from timelike DGLAP evolution at fixed $z$, where gluon self-splitting dominates at large $\mu_F$ without inducing significant distortions in the shape or normalization.

The constituent heavy quark channels, [$c \to \etc$] (left) and [$b \to \etb$] (right), lie slightly above or near the gluon-induced FFs and exhibit a mild logarithmic growth with $\mu_F$, driven by positive-valued $P_{qq}$ splitting kernels. 
In contrast, the nonconstituent channels ($u(\bar{u})$, $d(\bar{d})$, $s(\bar{s})$ for both panels, and $b(\bar{b})$ for $\etc$ or $c(\bar{c})$ for $\etb$) are suppressed by about 1 order of magnitude over the full $\mu_F$ spectrum.

Among these, the light-flavor FFs ($u$, $d$, $s$) appear nearly degenerate and indistinguishable, as expected from their negligible mass and universal NRQCD input. 
A more distinctive behavior is seen for the heavy nonconstituent contributions: the [$b \to \etc$] and [$c \to \etb$] FFs are clearly separated from the light-quark plateau and grow visibly with $\mu_F$. 
This trend arises from the explicit mass dependence of the short-distance coefficients, which account not only for the fragmenting quark mass but also for that of the constituent heavy quark within the bound state. 
Although this effect is negligible for $m_{u,d,s} \ll m_Q$, it becomes significant when both masses are heavy, especially for the [$b \to \etc$] channel.

An overall feature emerging from Fig.~\ref{fig:NRFF10_FFs-z} is the smoothness and radiative stability of the evolution profiles across all flavors. 
Such regular behavior, already observed in studies of heavy-light mesons~\cite{Celiberto:2021dzy}, vector quarkonia~\cite{Celiberto:2022dyf}, $B_c$-like systems~\cite{Celiberto:2022keu}, and exotic hadrons~\cite{Celiberto:2023rzw,Celiberto:2024mab,Celiberto:2025dfe,Celiberto:2025ziy}, enhances the theoretical predictivity of the fragmentation framework. 
In particular, the {\NRFF} FFs for pseudoscalar quarkonia inherit the same \emph{natural stabilization}
properties that are crucial for precision applications, especially when controlling the impact of scale uncertainties in high-energy observables.

\begin{figure*}[!t]
\centering

 \includegraphics[scale=0.52,clip]{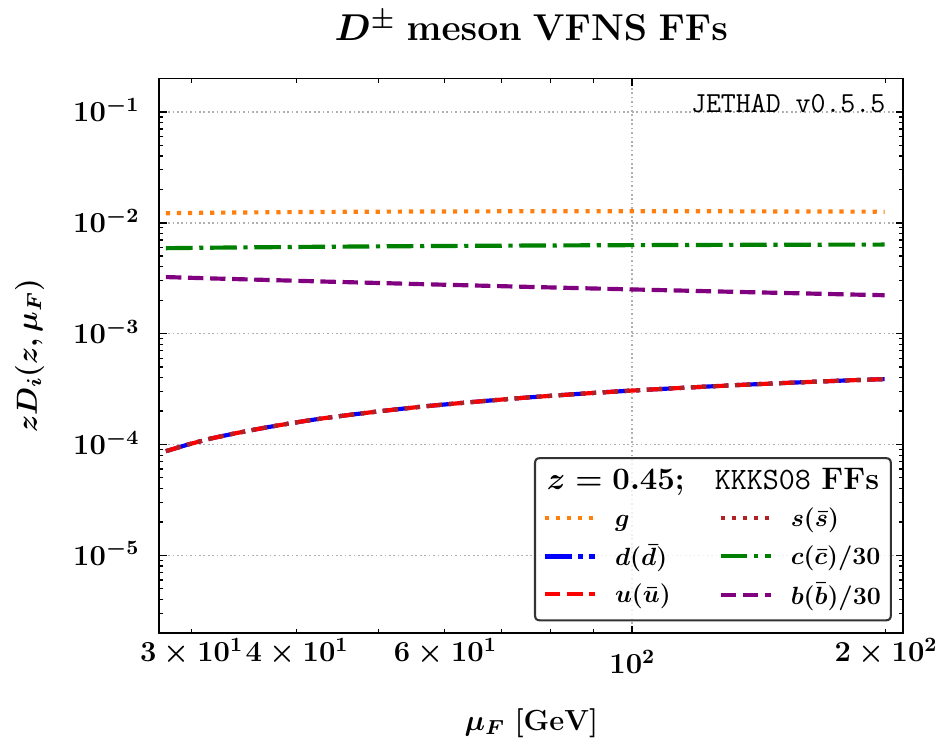}
 \includegraphics[scale=0.52,clip]{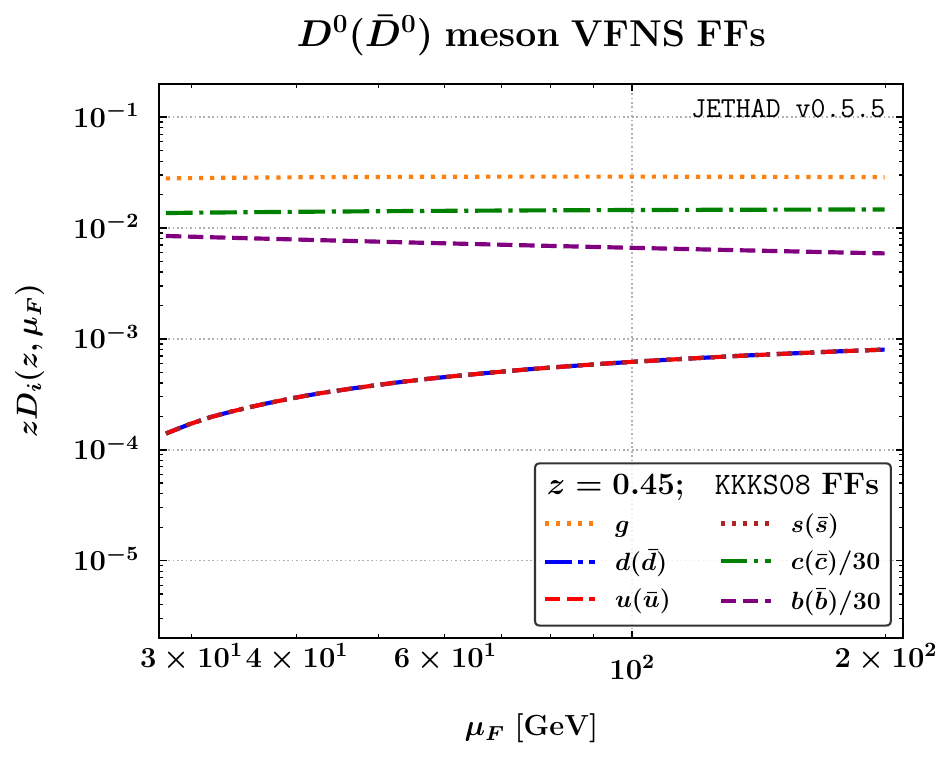}
 
 \vspace{0.35cm}
 
 \includegraphics[scale=0.52,clip]{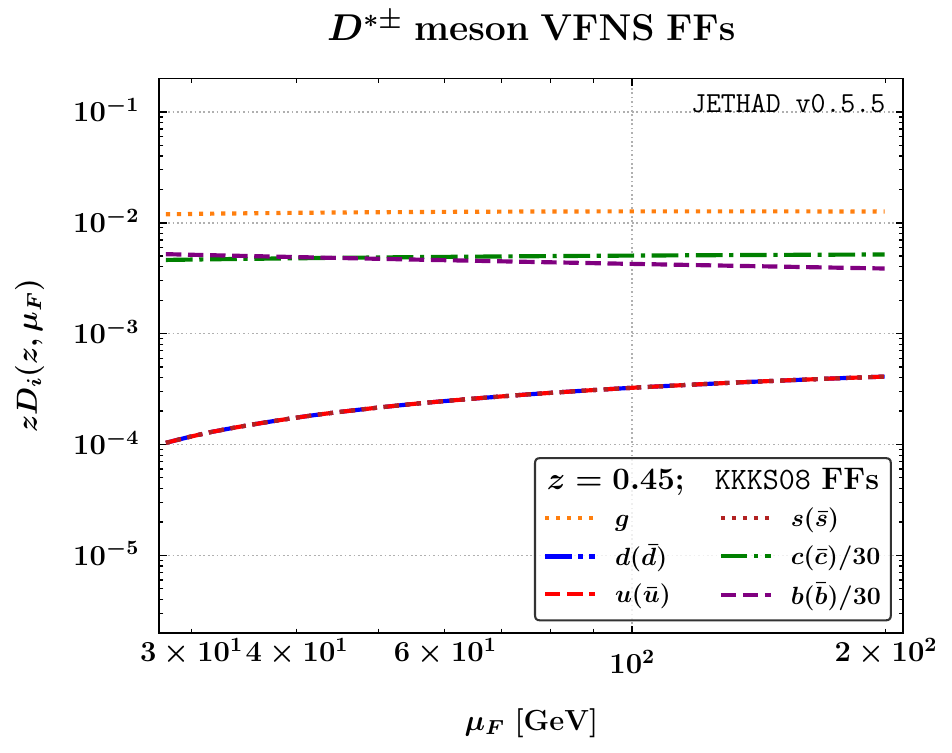}
 \includegraphics[scale=0.52,clip]{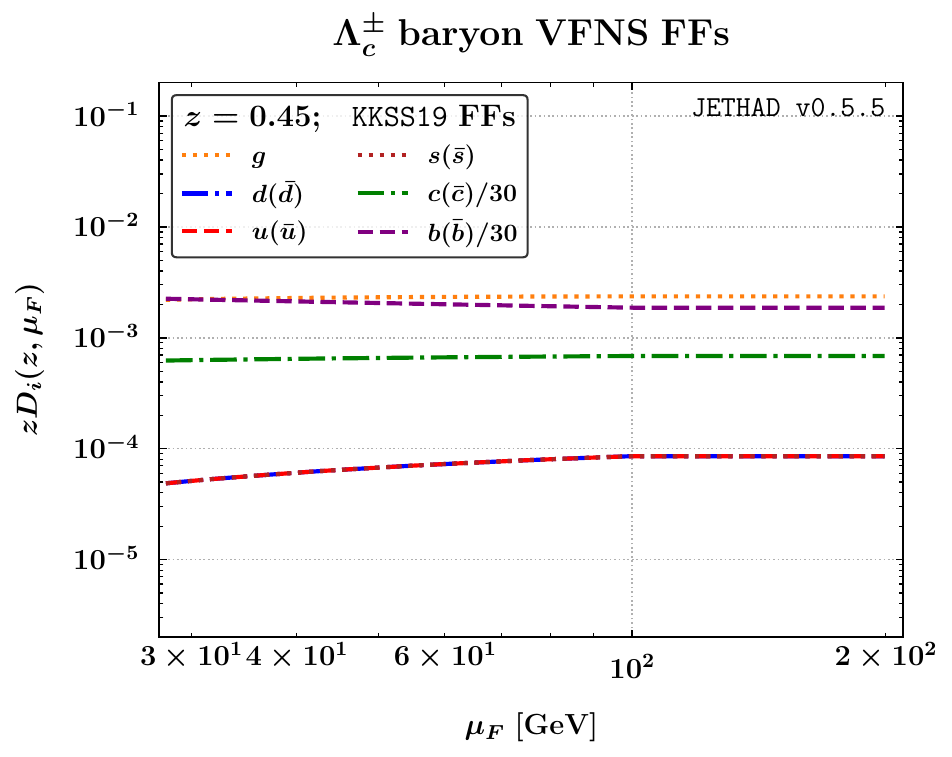}

\caption{Factorization scale dependence, at $z \equiv \langle z \rangle \simeq 4.5 \times 10^{-1}$, of {\tt KKKS08} NLO FFs~\cite{Kniehl:2005de,Kneesch:2007ey} depicting $D^\pm$ (upper left), $D^0(\bar{D}^0)$ (upper right), and $D^{* \pm}$ (lower left) meson emissions, and of {\tt KKSS19} NLO FFs~\cite{Kniehl:2020szu} describing $\Lambda_c^\pm$ baryon production (lower right).}
\label{fig:scH_FFs-z}
\end{figure*}

\begin{figure*}[!t]
\centering

\includegraphics[scale=0.52,clip]{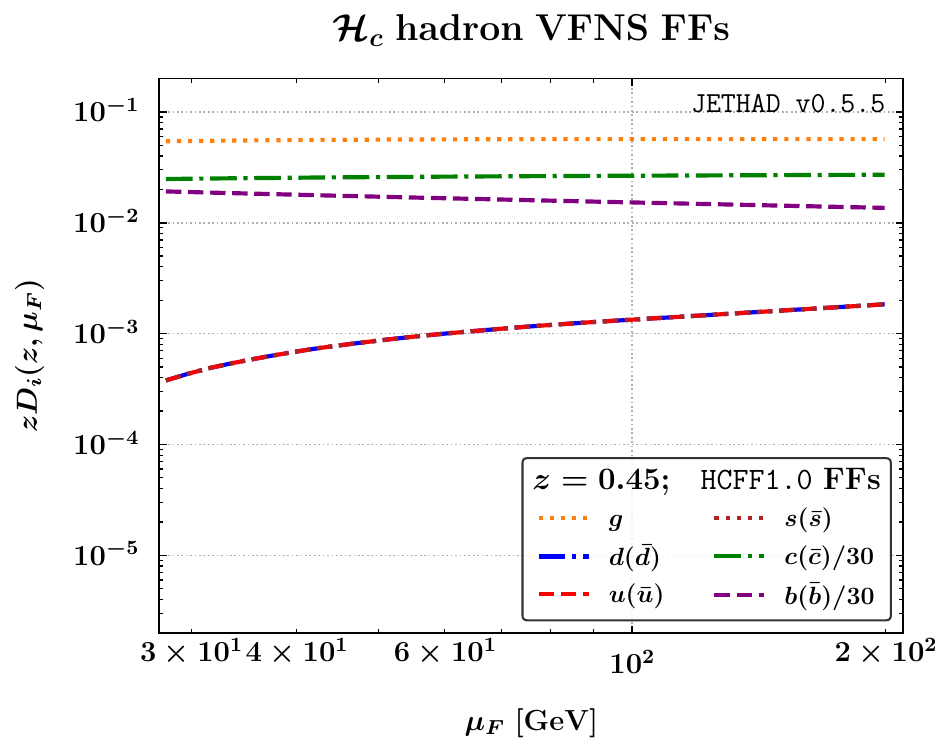}
\includegraphics[scale=0.52,clip]{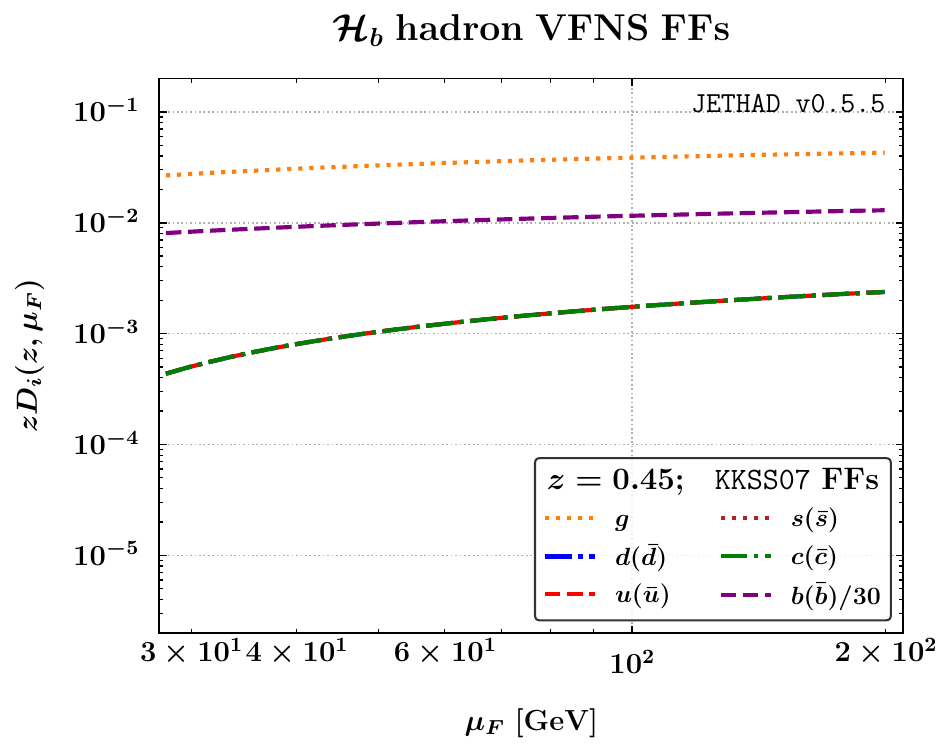}

\caption{Factorization scale dependence, at $z \equiv \langle z \rangle \simeq 4.5 \times 10^{-1}$, of {\HCFF} NLO FFs depicting $\Hc$ hadron emission (left), and of {\tt KKSS07} NLO FFs~\cite{Kniehl:2011bk,Kramer:2018vde} describing $\Hb$ hadron production (right).}
\label{fig:HCFF10_FFs-z}
\end{figure*}

\subsection{\HCFF: FFs for singly charmed hadrons}
\label{ssec:HCFF}

We describe the production of a singly charmed hadron, ${\cal H}_c$, in terms of a new NLO collinear FF parametrization, called {\HCFF}, which we have built in {\LHAPDF} format~\cite{Buckley:2014ana}.
The {\HCFF} set was obtained by combining the output of {\tt KKKS08} NLO FFs for $D$ mesons~\cite{Kniehl:2005de,Kneesch:2007ey}, together with the output of {\tt KKSS19} NLO FFs for $\Lambda_c$ baryons~\cite{Kniehl:2020szu}.
Vice versa, the production of a singly bottomed hadron, ${\cal H}_b$, is portrayed by {\tt KKSS07} NLO FFs~\cite{Kniehl:2011bk,Kramer:2018vde}.
The latter encode both the heavy-light, noncharmed $B$ meson and the $\Lambda_b$ channels.

All these FF determinations were provided by the same collaboration and they share the same extraction technology.
While the interested reader will find technical details in the corresponding references, we limit ourselves to say that these VFNS functions mainly rely upon Bowler~\cite{Bowler:1981sb}, Peterson~\cite{Peterson:1982ak}, or powerlike~\cite{Kartvelishvili:1985ac} based initial-scale inputs for heavy quark species, while light quarks are generated by DGLAP evolution.

Plots of Fig.~\ref{fig:scH_FFs-z} illustrate the $\mu_F$ behavior of NLO functions for $D^\pm$, $D^0(\bar{D}^0)$, $D^{* \pm}$, and $\Lambda_c^\pm$ particles, at $z \equiv \langle z \rangle \simeq 4.5 \times 10^{-1}$.
Starting from them, the {\HCFF} NLO set was constructed. 
The left (right) plot of Fig.~\ref{fig:HCFF10_FFs-z} exhibits $\mu_F$ trend of {\HCFF} ({\tt KKSS07}) functions describing the collinear emission of a $\Hc$ ($\Hb$) hadron.
As already discussed in our previous studies~\cite{Celiberto:2021dzy,Celiberto:2021fdp}, and in analogy to our findings for pseudoscalar quarkonia (see Section~\ref{sssec:NRFF_ns}), we remark that the smooth-behaved, nondecreasing pattern of the [$g \to \Hc$] function, as well as the moderately increasing one of the [$g \to \Hb$] one, represent the key ingredient for the \emph{natural stability}~\cite{Celiberto:2022grc} of the high-energy resummation via heavy-hadron fragmentation~\cite{Bolognino:2022paj,Celiberto:2022kza_Zenodo,Celiberto:2022qbh}.

\section{Theoretical setup: High-energy resummation}
\label{sec:HE_resummation}

Section~\ref{ssec:semihard} begins with a brief survey of recent phenomenological developments aimed at understanding QCD dynamics in the semihard regime.
Section~\ref{ssec:HyF} then introduces the $\NLLp$ HyF formalism and describes its application to the semi-inclusive production of a pseudoscalar quarkonium state accompanied by a jet in hadronic collisions.

\subsection{Semihard processes: A brief review}
\label{ssec:semihard}

Heavy-flavored hadron production offers valuable insight into the high-energy behavior of QCD, particularly in regimes where large logarithms in energy challenge the reliability of fixed-order perturbation theory. 
The Balitsky-Fadin-Kuraev-Lipatov (BFKL) framework~\cite{Fadin:1975cb,Kuraev:1977fs,Balitsky:1978ic} addresses this issue by resumming energy logarithms to all orders, including both leading-logarithmic (LL) terms of the form [$\alpha_s^n (\ln s)^n$ and NLL corrections [$\alpha_s^{n} (\ln s)^{n-1}$.

Within BFKL, cross sections are calculated through transverse momentum convolutions involving a universal NLO Green's function and process-specific emission functions, often referred to as forward impact factors. 
These functions incorporate collinear elements like parton distribution functions (PDFs) and FFs, resulting in a hybrid factorization scheme, named HyF, that blends high-energy and collinear QCD dynamics.

Over the years, the HyF formalism has been applied to a wide range of processes, such as Mueller-Navelet jets within full $\NLL$ accuracy~\cite{Mueller:1986ey,Ducloue:2013hia,Colferai:2015zfa,Celiberto:2015yba,Celiberto:2015mpa,Celiberto:2016ygs,Celiberto:2017ius,deLeon:2021ecb,Celiberto:2022gji,Baldenegro:2024ndr}, dihadron~\cite{Celiberto:2016hae,Celiberto:2017ptm,Celiberto:2017ius,Celiberto:2020rxb,Celiberto:2022rfj} and hadron-jet systems~\cite{Bolognino:2018oth,Bolognino:2019cac,Bolognino:2019yqj,Celiberto:2020wpk,Celiberto:2020rxb,Mohammed:2022gbk,Celiberto:2022kxx}, multijet configurations~\cite{Caporale:2016soq,Caporale:2016xku,Celiberto:2016vhn,Caporale:2016zkc,Celiberto:2017ius}, forward Higgs production~\cite{Nefedov:2019mrg,Hentschinski:2020tbi,Celiberto:2022fgx,Celiberto:2020tmb,Mohammed:2022gbk,Celiberto:2023rtu,Celiberto:2023uuk,Celiberto:2023eba,Celiberto:2023nym,Celiberto:2023rqp,Celiberto:2022zdg,Celiberto:2024bbv,DelDuca:2025vux}, Drell-Yan processes~\cite{Celiberto:2018muu,Golec-Biernat:2018kem}, and heavy-flavor observables~\cite{Boussarie:2017oae,Bolognino:2019ouc,Celiberto:2021dzy,Celiberto:2021fdp,Celiberto:2022dyf,Celiberto:2023fzz,Celiberto:2022grc,Bolognino:2022paj,Celiberto:2022keu,Celiberto:2022kza,Celiberto:2024omj,Celiberto:2025vra}.

Forward production observables have offered a window into small-$x$ gluon dynamics via the unintegrated gluon distribution (UGD), with key measurements at HERA~\cite{Besse:2013muy,Bolognino:2018rhb,Bolognino:2018mlw,Bolognino:2019bko,Bolognino:2019pba,Celiberto:2019slj,Bolognino:2021bjd,Luszczak:2022fkf} and prospects at the EIC~\cite{Bolognino:2021niq,Bolognino:2021gjm,Bolognino:2021bjd,Bolognino:2022uty,Bolognino:2022ndh}.
These developments have also enabled the construction of energy-resummed PDFs~\cite{Ball:2017otu,Abdolmaleki:2018jln,Bonvini:2019wxf,Silvetti:2022hyc,Silvetti:2023suu,Rinaudo:2024hdb,Celiberto:2025nnq} and small-$x$ resummed transverse mo\-men\-tum--dependent (TMD) distributions~\cite{Bacchetta:2020vty,Bacchetta:2024fci,Celiberto:2021zww,Bacchetta:2021oht,Bacchetta:2021lvw,Bacchetta:2021twk,Bacchetta:2022esb,Bacchetta:2022crh,Bacchetta:2022nyv,Celiberto:2022omz,Bacchetta:2023zir}.

In heavy-hadron emissions, observables such as ${\rm \Lambda}_c$~\cite{Celiberto:2021dzy} and $b$-hadron production~\cite{Celiberto:2021fdp} have revealed effective strategies to address the issue of unnatural scales that often affect semihard processes. 
Unlike their light-flavored counterparts, which exhibit enhanced sensitivity to threshold effects and large NLL corrections~\cite{Bolognino:2018oth,Celiberto:2020wpk}, heavy-flavored hadrons tend to exhibit a \emph{natural stabilization} pattern~\cite{Celiberto:2022grc}, driven by collinear fragmentation within a VFNS.

This behavior has inspired the development of VFNS FFs grounded in NRQCD-based initial conditions~\cite{Braaten:1993mp,Zheng:2019dfk,Braaten:1993rw,Chang:1992bb,Braaten:1993jn,Ma:1994zt,Zheng:2019gnb,Zheng:2021sdo,Feng:2021qjm,Feng:2018ulg}, spanning a wide range of final states---from vector quarkonia~\cite{Celiberto:2022dyf,Celiberto:2023fzz} to heavier mesons like $\BCs$ and $\Bss$~\cite{Celiberto:2022keu,Celiberto:2024omj}.
This stabilization pattern has paved the way for new investigations into rare exotic hadrons, supporting leading-power-fragmentation studies into triply heavy baryons \cite{Celiberto:2025ogy}, doubly and fully heavy tetraquarks~\cite{Celiberto:2023rzw,Celiberto:2024beg,Celiberto:2024mab,Celiberto:2025dfe,Celiberto:2025ziy}, and five-quark configurations such as pentacharm states~\cite{Celiberto:2025ipt}.

\subsection{HyF distributions at NLL/NLO$^+$}
\label{ssec:HyF}

The reaction under scrutiny is
\begin{equation}
\label{reaction}
{\rm p}(P_1) + {\rm p}(P_2) \rightarrow \etQ(q_1, y_1, \varphi_1) + {\cal X} + {\cal P}(q_2, y_2, \varphi_2) \;,
\end{equation}
where ${\rm p}(P_{1,2})$ denotes an incoming proton with momentum $P_{1,2}$, $\etQ(q_1, y_1, \varphi_1)$ represents a pseudoscalar quarkonium ($\etc$ or $\etb$) produced with momentum $q_1$ and rapidity $y_1$, while ${\cal P}(q_2, y_2, \varphi_2)$ inclusively refers to a singly heavy-flavored hadron $\HQ$ ($\Hc$ or $\Hb$) or a jet emitted with momentum $q_2$ and rapidity $y_2$, and ${\cal }$ accounts for all undetected final-state particles.
The azimuthal angle distance between $\etQ$ and ${\cal P}$ is $\varphi = \varphi_1 - \varphi_2$.
By $\Hc$, we consider an inclusive hadron state comprising the sum over fragmentation channels into singly charmed mesons ($D^\pm$, $D^0$, and $D^{*\pm}$) as well as $\Lambda_c^\pm$ hyperons.
Similarly, $\Hb$ denotes an inclusive configuration involving noncharmed $B$ mesons and $\Lambda_b^0$ baryons~\cite{Celiberto:2021fdp}.

To ensure a semihard diffractive configuration in the final state, the observed transverse momenta $|\vec q_{T_{1,2}}|$ must be large, and a sizable rapidity separation, $\DY = y_1 - y_2$, is required.
Additionally, the transverse momentum ranges must be sufficiently high to justify a description of the quarkonium production mechanism in terms of single-parton VFNS collinear fragmentation.

The momenta of the incoming protons are taken as Sudakov vectors satisfying the relations $P_1^2 = P_2^2 = 0$ and $2 (P_1 \cdot P_2) = s$.
The four-momenta of the final-state particles can then be decomposed as
\begin{equation}
\label{Sudakov}
 q_{1,2} = x_{1,2} \, P_{1,2} - \frac{q_{1,2 \perp}^2}{x_{1,2} s} \, P_{2,1} + q_{1,2 \perp} \;,
\end{equation}
with $q_{1,2 \perp}^2 \equiv - \vec q_{T_{1,2}}^{\,2}$, and the longitudinal momentum fractions of the outgoing particles, $x_{1,2}$, being related to their rapidities via
$y_{1,2} = \pm \ln(x_{1,2} \sqrt{s}/ |\vec q_{T_{1,2}}|)$.
This implies $\drv y_{1,2} = \pm \drv x_{1,2}/x_{1,2}$, and the rapidity separation is given by
\begin{equation}
\label{Delta_Y}
 \DY = y_1 - y_2 \equiv \ln \frac{x_1 x_2 s}{|\vec q_{T_1}| |\vec q_{T_2}|} \;.
\end{equation}

\begin{figure*}[!t]
\centering
\includegraphics[width=0.475\textwidth]{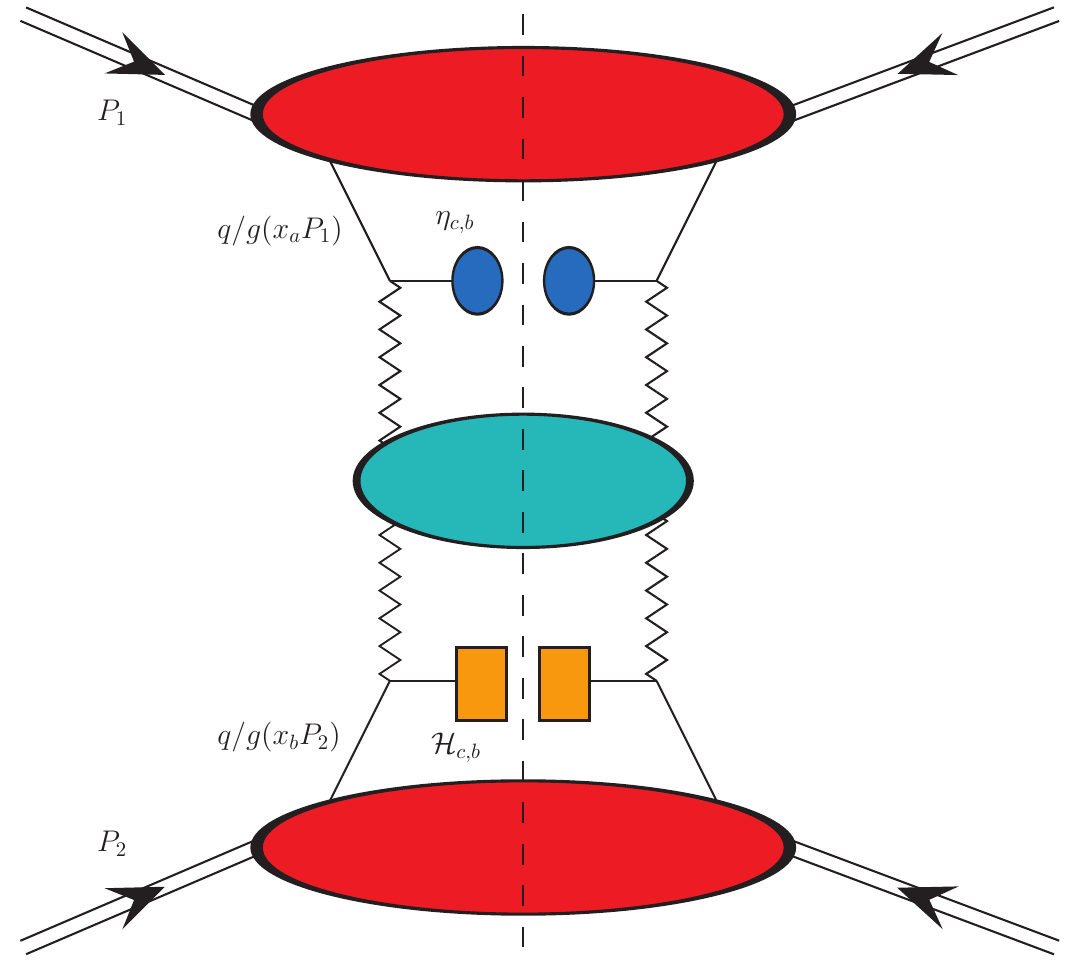}
\hspace{0.55cm}
\includegraphics[width=0.475\textwidth]{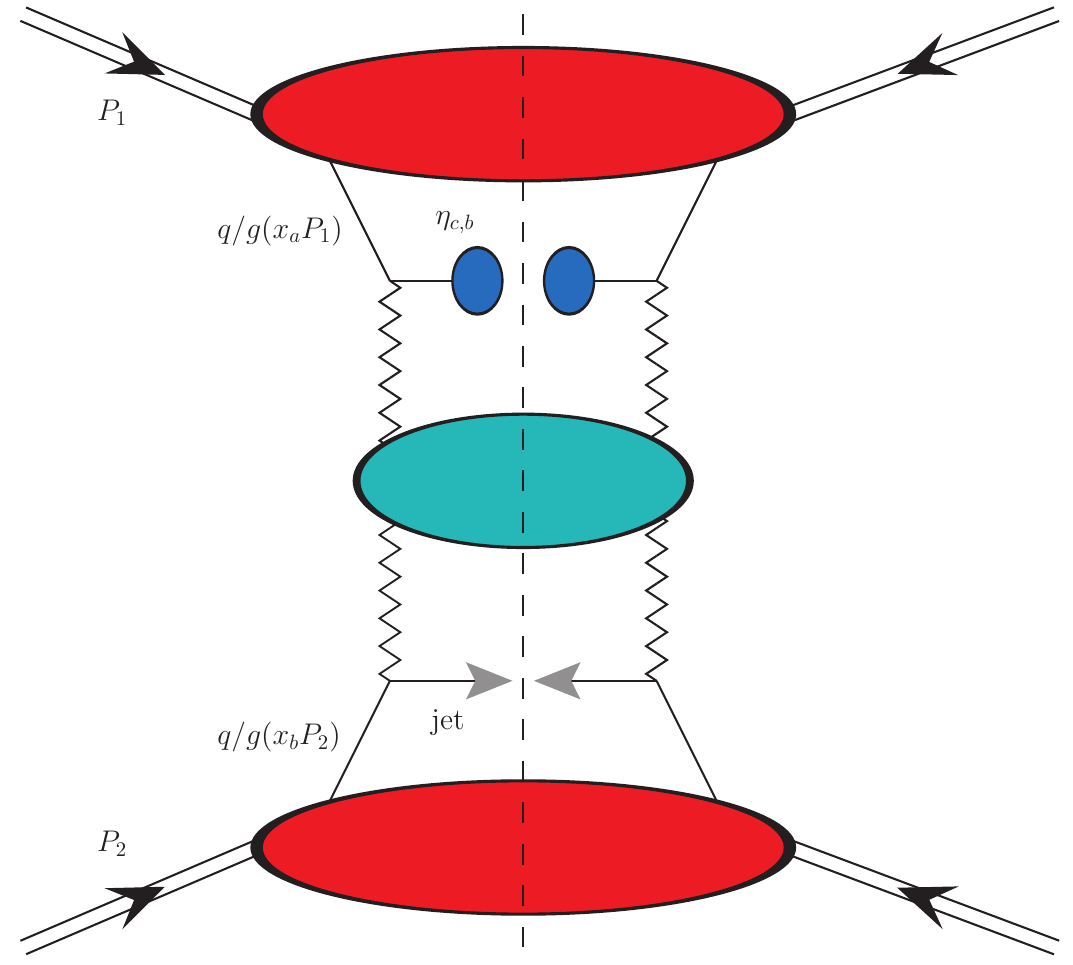}

\caption{Hybrid factorization for the  semi-inclusive hadroproduction of a $\etcb$ quarkonium accompanied by a $\Hcb$ hadron (left) or a light jet (right).
Blue ovals and orange squares respectively represent $\etcb$ and $\Hcb$ collinear FFs, while gray arrows portray jet emissions. 
Red ovals depict parent-proton collinear PDFs.
The BFKL Green's function (turquoise blob) is connected to the two singly off-shell emission functions by waggle Reggeon lines.}
\label{fig:reactions}
\end{figure*}

Within the pure collinear factorization framework, the differential LO cross section for the two-hadron process [$\etQ + \HQ$ in the left diagram of Fig.~\eqref{fig:reactions} can be expressed as a one-dimensional convolution involving the partonic hard-scattering factor, the proton PDFs, the $\etQ$ FFs and the $\HQ$ ones.
One has
\begin{eqnarray}
\label{sigma_collinear_QQ}
&&\hspace{-0.25cm}
\frac{\drv\sigma^{[{\rm pp} \,\to\, \etQ \,+\, {\rm jet}]}_{\rm LO \, [collinear]}}{\drv x_1\drv x_2\drv ^2\vec q_{T_1}\drv ^2\vec q_{T_2}}
= \hspace{-0.25cm} \sum_{i,j=q,{\bar q},g}\int_0^1 \hspace{-0.20cm} \drv x_a \!\! \int_0^1 \hspace{-0.20cm} \drv x_b\ f_i\left(x_a\right) f_j\left(x_b\right) 
\nonumber \\
&&\quad\times \, 
\int_{x_1}^1 \hspace{-0.15cm} \frac{\drv \xi_1}{\xi_1} \, D^{\etQ}_i\left(\frac{x_1}{\xi_1}\right) 
\int_{x_1}^2 \hspace{-0.15cm} \frac{\drv \xi_2}{\xi_2} \, D^{\HQ}_j\left(\frac{x_2}{\xi_2}\right)
\nonumber \\
&&\quad\times \, 
\frac{\drv {\hat\sigma}_{i,j}\left(\hat s\right)}
{\drv x_1\drv x_2\drv ^2\vec q_{T_1}\drv ^2\vec q_{T_2}}\;,
\end{eqnarray}
where the indices $i$ and $j$ run over all parton types, excluding the (anti)top quark, which does not hadronize.
Here, $f_{i,j}(x_{1,2}, \mu_F)$ are the proton PDFs.
Moreover, $D^{\etQ,\HQ}_{i,j}(x_{1,2}/\xi_{1,2} \equiv z_{1,2}, \mu_F)$ denotes $\etQ$ or $\HQ$ FFs.
The quantities $x_{1,2}$ and $\xi_{1,2}$ are the longitudinal momentum fractions of the incoming and fragmenting partons, respectively, while $\drv \hat\sigma_{ij}$ is the partonic cross section.

Analogously, the differential LO cross section for the single-hadron process [$\etQ \,+\, {\rm jet}$ in the right diagram of Fig.~\eqref{fig:reactions} reads as a one-dimensional convolution partonic hard-scattering factor, the proton PDFs and the $\etQ$ FFs.
One writes
\begin{eqnarray}
\label{sigma_collinear_QJ}
&&\hspace{-0.25cm}
\frac{\drv\sigma^{[{\rm pp} \,\to\, \etQ + \HQ]}_{\rm LO \, [collinear]}}{\drv x_1\drv x_2\drv ^2\vec q_{T_1}\drv ^2\vec q_{T_2}}
= \hspace{-0.25cm} \sum_{i,j=q,{\bar q},g}\int_0^1 \hspace{-0.20cm} \drv x_a \!\! \int_0^1 \hspace{-0.20cm} \drv x_b\ f_i\left(x_a\right) f_j\left(x_b\right) 
\nonumber \\
&&\quad\times \, 
\int_{x_1}^1 \hspace{-0.15cm} \frac{\drv \xi}{\xi} \, D^{\etQ}_i\left(\frac{x_1}{\xi}\right) 
\frac{\drv {\hat\sigma}_{i,j}\left(\hat s\right)}
{\drv x_1\drv x_2\drv ^2\vec q_{T_1}\drv ^2\vec q_{T_2}}\;.
\end{eqnarray}
For brevity, the explicit dependence on the factorization scale $\mu_F$ is omitted.

In our HyF framework, which combines high-energy and collinear factorization, the differential cross section is formulated as a transverse momentum convolution involving the BFKL Green's function and two singly off-shell impact factors.
The cross section is expressed in terms of a Fourier expansion over azimuthal angle coefficients, ${\cal C}_{n \, \ge \, 0}$, as follows
\begin{equation}
\label{dsigma_Fourier}
\hspace{-0.22cm}
\frac{(2\pi)^2 \, \drv \sigma}{\drv y_1 \drv y_2 \drv |\vec q_{T_1}| \drv |\vec q_{T_2}| \drv \varphi_1 \drv \varphi_2} =
\left[ {\cal C}_0 + 2 \sum_{n=1}^\infty \cos (n \varphi) \,
{\cal C}_n \right] \,,
\end{equation}
where $\varphi \equiv \varphi_1 - \varphi_2 - \pi$.
In the $\overline{\text{MS}}$ renormalization scheme~\cite{PhysRevD.18.3998}, and using the HyF formalism, the azimuthal coefficients read
\begin{equation}
\label{Cn_NLLp_MSb}
\CnNLLp =
\frac{e^{\DY}}{s}
\int_{-\infty}^{+\infty} \drv \nu \, e^{{\DY} \bar \alpha_s(\mu_R)
\chi^{\rm NLO}(n,\nu)}
\end{equation}
\[
\times \, \alpha_s^2(\mu_R) \biggl\{ \E_1^{\rm NLO}(n,\nu,|\vec q_{T_1}|, x_1)[\E_2^{\rm NLO}(n,\nu,|\vec q_{T_2}|,x_2)]^*
\]
\[
+ \!
\left.
\alpha_s^2(\mu_R) \DY \frac{\beta_0}{4 \pi} 
\chi(n,\nu) \!
\left[\ln\left(|\vec q_{T_1}| |\vec q_{T_2}|\right) \!+\! \frac{i}{2} \frac{\drv}{\drv \nu} \! \ln\frac{\E_1}{\E_2^*}\right]
\right\}
.
\]

In this formula, $\bar \alpha_s(\mu_R) \equiv \alpha_s(\mu_R) N_c/\pi$ is the QCD running coupling and $\beta_0 = 11N_c/3 - 2 n_f/3$ is the first coefficient of the QCD $\beta$-function.
We employ a two-loop evolution for the coupling, with $\alpha_s(m_Z)=0.118$ and a dynamic number of active flavors $n_f$.

The high-energy kernel exponentiated in Eq.~\eqref{Cn_NLLp_MSb} resums energy logarithms at LL and NLL:
\begin{eqnarray}
\label{chi}
\chi^{\rm NLO}(n,\nu) = \chi(n,\nu) + \bar\alpha_s \hat \chi(n,\nu) \;,
\end{eqnarray}
where $\chi(n,\nu)$ is the eigenvalue of the LO BFKL kernel
\begin{eqnarray}
\label{kernel_LO}
\chi\left(n,\nu\right) = -2 \gamma_{\rm E} - 2 \, {\rm Re} \left\{ \psi\left(\frac{1+n}{2} + i \nu \right) \right\} \,,
\end{eqnarray}
with $\gamma_{\rm E}$ denoting the Euler-Mascheroni constant, and $\psi(z) \equiv \Gamma^\prime(z)/\Gamma(z)$ being the digamma function.
The term $\hat\chi(n,\nu)$ in Eq.~\eqref{chi} encodes the NLO correction to the BFKL kernel
\begin{eqnarray}
\label{chi_NLO}
&\hat \chi&\left(n,\nu\right) = \bar\chi(n,\nu)+\frac{\beta_0}{8 N_c}\chi(n,\nu)
\\ \nonumber &\times& \left\{\chi(n,\nu)+\frac{10}{3}+2\ln\left[\mu_R^2/(|\vec q_{T_1}| |\vec q_{T_2}|)\right]\right\} \;,
\end{eqnarray}
where $\bar\chi(n,\nu)$ is the characteristic NLO correction function~\cite{Kotikov:2000pm}.

The two functions
\begin{eqnarray}
\label{EFs}
\E_{\rm 1,2}^{\rm NLO}(n,\nu,|\vec q_{T_{1,2}}|,x_{1,2}) =
\E_{1,2} +
\alpha_s(\mu_R) \, \hat \E_{1,2}
\end{eqnarray}
represent the NLO impact factors depicting the forward or backward emission of the final-state particles.
These emission functions are computed in Mellin space and projected onto the eigenfunctions of the LO.
For the production of $\etQ$ and $\HQ$ hadrons, we employ the NLO result from Ref.~\cite{Ivanov:2012iv}.
Although originally derived for light-flavored hadrons, this expression is compatible with our VFNS description of heavy baryons, provided that the transverse momenta exceed the heavy quark thresholds relevant for DGLAP evolution.

At LO, the hadron impact factor reads
\begin{equation}
\begin{split}
\label{LOHEF}
\E_H(n,\nu,|\vec q_{T_1}|,x) = \delta_c \, |\vec q_{T_1}|^{2i\nu-1} \int_x^1 \frac{\drv \xi}{\xi} \; \hat{x}^{1-2i\nu} 
\\
\times \, \Big[\rho_c f_g(\xi)D_g^H\left(\hat{x}\right)
\sum_{i=q,\bar q}
+
f_i(\xi)D_i^H\left(\hat{x}\right)\Big] \;,
\end{split}
\end{equation}
where $\hat{x} = x/\xi$, $\delta_c = 2 \sqrt{C_F/C_A}$, and $\rho_c = C_A/C_F$.
Here, $C_F = (N_c^2-1)/(2N_c)$ and $C_A = N_c$ are the usual QCD Casimir operators, corresponding to gluon emissions from quarks and gluons, respectively.
The full expression for $\E_H^{\rm NLO}$ can be found in Ref.~\cite{Ivanov:2012iv}.

The LO emission function for the jet reads
\begin{equation}
\begin{split}
\label{LOJEF}
\E_J(n,\nu,|\vec q_{T_2}|,x) &= 
\delta_c 
\,
|\vec q_{T_2}|^{2i\nu-1} 
\\
&\times \,
\Big[\rho_c f_g(x)
\sum_{j=q,\bar q}
f_j(x)\Big] \,.
\end{split}
\end{equation}
The NLO correction to the jet impact factor was calculated in Refs.~\cite{Ivanov:2012iv,Ivanov:2012ms} and has been adapted for numerical implementation.
It relies on a \emph{small-cone} approximation for jet definition~\cite{Furman:1981kf,Aversa:1988vb}, using a \emph{cone-type} jet algorithm as outlined in Ref.~\cite{Colferai:2015zfa}.

Equations~(\ref{Cn_NLLp_MSb}),(\ref{LOHEF}), and(\ref{LOJEF}) illustrate the structure of our hybrid high-energy and collinear factorization framework.
Within this approach, the cross section is formulated through the BFKL formalism, with the Green's function and the emission functions serving as the principal components.
The Green's function governs the resummation of large logarithmic contributions arising in the high-energy limit, while the emission functions encode the parton distribution functions (PDFs) and FFs, thereby linking collinear factorization with high-energy dynamics.

The `$+$' superscript in $\CnNLLp$ denotes that the expression for the azimuthal coefficients in Eq.~(\ref{Cn_NLLp_MSb}) includes corrections that go beyond strict next-to-leading logarithmic (NLL) accuracy.
These refinements originate from two primary sources: the exponentiation of NLO corrections to the BFKL kernel and the contributions arising from the NLO terms in the impact factors.
Consequently, the resulting azimuthal coefficients offer a more accurate and detailed description of the dynamics, incorporating subtle effects that are crucial for precision predictions in processes sensitive to both high-energy and collinear logarithms.

By neglecting all NLO contributions in Eq.~\eqref{Cn_NLLp_MSb}, one recovers the pure LL limit of the azimuthal coefficients.
In this regime, the expression becomes
\begin{equation}
\begin{split}
\label{Cn_LL_MSb}
&\CnLL = 
\frac{e^{\DY}}{s}
\int_{-\infty}^{+\infty} \drv \nu \, e^{{\DY} \bar \alpha_s(\mu_R)\chi(n,\nu)}
\\[0.18cm] &\hspace{0.15cm}
\times \, \alpha_s^2(\mu_R) \, \E_1(n,\nu,|\vec q_{T_1}|, x_1)[\E_2(n,\nu,|\vec q_{T_2}|,x_2)]^* \;.
\end{split}
\end{equation}

In our notation, the $\NLLpp$ label indicates the perturbative accuracy of the BFKL kernel used in resumming high-energy logarithms. 
As mentioned earlier, the $\LL$ level corresponds to resumming all terms [$\alpha_s^n (\ln s)^n$] using the LO kernel, while the $\NLLpp$ level accounts for next-to-leading terms [$\alpha_s^{n} (\ln s)^{n-1}$] via the NLO kernel.
This kernel is convoluted with process-specific impact factors, computed at LO or NLO to match the fixed-order accuracy of the observable.
Focusing on the kernel accuracy provides a universal prescription that avoids ambiguities from process-dependent variations in the impact factors, whose leading $\alpha_s$ behavior can differ---for instance, between Higgs and jet or hadron production~\cite{Celiberto:2020tmb}.

To enable a meaningful comparison between high-energy resummed predictions and those obtained from purely collinear, DGLAP-inspired frameworks, it is essential to evaluate observables in both our hybrid factorization formalism and fixed-order approaches.
However, current computational limitations prevent the evaluation of fixed-order distributions at NLO accuracy for inclusive semihard hadron-plus-jet production.
To overcome this constraint and quantify the effects of high-energy resummation relative to DGLAP-based predictions, we adopt an alternative methodology.

Our approach---originally introduced for the study of azimuthal correlations in Mueller-Navelet~\cite{Celiberto:2015yba,Celiberto:2015mpa} and hadron-jet~\cite{Celiberto:2020wpk} configurations---relies on truncating the high-energy series expansion at the NLO level.
This allows us to emulate how high-energy dynamics would manifest within a purely fixed-order context.
In particular, we truncate the azimuthal coefficients at ${\cal O}(\alpha_s^3)$, thereby constructing a high-energy fixed-order approximation, denoted $\HENLOp$.
This scheme offers a practical means to assess the role of BFKL resummation by contrasting it with its fixed-order counterpart in the high-energy domain.
The $\HENLOp$ azimuthal coefficients, computed within the $\MSb$ scheme, take the form
\begin{align}
\label{Cn_HENLOp_MSb}
&\CnHENLOp = 
\frac{e^{\DY}}{s}
\int_{-\infty}^{+\infty} \drv \nu \,
\alpha_s^2(\mu_R)
\nonumber \\[0.75em]
&\hspace{0.50cm}\times \,
\left[ 1 + \bar \alpha_s(\mu_R) \DY \chi(n,\nu) \right]
\\[0.75em] \nonumber
&\hspace{0.50cm}\times \,
\E_1^{\rm NLO}(n,\nu,|\vec q_{T_1}|, x_1)[\E_2^{\rm NLO}(n,\nu,|\vec q_{T_2}|,x_2)]^* \;.
\end{align}

In our analysis, the factorization and renormalization scales are set to \emph{natural} values dictated by the kinematics of the final state.
Specifically, we adopt $\mu_F = \mu_R \equiv \mu_N$, where the natural scale is defined as $\mu_N \equiv m_{1, \perp} + m_{2, \perp}|$.
Here, $m_{\etQ,\HQ, \perp} = \sqrt{m_{\etQ,\HQ}^2 + |\vec q_{T_1,2}|^2}$ represents the transverse mass of the detected hadron, with: $m_{\etc} = 2.9834$~GeV, $m_{\etb} = 9.3909$~GeV, $m_{\Hc} \equiv m_{\Lambda_c^\pm} = 2.28646$~GeV, and $m_{\Hb} \equiv m_{\Lambda_b^0} = 5.6202$~GeV.
The transverse mass of the jet coincides with its transverse momentum, $|\vec q_{T_2}|$.

While two-particle emission naturally introduces distinct energy scales for each final-state object, we simplify the calculation by consolidating them into a single reference scale, $\mu_N$.
This prescription---defined as the sum of the transverse masses---aligns with widely adopted strategies in precision QCD studies and implementations~\cite{Alioli:2010xd,Campbell:2012am,Hamilton:2012rf}, allowing for a consistent comparison with predictions from other approaches.

To explore the impact of MHOUs, we perform a scale variation analysis by independently varying $\mu_F$ and $\mu_R$ over the range $\mu_N/2$ to $2\mu_N$, controlled by the scale factor $C_\mu$.
This enables us to gauge the theoretical uncertainty and assess the sensitivity of our predictions to the choice of scales.

\section{Phenomenological analysis}
\label{sec:results}

All numerical results shown in this work were obtained using the multimodular {\Jethad} interface, developed in \textsc{Python} and \textsc{Fortran}~\cite{Celiberto:2020wpk,Celiberto:2022rfj,Celiberto:2023fzz,Celiberto:2024mrq,Celiberto:2024swu}.

Incoming protons are described in terms of collinear PDFs as provided by the novel {\tt NNPDF4.0} NLO determination~\cite{NNPDF:2021uiq,NNPDF:2021njg}.
This PDF set was obtained by means of global fits and through the Monte Carlo \emph{replica} method, introduced in Ref.~\cite{Forte:2002fg} in the context of neural network analyses.
In principle, a potentially relevant source of uncertainty comes from the choice of different PDFs, as well as different replica members inside the same set.\footnote{We refer to Ref.~\cite{Ball:2021dab} for a comprehensive discussion on ambiguities in the \emph{correlations} among different PDF determinations.}
It was shown, however, that the choice of distinct PDF parametrizations or members does not generate any relevant uncertainty~\cite{Bolognino:2018oth,Celiberto:2020wpk,Celiberto:2021fdp,Celiberto:2022rfj}).
Therefore, we make use of just the central member the {\tt NNPDF4.0} set, and we will focus on other relevant sources of uncertainties, as explained hereafter. 

Shaded bands in the plots represent the combined effect of hard-factor perturbative MHOUs, nonperturbative LDME variations, and numerical uncertainties from multidimensional integrations, the latter kept consistently below 1\% by the {\Jethad} integration routines.
While variations of LDMEs are fully propagated to the final observables, we do not include in this work the impact of perturbative uncertainties affecting the initial FFs (see Sec.~\ref{sssec:NRFF_ns}), such as those stemming from F-MHOUs. 
This choice allows us to isolate other uncertainty sources entering at the cross section level. 
Nevertheless, quantifying the impact of FF-related scale variations at the observable level is an important task, and we plan to address it in detail in a forthcoming study.
We set the center-of-mass energy at set to $\sqrt{s} = 13$ TeV.

\subsection{Rapidity distributions}
\label{ssec:pheno_DY}

The first observable examined in our phenomenological analysis is the rapidity distribution, that is, the cross section differential with respect to the rapidity interval $\DY \equiv y_1 - y_2$ between the two outgoing objects
\begin{equation}
\begin{split}
\label{DY_distribution}
 \hspace{-0.12cm}
 \frac{\drv \sigma(\DY, s)}{\drv \DY} &=
 \int_{y_1^{\rm min}}^{y_1^{\rm max}} \!\!\!\!\! \drv y_1
 \int_{y_2^{\rm min}}^{y_2^{\rm max}} \!\!\!\!\! \drv y_2
 \, \,
 \delta (\DY- (y_1 - y_2))
 \\[0.10cm]
 &\times \,
 \int_{|\vec q_{T_1}|^{\rm min}}^{|\vec q_{T_1}|^{\rm max}} 
 \!\!\!\!\! \drv |\vec q_{T_1}|
 \int_{|\vec q_{T_2}|^{\rm min}}^{|\vec q_{T_2}|^{\rm max}} 
 \!\!\!\!\! \drv |\vec q_{T_2}|
 \, \,
 {\cal C}_{0}^{\rm [level]}
 \;,
\end{split}
\end{equation}
where ${\cal C}_0$ denotes the $\varphi$-summed angular coefficient defined in Section~\ref{ssec:HyF}.
The superscript `${\rm [level]}$' identifies the perturbative precision used in the calculation, \emph{i.e.}, $\NLLp$, $\LL$, or $\HENLOp$.

The transverse momentum range for the quarkonium spans from $60$ to $120$~GeV, while that for the singly heavy hadron (jet) extends from $30$ ($50$) to $120$~GeV.
These intervals are consistent with current and forthcoming LHC analyses involving hadrons and jets~\cite{Khachatryan:2016udy,Khachatryan:2020mpd}.
The adoption of \emph{asymmetric} transverse momentum windows is expected to magnify high-energy resummation effects relative to the fixed-order background~\cite{Celiberto:2015yba,Celiberto:2015mpa,Celiberto:2020wpk}.

Our choice of rapidity coverage follows the experimental constraints adopted in ongoing LHC measurements.
Hadrons are assumed to be detected in the barrel calorimeter region, as in the CMS setup~\cite{Chatrchyan:2012xg}, restricting their rapidities to be within the interval between $-2.5$ and $+2.5$.
Jets, which can also be reconstructed in the endcap regions~\cite{Khachatryan:2016udy}, are allowed a broader rapidity range of $|y_2| < 4.7$.

\begin{figure*}[!t]
\centering

   \hspace{0.00cm}
   \includegraphics[scale=0.390,clip]{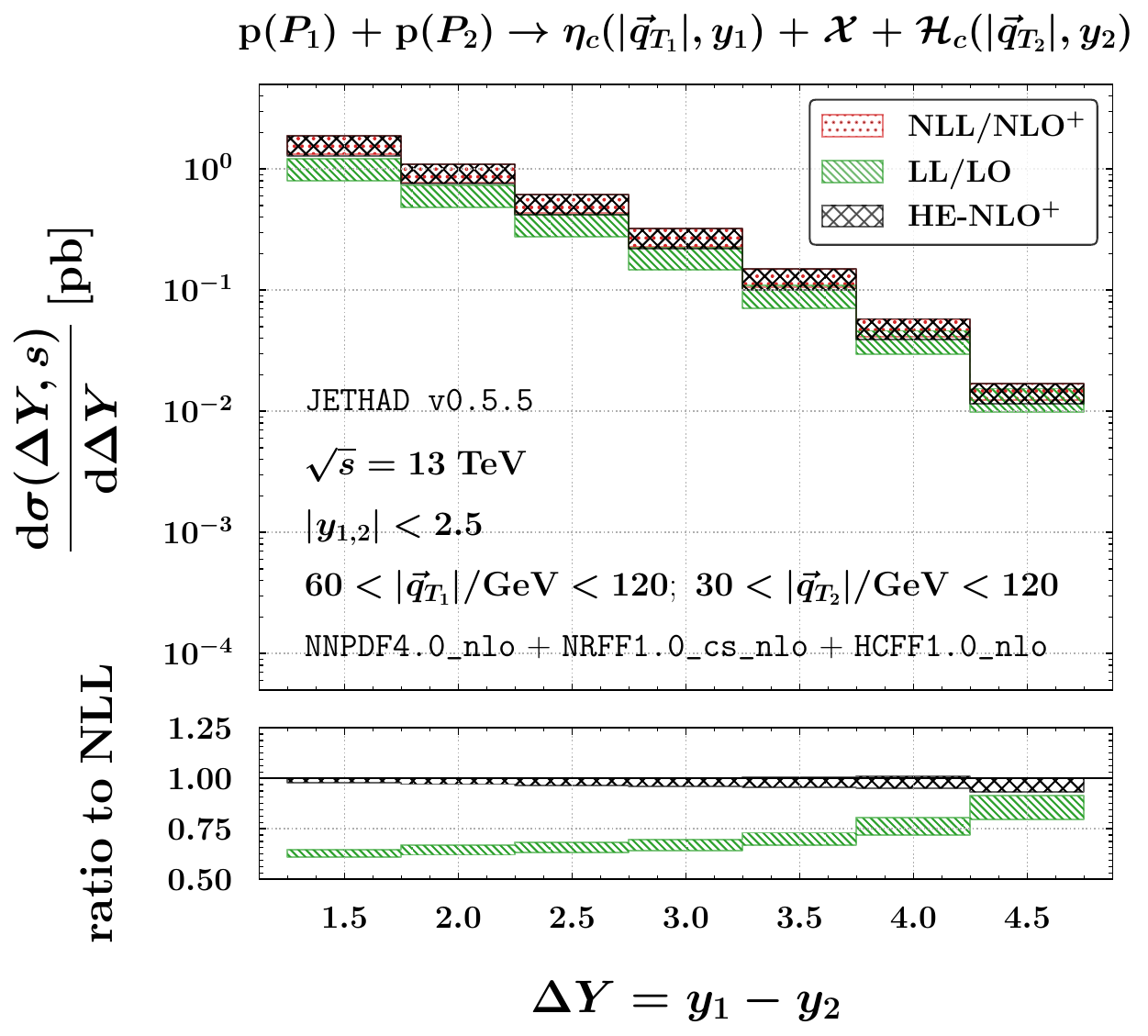}
   \hspace{-0.00cm}
   \includegraphics[scale=0.390,clip]{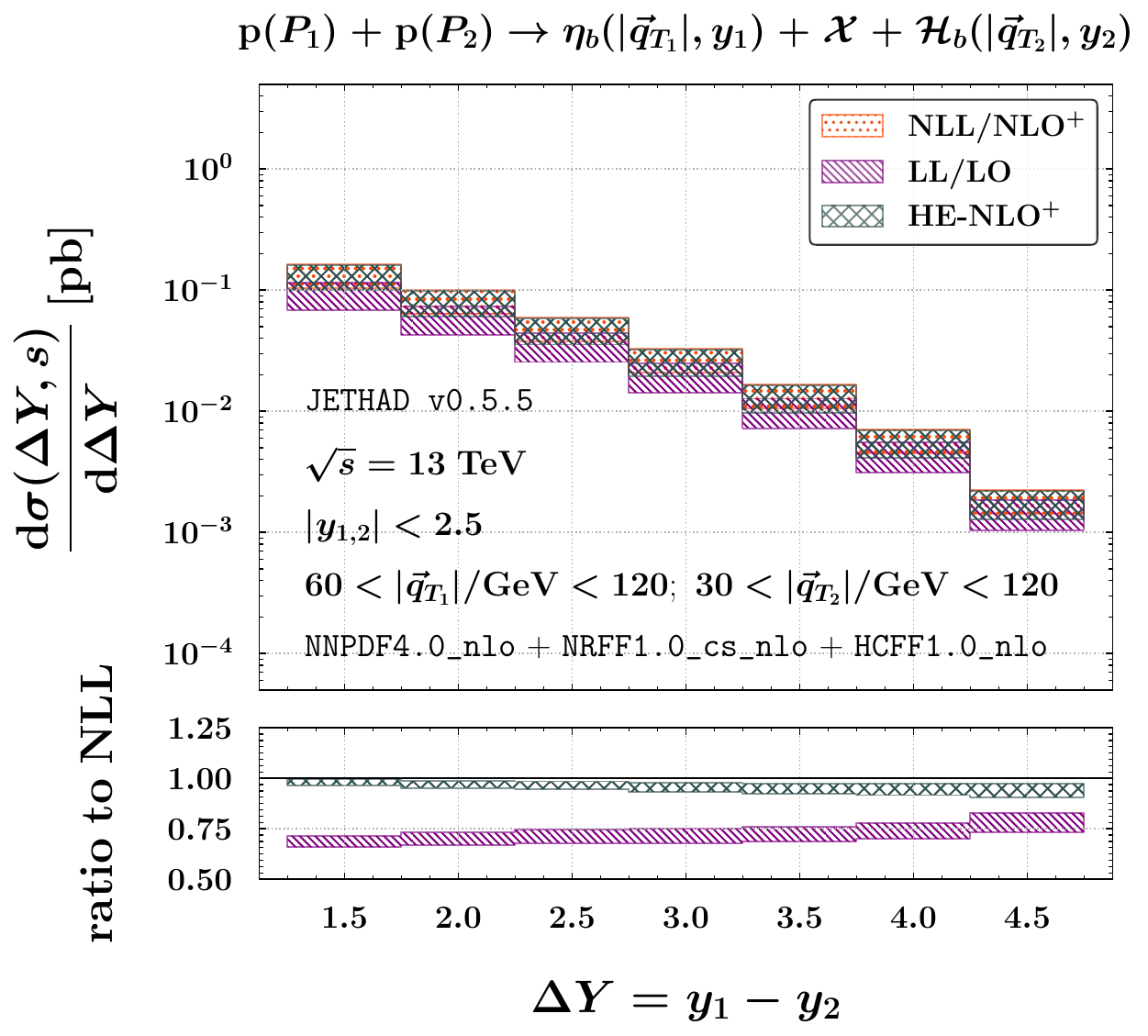}

\caption{Rapidity distributions for the semi-inclusive hadroproduction of an $\etc$ (left) or $\etb$ (right) meson in association with a heavy-light hadron, $\Hc$ or $\Hb$, respectively, at $\sqrt{s} = 13$~TeV.
Shaded bands in the main panels represent the combined uncertainty from MHOUs, LDMEs, and multidimensional phase-space integration.
The ancillary panels below the main plots show the ratio of $\LL$ and $\HENLOp$ predictions to $\NLLp$, with bands capturing MHOUs only.}
\label{fig:DY_HQ}
\end{figure*}

\begin{figure*}[!t]
\centering

   \hspace{0.00cm}
   \includegraphics[scale=0.390,clip]{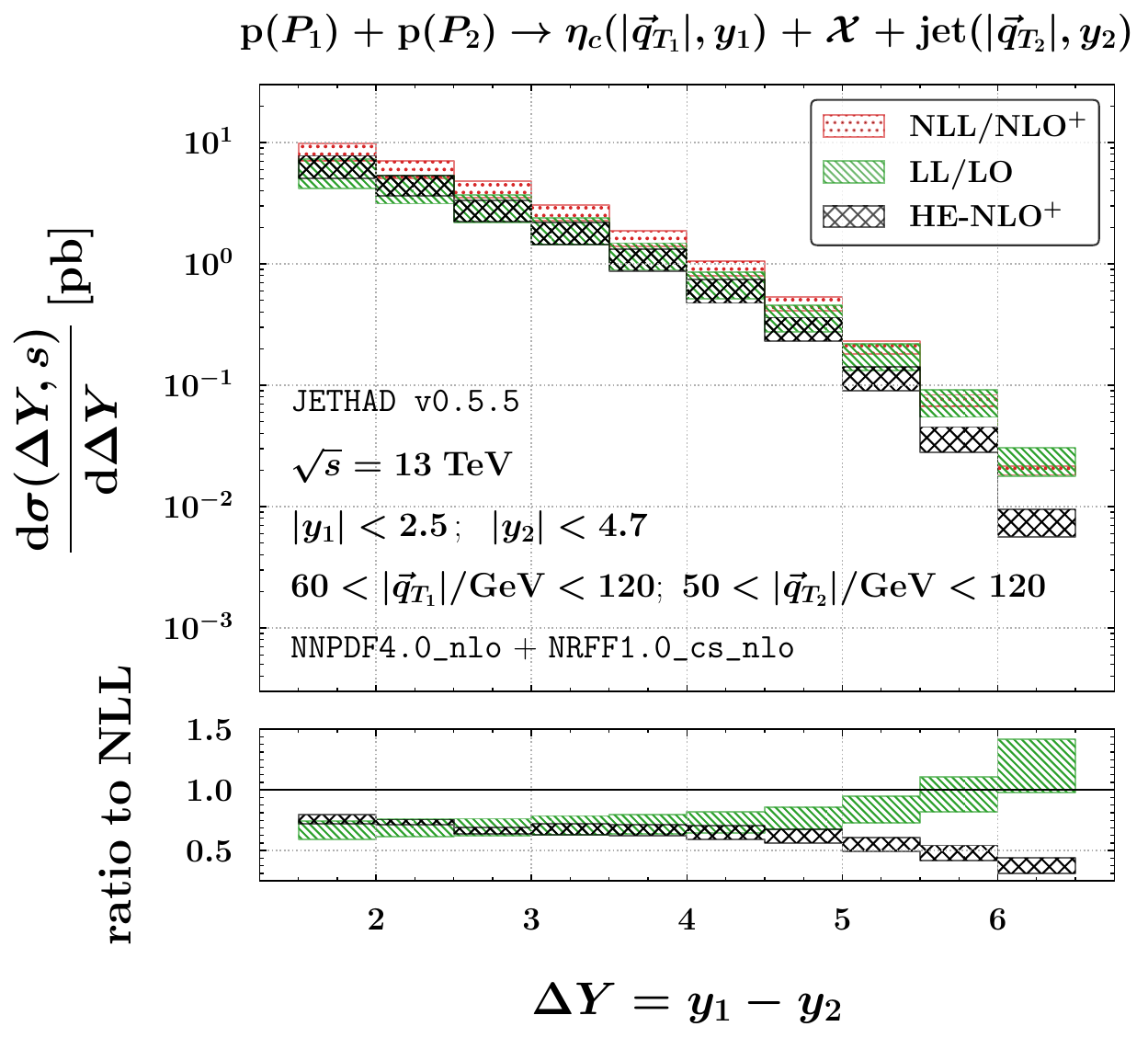}
   \hspace{-0.00cm}
   \includegraphics[scale=0.390,clip]{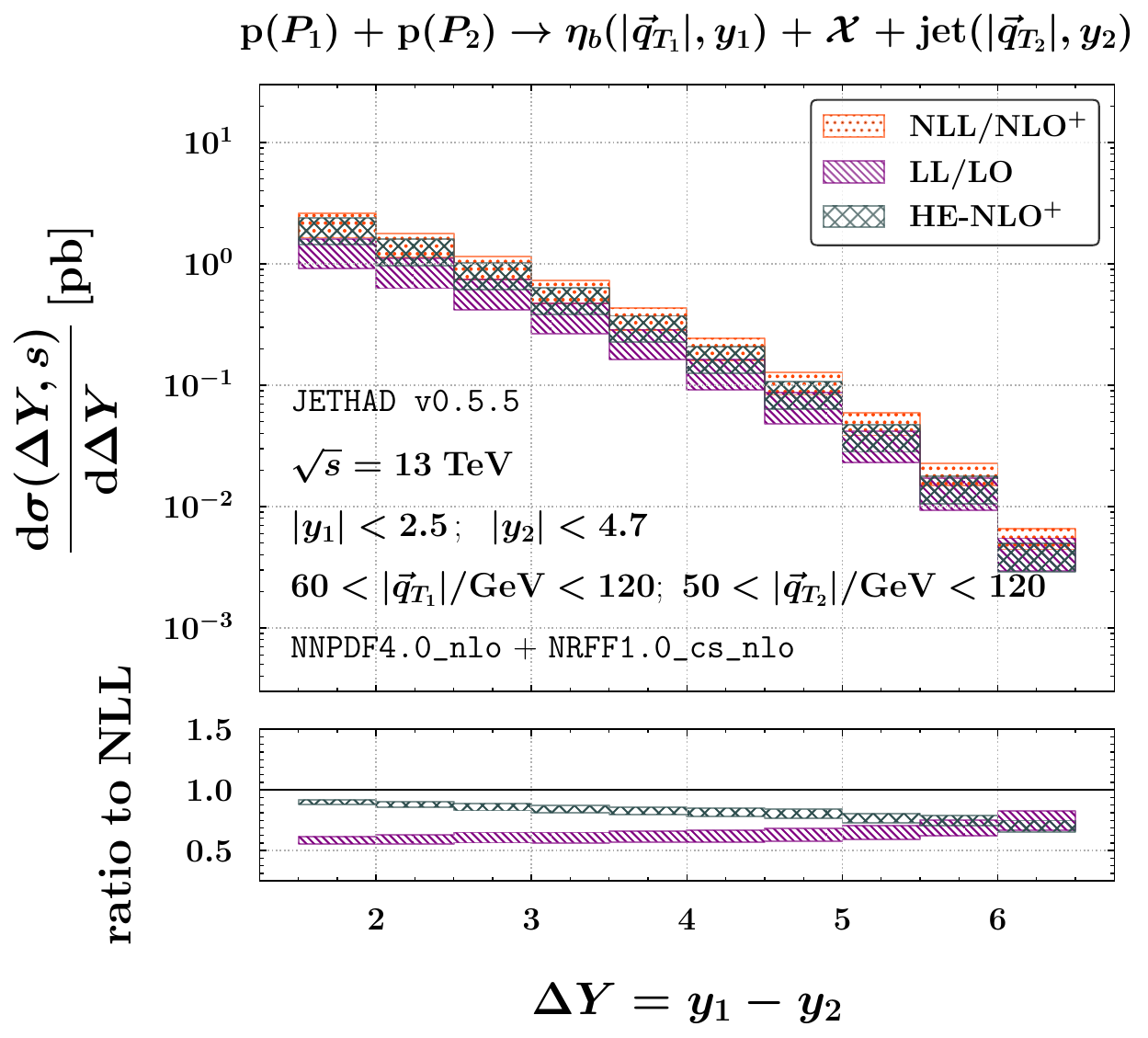}

\caption{Rapidity distributions for the semi-inclusive hadroproduction of an $\etc$ (left) or $\etb$ (right) meson in association with jet, at $\sqrt{s} = 13$~TeV.
The shaded bands in the main panels represent the combined uncertainty from MHOUs, LDMEs, and multidimensional phase-space integration.
The ancillary panels below the main plots display the ratio of the $\LL$ and $\HENLOp$ predictions to the $\NLLp$ result, with bands capturing MHOUs only.}
\label{fig:DY_jet}
\end{figure*}

Figures~\ref{fig:DY_HQ} and~\ref{fig:DY_jet} show the rapidity distributions for the semi-inclusive hadroproduction of pseudoscalar quarkonia, either $\etc$ or $\etb$, in association with a singly heavy-flavored hadron ($H_Q$) or a light jet, respectively.
Left (right) panels correspond to $\etc$ ($\etb$) production.
Ancillary panels below the main plots display the ratio of $\LL$ and $\HENLOp$ predictions to the $\NLLp$ baseline.

All distributions exhibit a marked decrease as $\DY$ increases, with cross section values spanning two to three orders of magnitude.
This suppression reflects the lower probability of producing two hard final-state objects at large rapidity separation, resulting from the interplay between two competing effects:
while BFKL dynamics leads to a growth with energy at the partonic level, this trend is mitigated by the convolution with collinear PDFs and FFs in the impact factors, which uniformly dampen the rise.

A notable aspect of our results is the pronounced stability of the predictions against theoretical uncertainties.
At $\NLLp$ accuracy, uncertainty bands from MHOUs, LDMEs, and multidimensional integration remain tightly constrained throughout the $\DY$ spectrum.
Furthermore, the $\NLLp$ bands are almost entirely nested within the $\LL$ ones---par\-ti\-cu\-lar\-ly in the [$\etQ + H_Q$ channel, where both final states originate from heavy-flavor fragmentation.

This dual-fragmentation scenario enhances the stabilizing role of the collinear framework, underscoring how heavy quark thresholds and DGLAP evolution act as intrinsic regulators of high-energy logarithmic behavior.
The impact of $\NLLp$ corrections relative to $\LL$ predictions is typically limited to 10\% or less and becomes even milder at large $\DY$, where BFKL resummation is expected to dominate.

The behavior of the $\LL$ to $\NLLp$ ratio further confirms this picture.
In both channels, the ratio approaches unity as $\DY$ increases, signaling the stabilizing role of NLL resummation.
This convergence is more rapid and uniform in the double heavy-hadron case (see Fig.~\ref{fig:DY_HQ}), once again demonstrating that dual heavy-flavor fragmentation suppresses the influence of subleading contributions.

This emerging pattern exemplifies the \emph{natural stability}~\cite{Celiberto:2022grc} observed in high-energy QCD observables involving heavy-flavor production.
As noted in previous studies on ordinary heavy hadrons\cite{Celiberto:2021dzy,Celiberto:2021fdp,Celiberto:2022zdg,Celiberto:2022dyf,Celiberto:2022keu,Celiberto:2024omj} and exotic matter~\cite{Celiberto:2023rzw,Celiberto:2024mab,Celiberto:2024beg,Celiberto:2025dfe,Celiberto:2025ziy,Celiberto:2025ipt}, the presence of heavy-flavored FFs---especially in both emission functions---naturally suppresses logarithmic instabilities and reduces perturbative sensitivity.
This makes [$\etQ + H_Q$ production one of the most reliable channels to study BFKL-resummed dynamics in collider environments.

Closer examination of the ancillary plots in Figs.~\ref{fig:DY_HQ} and~\ref{fig:DY_jet} reveals that the discriminating power of $\DY$ distributions is comparatively modest, especially when contrasted with that of transverse momentum observables (see Section~\ref{ssec:pheno_qT}).
In both [$\etQ + H_Q$ and [$\etQ + \text{jet}$ configurations, the $\HENLOp$ to $\NLLp$ ratio remains within a narrow range around unity, showing deviations no larger than 20\% over the full $\DY$ interval.
This confirms that $\DY$ distributions are generally less sensitive to resummation structures due to their reduced response to logarithmic enhancements.

This behavior is not surprising.
It aligns with earlier findings in high-energy QCD involving final-state heavy-flavor fragmentation~\cite{Celiberto:2022rfj,Celiberto:2022keu,Celiberto:2024omj,Celiberto:2025dfe,Celiberto:2025ziy}, where the smoothness of the emission functions and the collinear robustness of the FF inputs suppress the contrast between fixed-order and resummed predictions.

A notable exception to this trend was found in the study of rare baryon production, where the [$\Omega_{3c} + \text{jet}$ channel exhibited a surprisingly high discriminating power even in $\DY$ distributions~\cite{Celiberto:2025ogy}.
This remains an open topic of investigation and may be linked to the unique interplay between multiheavy-quark fragmentation and high-energy effects in the semihard regime.

\subsection{Angular multiplicities}
\label{ssec:pheno_phi}

The second key observable in our investigation is the angular distribution.
We analyze the following normalized multiplicities
\begin{equation}
\label{angular_multiplicity}
\frac{1}{\sigma} \frac{\drv \sigma(\varphi, s)}{\drv \varphi} = \frac{1}{2 \pi} + \frac{1}{\pi} \sum_{n=1}^\infty
\langle \cos(n \varphi) \rangle \, \cos (n \varphi) \;,
\end{equation}
where $\varphi = \varphi_1 - \varphi_2 - \pi$, and the mean values $\langle \cos(n \varphi) \rangle = C_n^{\rm [level]}/C_0^{\rm [level]}$ are the azimuthal correlation moments.
Here, $C_{n \geq 0}^{\rm [level]}$ are the azimuthal coefficients integrated over the full rapidity and transverse momentum ranges described in Section~\ref{ssec:pheno_DY}, and integrated over fixed bins of $\DY$.
For simplicity, we restrict our analysis to the quarkonium plus jet channel (Fig.~\eqref{fig:reactions}, right diagram), considering the same rapidity intervals adopted throughout, namely $|y_1| < 2.5$ for the $\etQ$ and $|y_2| < 4.7$ for the jet.

Originally introduced to study the azimuthal decorrelation of light two-jet systems~\cite{Marquet:2007xx}, these angular multiplicities have emerged as highly promising observables for exploring high-energy QCD dynamics.
They carry information from all azimuthal harmonics and provide a robust probe of the underlying partonic structure.

Moreover, their differential form in $\varphi$ is particularly advantageous for experimental studies, as it facilitates direct comparison with data from detectors that may not have uniform acceptance over the full azimuthal range ($2\pi$).
A recent study on double-jet multiplicities revealed two notable benefits~\cite{Celiberto:2022gji}.
First, it helps mitigate longstanding issues in modeling light-flavored final states at natural energy scales.
Second, it improves agreement with the CMS measurements at $\sqrt{s} = 7$~TeV~\cite{Khachatryan:2016udy}.

\begin{figure*}[!t]
\centering

 \hspace{0.00cm}
 \includegraphics[scale=0.395,clip]{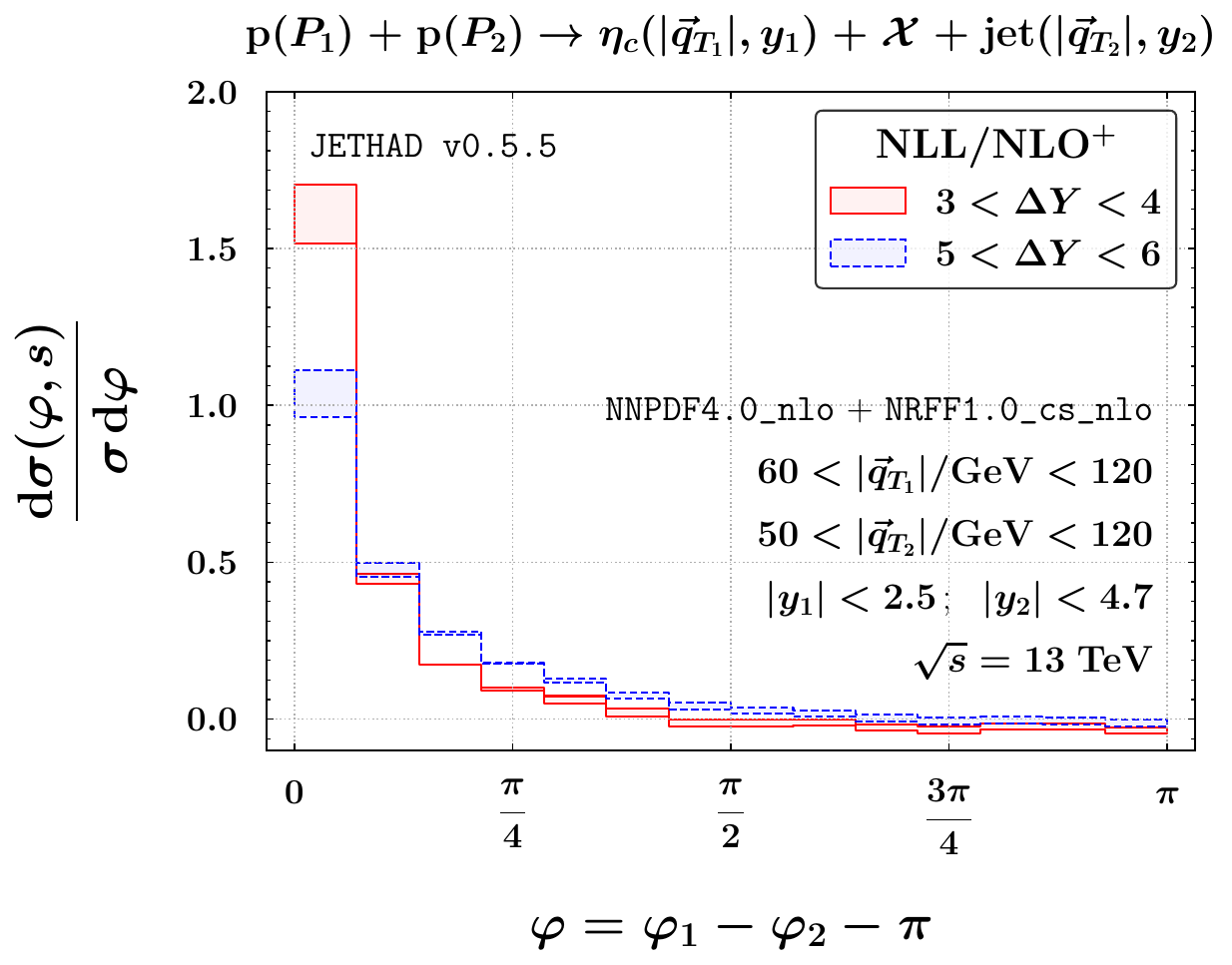}
 \hspace{-0.00cm}
 \includegraphics[scale=0.395,clip]{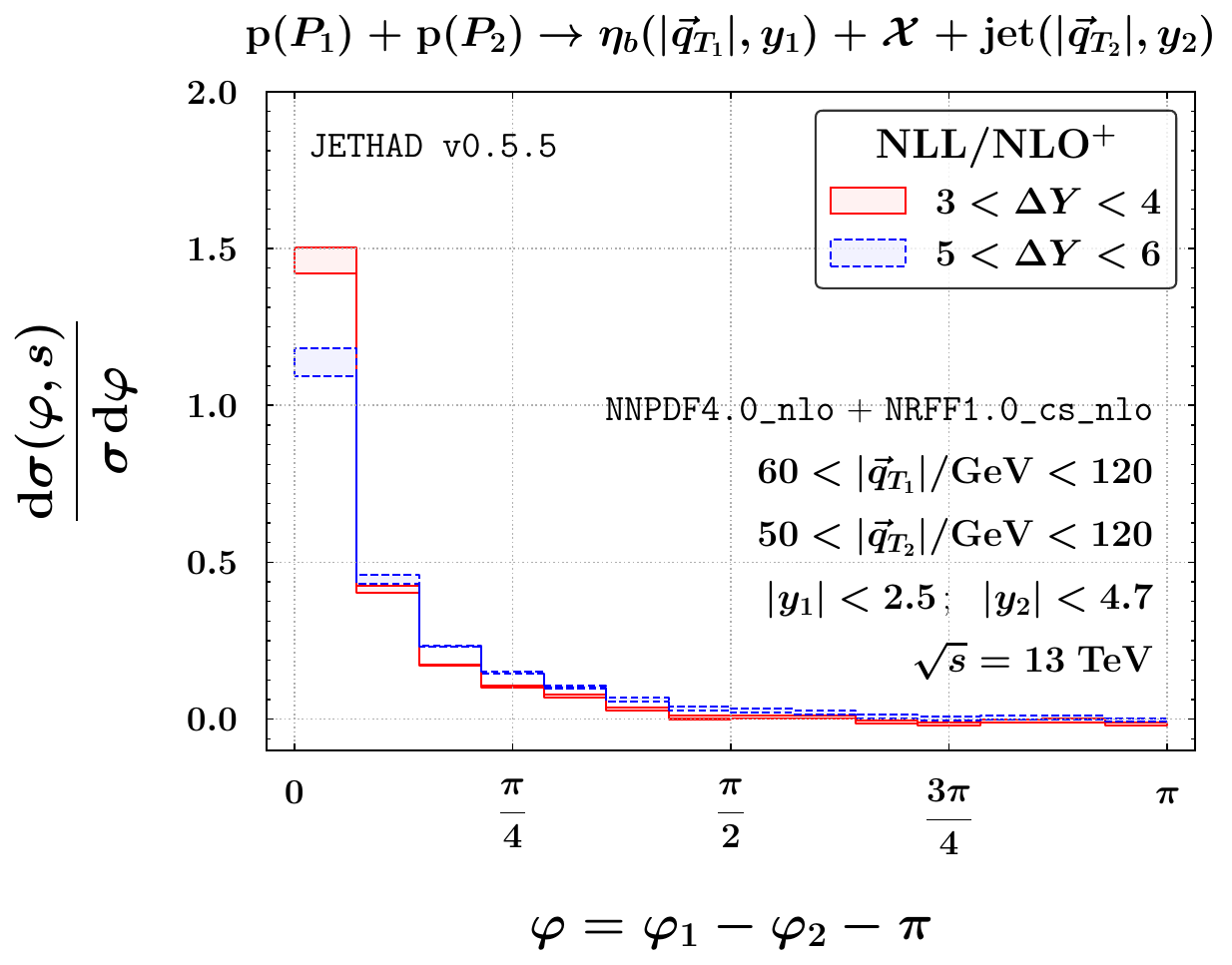}
 
 \vspace{0.35cm}
 
 \hspace{0.00cm}
 \includegraphics[scale=0.395,clip]{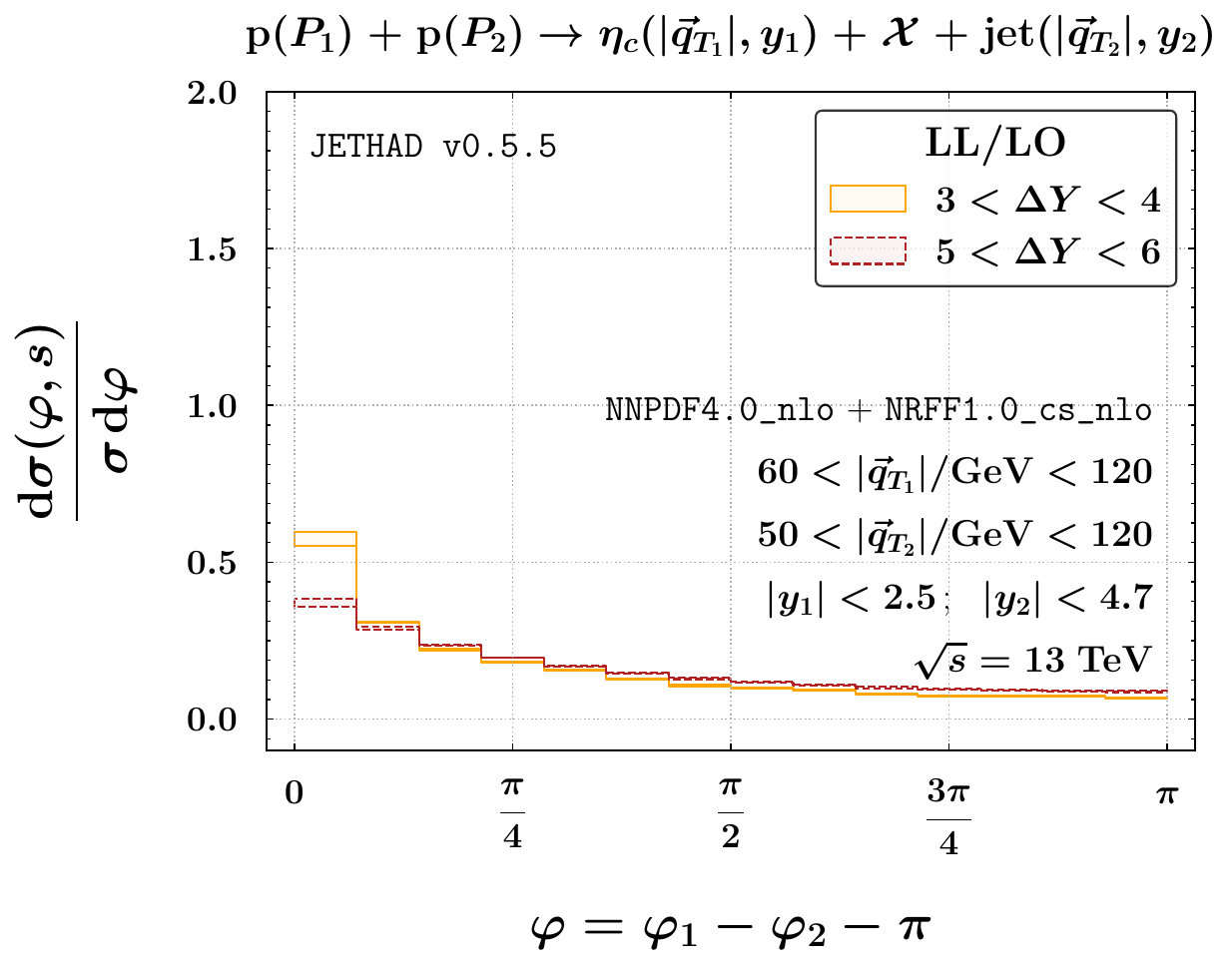}
 \hspace{-0.00cm}
 \includegraphics[scale=0.395,clip]{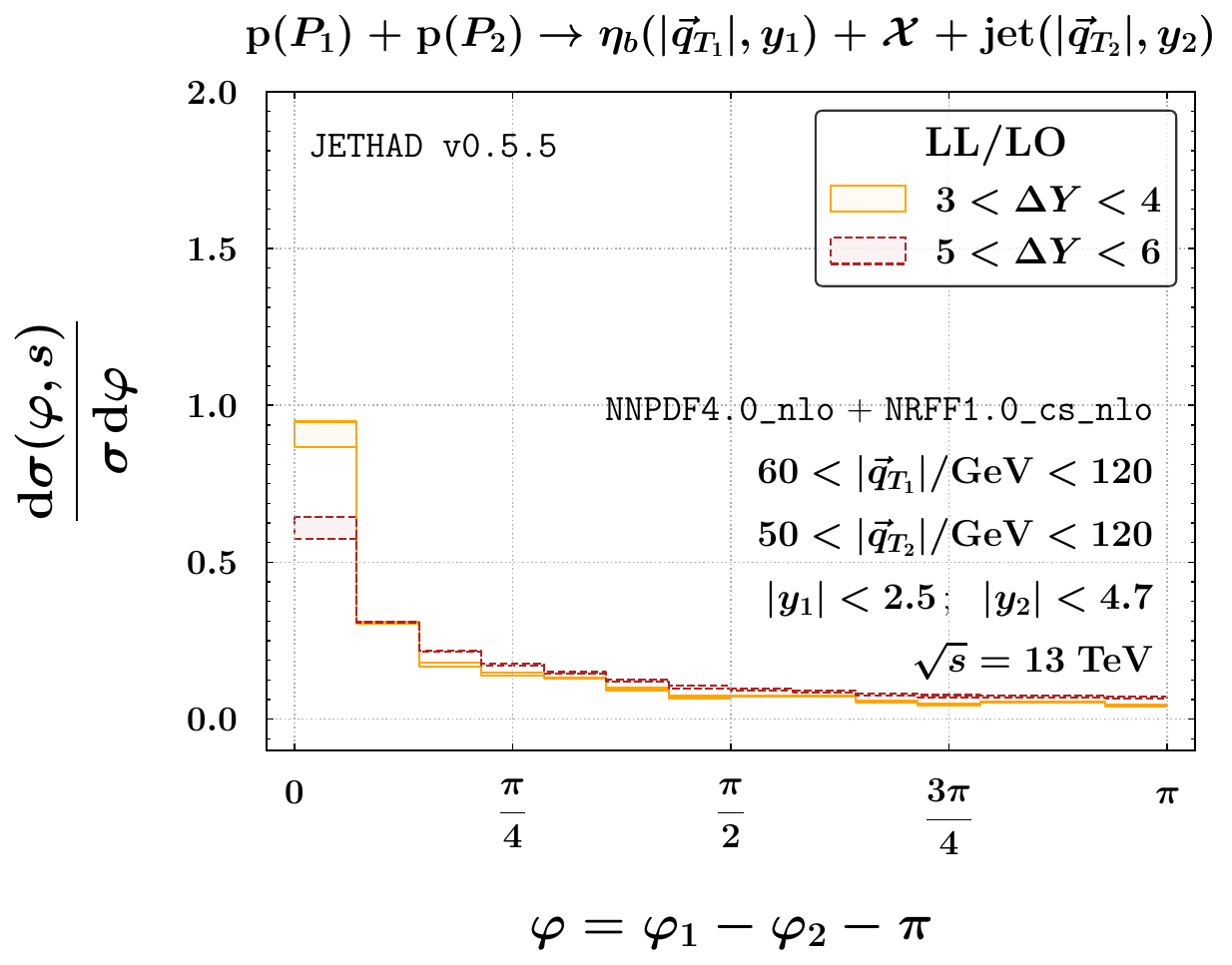}

\caption{Angular multiplicities for the semi-inclusive hadroproduction of an $\etc$ (left) or $\etb$ (right) meson in association with the jet, at $\sqrt{s} = 13$~TeV.
Predictions in the upper (lower) plots are computed at $\NLLp$ ($\LL$) accuracy within the hybrid factorization framework.
The shaded bands in the main panels represent the combined uncertainty from MHOUs and multidimensional phase-space integration.}
\label{fig:I-phi}
\end{figure*}

Predictions $\sqrt{s} = 13$~TeV and within $\NLLp$ accuracy (see upper panels of Fig.\ref{fig:I-phi}) are compared to those computed at $\LL$ accuracy (see lower panels of Fig.~\ref{fig:I-phi}).
The left and right columns correspond to emissions of pseudoscalar charmonium and bottomonium, respectively.
The rapidity difference $\DY$ is varied within the range of three to six units, and results are presented for two representative, nonoverlapping unit-length bins.
By definition, the azimuthal distributions described by Eq.~\eqref{angular_multiplicity} are symmetric under the transformation $\varphi \to -\varphi$.
As a result, we display the distributions only in the interval $0 < \varphi < \pi$.
For easier comparison with future measurements, distributions are computed as averages over uniform $\varphi$ bins.

Our multiplicity distributions feature a pronounced peak around $\varphi = 0$, which corresponds kinematically to configurations in which the detected particles are nearly back-to-back.
This peak is most prominent in the lowest rapidity difference interval, $3 < \DY < 4$.
As $\DY$ increases, the height of the peak gradually decreases.
This behavior signals the growing relevance of high-energy dynamics in the large $\DY$ regime.
In this domain, the contribution of secondary gluon emissions, effectively described by our high-energy resummation, becomes increasingly significant.
As a result, the number of nearly back-to-back events is reduced, leading to a progressive weakening of the azimuthal correlation between the two final-state particles.

This progressive flattening of the distribution with increasing $\DY$ reflects the onset of gluon-induced decorrelation effects.
The qualitative similarity between the $\etc$ and $\etb$ panels confirms that the trend is primarily driven by kinematics and resummation dynamics, rather than by the mass or flavor of the quarkonium.

\subsection{Transverse momentum distributions}
\label{ssec:pheno_qT}

Although rapidity and angular distributions are key to exploring high-energy QCD effects, they do not fully capture dynamics in regimes where additional logarithmic enhancements arise. 
In particular, when the detected particles carry large transverse momenta or exhibit a strong momentum imbalance, one encounters enhanced collinear DGLAP logarithms and soft threshold logarithms~\cite{Sterman:1986aj,Catani:1989ne,Catani:1996yz,Bonciani:2003nt,deFlorian:2005fzc,Ahrens:2009cxz,deFlorian:2012yg,Forte:2021wxe,Mukherjee:2006uu,Bolzoni:2006ky,Becher:2006nr,Becher:2007ty,Bonvini:2010tp,Ahmed:2014era,Muselli:2017bad,Banerjee:2018vvb,Duhr:2022cob,Shi:2021hwx,Wang:2022zdu,Bonvini:2023mfj}. 
These logarithmic structures, unrelated to those resummed at high energy, demand a dedicated resummation strategy.

Performing a joint resummation of threshold and high-energy logarithms remains a complex task. 
In inclusive Higgs production via gluon fusion, a breakthrough came from isolating small- and large-$x$ logarithms in Mellin space~\cite{Bonvini:2018ixe}, enabling separate resummation of the small-$N$ and large-$N$ contributions~\cite{Ball:2013bra,Bonvini:2014joa}. 
However, the semi-inclusive and rapidity-sensitive nature of our observables prevents us from applying this method directly, and a unified resummation scheme is still lacking.

Conversely, in the low transverse momentum region, Sudakov logarithms become large and fall outside the scope of hybrid factorization. 
Here, a significant influence is also expected from the so-called diffusion pattern~\cite{Bartels:1993du,Caporale:2013bva,Ross:2016zwl}. 
The most reliable technique to resum these effects is the transverse momentum resummation framework~\cite{Catani:2000vq,Bozzi:2005wk,Bozzi:2008bb,Catani:2010pd,Catani:2011kr,Catani:2013tia,Catani:2015vma,Duhr:2022yyp}.

Semi-inclusive final states involving two tagged particles in hadronic collisions, such as photon pairs~\cite{Cieri:2015rqa,Alioli:2020qrd,Becher:2020ugp,Neumann:2021zkb}, Higgs~\cite{Ferrera:2016prr}, $W^\pm$ bosons~\cite{Ju:2021lah}, or combinations like Higgs plus jet~\cite{Monni:2019yyr,Buonocore:2021akg} and $Z$ plus photon~\cite{Wiesemann:2020gbm}, are valuable for probing transverse momentum resummation mechanisms.
Recent studies achieved third-order resummation in Drell-Yan and Higgs channels \cite{Ebert:2020dfc,Re:2021con,Chen:2022cgv,Neumann:2022lft,Bizon:2017rah,Billis:2021ecs,Caola:2022ayt}, and Ref.~\cite{Monni:2019yyr} performed joint resummation of transverse logarithms for Higgs plus jet observables using {\RadISH}~\cite{Bizon:2017rah}.
Analogous distributions were examined for $W^+W^-$ leptonic decays at the LHC in~\cite{Kallweit:2020gva}.

Complications arise when heavy-flavor observables are involved.
As the transverse momentum of a heavy hadron decreases, its transverse mass approaches the quark mass thresholds that control the DGLAP evolution.
This undermines the reliability of purely VFNSs, especially near or below threshold.

To explore synergies between the $\NLLp$ hybrid formalism and other resummation strategies, singly and doubly differential transverse momentum distributions have been analyzed in jet~\cite{Bolognino:2021mrc}, Higgs~\cite{Celiberto:2020tmb}, $b$-hadron~\cite{Celiberto:2021fdp}, cascade baryon~\cite{Celiberto:2022kxx}, and exotic matter~\cite{Celiberto:2024mab,Celiberto:2024beg,Celiberto:2025ipt} production.
Here, we focus on the high-energy limit of cross sections differential in the $\etQ$ transverse momentum, $|\vec q_{T_1}|$, integrating over $|\vec q_{T_2}|$ in the 40 to 120~GeV jet range and across two bins in $\DY$.
We adopt the same rapidity intervals as in previous sections, namely $|y_1| < 2.5$ for the quarkonium and $|y_2| < 4.7$ for the jet.

To the best of our knowledge, no dedicated analysis of transverse momentum distributions in quarkonium-jet systems has been performed so far within the high-energy resummation framework.
Therefore, our analysis presented here represents a novel and original contribution, opening the way to deeper investigations of transverse observables in quarkonium-associated production.

The analytical expression for our transverse momentum rate reads
\begin{equation}
\begin{split}
\label{TM_distribution}
 &\frac{\drv \sigma(|\vec q_{T_1}|, s)}{\drv |\vec q_{T_1}|} = 
 \int_{\DY^{\rm min}}^{\DY^{\rm max}} \!\! \drv \DY
 \int_{y_1^{\rm min}}^{y_1^{\rm max}} \drv y_1
 \int_{y_2^{\rm min}}^{y_2^{\rm max}} \drv y_2
 \\
 &\hspace{0.55cm} \times \,
 \delta (\DY - (y_1 - y_2))
 \int_{|\vec q_{T_2}|^{\rm min}}^{|\vec q_{T_2}|^{\rm max}} 
 \!\!\drv |\vec q_{T_2}|
 \, \,
 {\cal C}_{0}^{\rm [level]}
 \;.
\end{split}
\end{equation}

\begin{figure*}[!t]
\centering

   \hspace{0.00cm}
   \includegraphics[scale=0.390,clip]{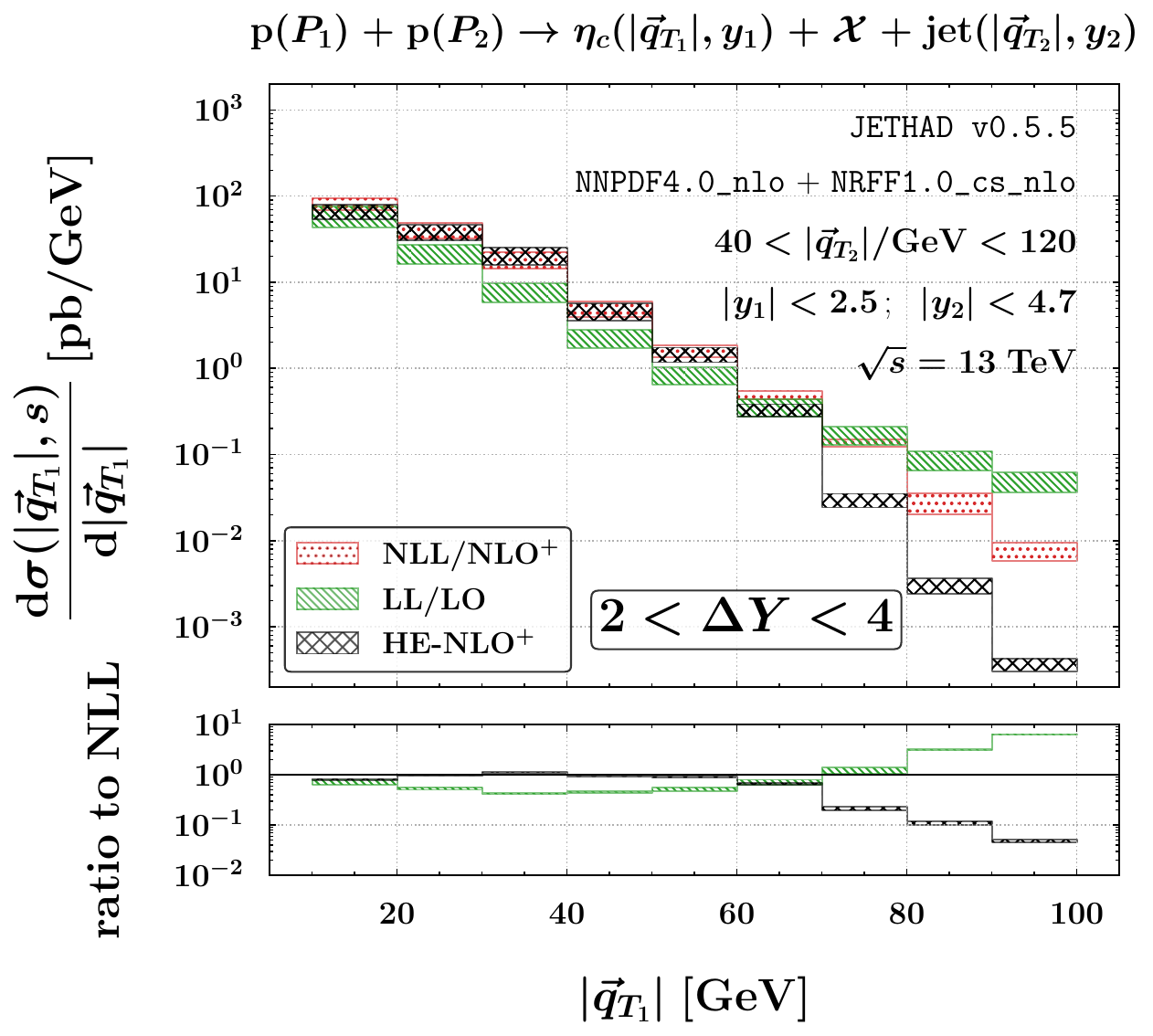}
   \hspace{-0.00cm}
   \includegraphics[scale=0.390,clip]{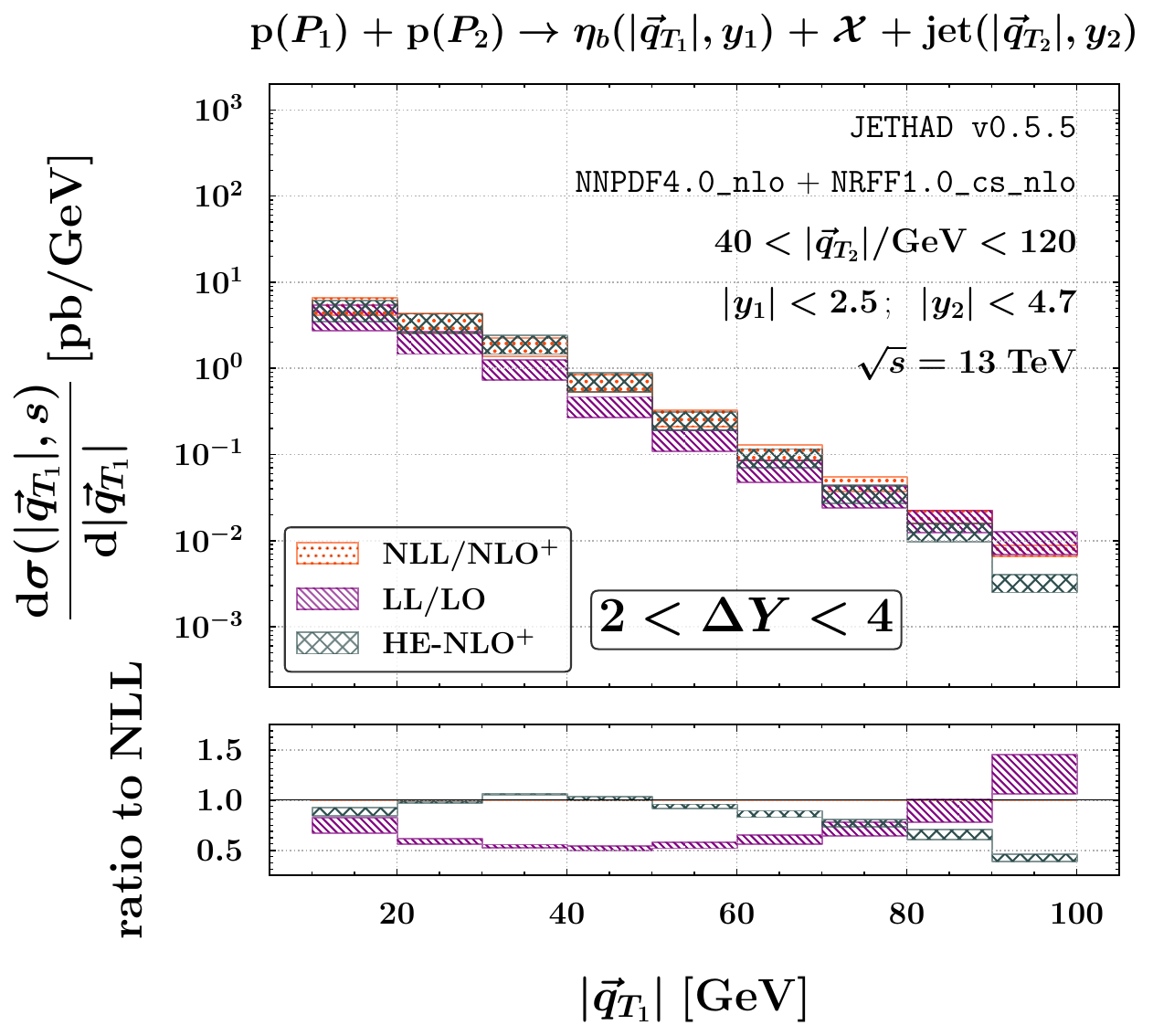}

   \vspace{0.35cm}

   \hspace{0.00cm}
   \includegraphics[scale=0.390,clip]{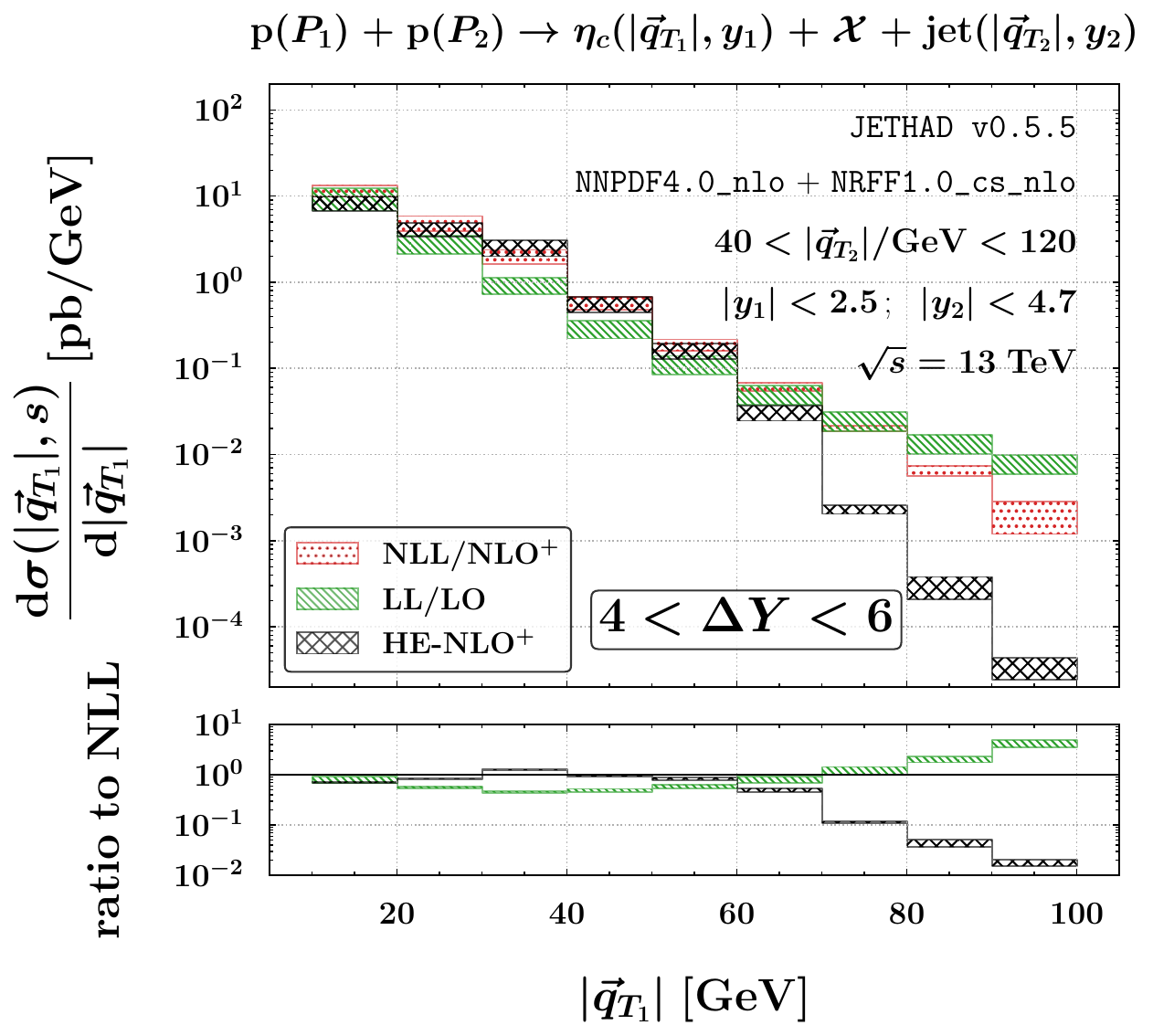}
   \hspace{-0.00cm}
   \includegraphics[scale=0.390,clip]{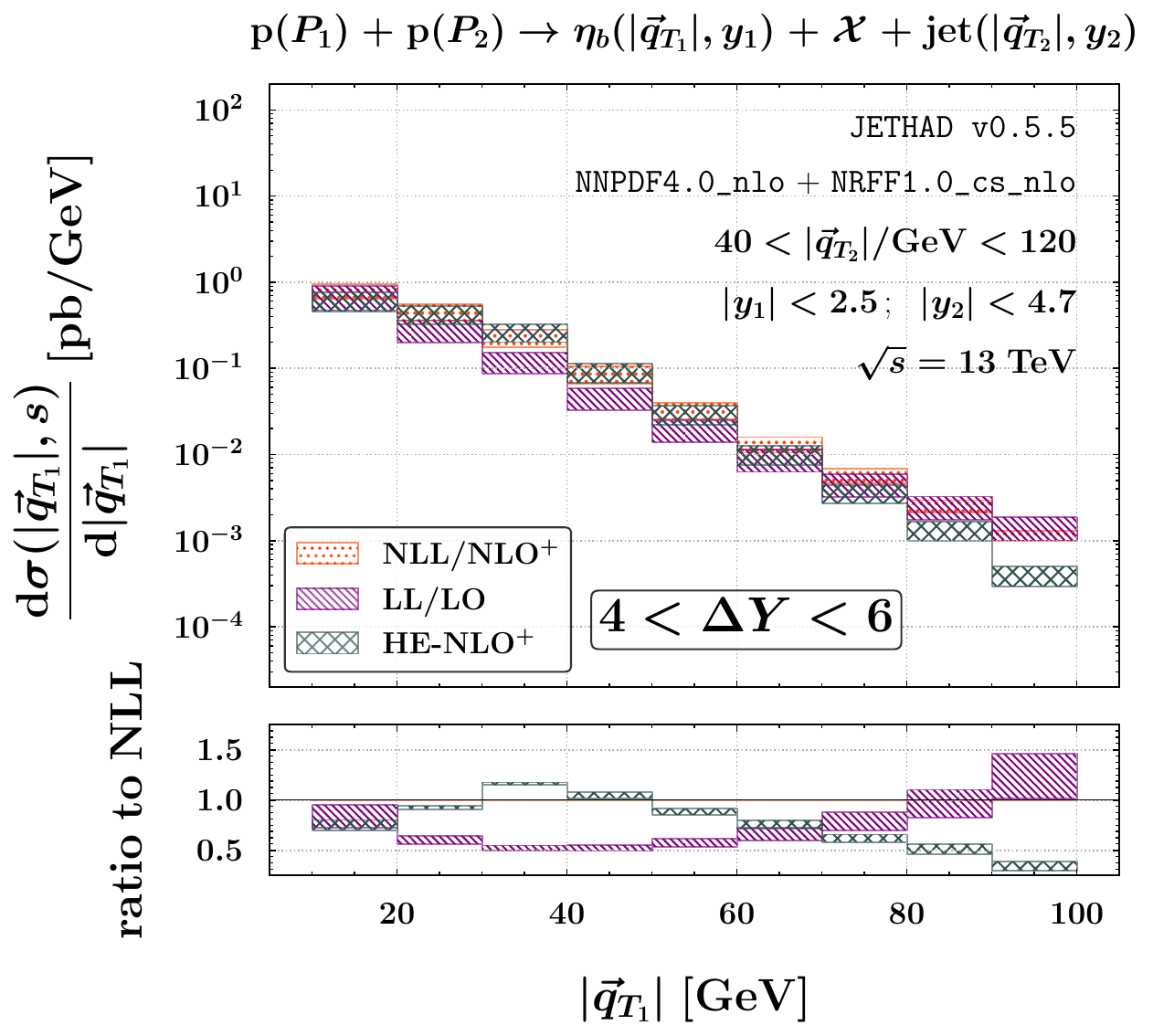}

\caption{Transverse momentum distributions for the semi-inclusive hadroproduction of an $\etc$ (left) or $\etb$ (right) meson in association with a jet, at $\sqrt{s} = 13$~TeV, and for $2 < \DY < 4$ (upper) or $4 < \DY < 6$ (lower) rapidity separation intervals.
The shaded bands in the main panels represent the combined uncertainty from MHOUs, LDMEs, and multidimensional phase-space integration.
The ancillary panels below the main plots display the ratio of the $\LL$ and $\HENLOp$ predictions to the $\NLLp$ result, with bands capturing MHOUs only.}
\label{fig:pT_jet}
\end{figure*}

Figure~\ref{fig:pT_jet} shows the transverse momentum distributions of the $\etQ$ meson in semi-inclusive $\etQ$ plus jet production at $\sqrt{s} = 13$~TeV.
Left (right) panels refer to $\etc$ ($\etb$), while the upper and lower plots correspond to the rapidity intervals $2 < \Delta Y < 4$ and $4 < \Delta Y < 6$, respectively.
Distributions are averaged over uniform $|\vec{q}_{T_1}|$ bins of 10~GeV.
We use the same rapidity ranges as in previous sections, namely $|y_1| < 2.5$ for the quarkonium and $|y_2| < 4.7$ for the jet.
The ancillary panels below the main plots display the ratio of $\LL$ and $\HENLOp$ predictions to the $\NLLp$ result.

As expected, the overall cross section decreases as $\Delta Y$ increases, reflecting the suppression of phase space.
The $|\vec{q}_{T_1}|$ spectra fall off with increasing transverse momentum, without any pronounced peaks.

The general trend shared by all our distributions is a steep decline as $|\vec q_{T_1}|$ increases. 
The results exhibit strong stability against MHOUs, with uncertainty bands showing a maximum width of 30\%.
We observe that $\HENLOp$ to $\NLLp$ ratios generally remain below 1, diminishing as $\vec q_{T_1}$ grows. 
In contrast, the $\LL$ to $\NLLp$ ratio shows an almost opposite pattern: it starts below one in the low-$|\vec q_{T_1}|$ region but steadily grows as $|\vec q_{T_1}|$ rises, eventually reaching a maximum value between 1.5 and 2.
Explaining these trends is challenging, as they result from a combination of several interacting effects.

On the one hand, previous studies on semihard processes have revealed that the behavior of the NLL-resummed signal relative to its NLO high-energy background in 
transverse momentum rates tends to vary depending on the process under consideration. 
For instance, the $\HENLOp$ to $\NLLp$ ratio for the cascade baryon plus jet channel consistently exceeds one, as shown in Fig.~7 of Ref.~\cite{Celiberto:2022kxx}. 
However, preliminary analyses of Higgs plus jet distributions, performed within a partially NLL-to-NLO matched accuracy, have exhibited a more complex pattern~\cite{Celiberto:2023dkr}.

The observation that the $\HENLOp$ to $\NLLp$ ratios are less than 1 in the context of $\etQ$ plus jet tags appears to be a distinctive characteristic of this process, and it is shared also by the same differential observables sensitive to the production of heavy-flavored rare baryons~\cite{Celiberto:2025ogy} and exotics~\cite{Celiberto:2024mab,Celiberto:2024beg,Celiberto:2025ipt}. 
This suggests that the dynamics governing these emissions are different from those observed in other semihard reactions, highlighting the unique interplay of high-energy and NLL resummation effects in these channels.

On the other hand, the ratio between the pure $\LL$ limit and full $\NLLp$ resummation is shaped by a nontrivial interplay of NLO corrections affecting emission functions.
For jets, NLO terms are typically negative~\cite{Bartels:2001ge,Ivanov:2012ms,Colferai:2015zfa}, while hadronic emissions mix positive contributions from the $C_{gg}$ coefficient with negative ones from subleading terms~\cite{Ivanov:2012iv}.
Depending on transverse momentum, these effects can partially offset, producing process-dependent patterns in the $\LL$ to $\NLLp$ ratio.

As an example cascade baryon plus jet production, the ratio exceeds 1~\cite{Celiberto:2022kxx}, whereas for charmed tetraquarks~\cite{Celiberto:2023rzw,Celiberto:2024beg}, the enhancement is milder.
Such variations reflect the differing balance between LL and NLL dynamics across channels.
The results in Fig.~\ref{fig:pT_jet} confirm the stability of our spectra under MHOUs and NLL effects, underscoring the predictive power of transverse momentum observables in quarkonium production within the VFNS framework.

This trend is a further manifestation of the \emph{natural stability}~\cite{Celiberto:2022grc} that becomes manifest across the full $|\vec q_{T_1}|$ range.
All this highlights the potential of quarkonium transverse momentum rates to cleanly isolate high-energy resummation from fixed-order contributions, offering a valuable window into semihard QCD dynamics.

\section{Toward new directions}
\label{sec:conclusions}

We have investigated the inclusive hadroproduction of pseudoscalar heavy quarkonia, $\etc$ and $\etb$ mesons, in high-energy proton collisions at moderate to large transverse momentum.

To this end, we have developed a new set of collinear FFs, denoted as {\NRFF}, tailored to describe quarkonium production at moderate to large transverse momentum. 
These functions build on the single-parton fragmentation approximation and evolve according to standard DGLAP equations in a VFNS. 
The underlying framework, known as the {\HFNRevo} scheme~\cite{Celiberto:2024mex,Celiberto:2024bxu,Celiberto:2024rxa,Celiberto:2025xvy}, is specifically designed to evolve heavy-hadron FFs from nonrelativistic NRQCD initial conditions, while incorporating heavy quark threshold effects in a consistent and systematic way. 
Initial conditions for all relevant partonic channels are computed at NLO in NRQCD. 
To the best of our knowledge, the {\NRFF} sets constitute the first public release of collinear FFs for heavy quarkonia that include all partonic channels within a fully consistent collinear factorization framework.

This work marks a significant advancement over previous determinations such as the {\tt ZCW19$^+$} sets~\cite{Celiberto:2022dyf,Celiberto:2023fzz} for vector quarkonia, which do not incorporate a consistent treatment of flavor thresholds across all partonic channels. 
In contrast, our {\NRFF} functions for pseudoscalar mesons $\etc$ and $\etb$ are constructed from fully NLO-accurate NRQCD inputs~\cite{Artoisenet:2014lpa,Zhang:2018mlo,Zheng:2021mqr,Zheng:2021ylc} and evolved using the {\HFNRevo} scheme~\cite{Celiberto:2024mex,Celiberto:2024bxu,Celiberto:2024rxa,Celiberto:2025xvy}, which systematically accounts for heavy quark threshold effects within DGLAP evolution. 
The pseudoscalar sector offers a particularly suitable testing ground for this methodology, since all partonic fragmentation channels are simultaneously active at NLO accuracy, unlike the vector case where the nonconstituent quark contribution vanishes.

In addition, theoretical expectations and experimental results both point to a strong suppression of color-octet mechanisms for $\etQ$ production~\cite{Braaten:1993rw,Han:2014jya,LHCb:2014oii,LHCb:2019zaj}, making the singlet-only approach adopted here both safe and well motivated. On the phenomenological side, upcoming facilities such as the HL-LHC~\cite{Chapon:2020heu,LHCspin:2025lvj}, EIC~\cite{Boer:2024ylx,AbdulKhalek:2021gbh,Khalek:2022bzd,Hentschinski:2022xnd,Amoroso:2022eow,Abir:2023fpo,Allaire:2023fgp}, NICA-SPD~\cite{Arbuzov:2020cqg,Abazov:2021hku}, IMC~\cite{Accettura:2023ked,InternationalMuonCollider:2024jyv,MuCoL:2024oxj,MuCoL:2025quu,Black:2022cth,InternationalMuonCollider:2025sys}, and FCC~\cite{FCC:2025lpp,FCC:2025uan,FCC:2025jtd} will offer unprecedented opportunities to investigate the pseudoscalar channel through semi-inclusive observables and quarkonium-in-jet studies. The release of the {\NRFF} sets thus fills a key gap in the toolkit for heavy-flavor phenomenology and provides a solid foundation for precision studies of quarkonium fragmentation in the high-energy limit.

To support phenomenological implications, we have analyzed the semi-inclusive hadroproduction of a pseudoscalar quarkonium state $\etQ$ accompanied by either a singly heavy-flavored hadron $\HQ$ or a jet, with both final-state objects well separated in rapidity. 
This process has been studied within the $\NLLp$ HyF formalism, which combines NLO collinear factorization with the resummation of high-energy logarithms beyond NLL accuracy. 
Predictions have been obtained using the {\Jethad} numerical interface in synergy with the {\symJethad} symbolic engine~\cite{Celiberto:2020wpk,Celiberto:2022rfj,Celiberto:2023fzz,Celiberto:2024mrq,Celiberto:2024swu}, enabling precise computations of forward high-energy observables sensitive to heavy quarkonium final states at the 13~TeV LHC.

The results reveal a remarkable perturbative stability across the full transverse momentum spectrum, reinforcing the \emph{natural stabilization} pattern~\cite{Celiberto:2022grc} previously observed in high-energy observables involving heavy-flavor production. 
As shown in earlier studies on ordinary heavy hadrons~\cite{Celiberto:2021dzy,Celiberto:2021fdp,Celiberto:2022zdg,Celiberto:2022dyf,Celiberto:2022keu,Celiberto:2024omj} and exotic bound states~\cite{Celiberto:2023rzw,Celiberto:2024mab,Celiberto:2024beg,Celiberto:2025dfe,Celiberto:2025ziy,Celiberto:2025ipt}, the presence of heavy-flavored FFs suppresses large logarithmic corrections and reduces the sensitivity to higher-order perturbative uncertainties. 
In this context, $\etQ$ production emerges as one of the cleanest and most robust channels to investigate BFKL-resummed dynamics in collider environments.

Our study provides an additional and complementary handle to isolate high-energy resummation effects from the fixed-order background, in a way that is both innovative and synergistic with more traditional approaches based on dijet correlations or forward light-hadron emissions. 
The theoretical cleanliness and numerical control offered by the [$\etQ + \HQ$ or [$\etQ + {\rm jet}$ systems highlight the potential of pseudoscalar quarkonia to serve as precision probes of high-energy QCD.

As a natural outlook, we plan to further develop the {\NRFF} sets by performing a systematic uncertainty analysis, with particular emphasis on the role of model-dependent inputs and their associated MHOU effects~\cite{Kassabov:2022orn,Harland-Lang:2018bxd,Ball:2021icz,McGowan:2022nag,NNPDF:2024dpb}. 
This effort will lay the groundwork for precision benchmarking of heavy quarkonium fragmentation and its interplay with high-energy dynamics.

In parallel, our aim is to extend the present study to vector quarkonia such as $\Jpsi$ and $\Yps$, for which the inclusion of color octet contributions is crucial. 
The ability to consistently quantify the impact of both singlet and octet channels will represent a significant step toward a unified and realistic description of quarkonium production at large transverse momentum.

Furthermore, the collinear fragmentation of quarkonia inside jets offers an additional window into the dynamics of hadronization in the presence of hard substructure. 
Future analyses will address this observable, both as a testing ground for the {\NRFF} sets and as a bridge to more differential studies, including TMD factorization and soft-collinear effective approaches~\cite{Ernstrom:1996am,Baumgart:2014upa,Kang:2016ehg,Kang:2017yde,Bain:2016clc,Bain:2017wvk,Makris:2018npl,Makris:2017sfs,Cooke:2023ukz,Gambhir:2025afb,Celiberto:2024rxa}.

From a phenomenological standpoint, the extension and application of the {\NRFF} functions within a purely collinear framework is of paramount importance. 
Fixed-order calculations in collinear factorization remain the most powerful and versatile tool to describe quarkonium production across a wide kinematic range, including both central and forward rapidities, and spanning a broad interval in transverse momentum and hard scales.

In particular, single-parton fragmentation, being applicable even at moderate transverse momentum, can act as a ``thermometer'' to determine the energy scales at which factorization instabilities and small-$x$ logarithmic enhancements start to become significant. 
Its flexibility makes it a valuable probe for delineating the transition between regions where fixed-order collinear predictions are reliable and those where resummation effects become necessary.

Beyond the present analysis, the HyF framework can be extended to cover single-inclusive emissions of quarkonium states in the forward region---unlike the two-particle final states at large rapidity separation considered here. 
It is widely recognized that forward quarkonium production at small Bjorken-$x$ and moderate scales $Q$ serves as a sensitive test bed for the perturbative stability and accuracy of collinear gluon PDFs in the low-$x$ domain. 
Recent studies~\cite{Lansberg:2020ejc,Lansberg:2023kzf} have demonstrated that enhancing collinear factorization by including high-energy resummation offers a promising path to curing the instabilities observed in this regime, leading to more robust predictions for quarkonium production at small $x$ and $Q$.

The recently reported evidence for the existence of intrinsic charm in the proton~\cite{Ball:2022qks}---building upon longstanding theoretical expectations~\cite{Brodsky:1980pb} and supported by several phenomenological studies~\cite{Jimenez-Delgado:2014zga,Ball:2016neh,Hou:2017khm,Guzzi:2022rca} as well as by new determinations of its valence component~\cite{NNPDF:2023tyk}---has reignited interest in the nonperturbative structure of the proton at moderate to large $x$. 
This breakthrough opens timely and compelling avenues for investigating how intrinsic charm manifests itself in high-energy hadronic processes.

In this context, the use of collinear fragmentation frameworks, and in particular the novel {\NRFF} functions, offers an ideal approach to probe the hadronization of heavy quarkonia in the presence of intrinsic charm~\cite{Flore:2020jau}. 
Their flexibility, theoretical consistency, and high-precision evolution enable detailed studies of how intrinsic charm may alter production rates, angular distributions, and flavor-tagged correlations in quarkonium-enriched final states. 
These developments place collinear fragmentation at the forefront of ongoing efforts to unravel the role of intrinsic heavy flavors in proton structure.

\section*{Acknowledgments}
\label{sec:acknowledgments}
\addcontentsline{toc}{section}{Acknowledgments}

We thank colleagues of the Quarkonia As Tools series of workshops for fruitful conversations and for the inspiring atmosphere.
We would like to express our gratitude to Alessandro Papa, Jean-Philippe Lansberg, Valerio Bertone, Kate Lynch, Andrea Signori, Charlotte Van Hulse, Ignazio Scimemi, Dani\"el Boer, Luca Maxia, Hua-Sheng Shao, Marco~Bonvini, Matteo~Cacciari, Terry~Generet, Felix~Hekhorn, and Hongxi~Xing for insightful discussions.
The Authors received support from the Atracci\'on de Talento Grant n. 2022-T1/TIC-24176 of the Comunidad Aut\'o\-no\-ma de Madrid, Spain.

\vspace{-0.25cm}
\section*{Data availability}
\label{sec:data}
\addcontentsline{toc}{section}{Data availability}
\vspace{-0.25cm}
The central sets of {\NRFF} FFs for $\etc$ and $\etb$ pseudoscalar quarkonium states~\cite{Celiberto:2025_NRFF10_cs_eQs} can be \textbf{}accessed at: \url{https://github.com/FGCeliberto/Collinear_FFs/}.
Since the FFs scale linearly with the LDME [Eqs.~\eqref{FFs_NRQCD}], users can simply rescale them using the LDME uncertainty ranges in Eqs.~\eqref{R0_etc} and~\eqref{R0_etb}, making dedicated error sets unnecessary.
The {\HCFF} functions for singly charmed hadrons are not publicly available. 
They can be provided under a reasonable request.
\onecolumn

\begin{appendices}

\newcommand{\appsubsection}[2]{
  \refstepcounter{subsection}
  \subsection*{\thesubsection\quad #1}
  \label{#2}
}

\setcounter{appcnt}{0}
\hypertarget{app:onium_puzzle}{
\section{On the quarkonium production puzzle}}
\label{app:onium_puzzle}
\counterwithout{table}{section}
\renewcommand{\thetable}{\arabic{table}}

In this Appendix, we present a focused discussion of the so-called quarkonium production puzzle, a longstanding issue in heavy quarkonium phenomenology that challenges the completeness of the NRQCD factorization framework in (semi-)inclusive processes. 
Our aim is to provide a basic and structured overview of the key mechanisms, spin configurations, and production environments where the puzzle manifests or, conversely, appears to be absent. 
Particular attention is given to the comparative role of color-singlet and color-octet channels in vector and pseudoscalar quarkonia, as well as to the process-dependent nature of the theoretical tension. 
The discussion is tailored to offer a coherent reference for future studies, especially in view of the increasing relevance of NRQCD-based fragmentation frameworks for both standard and exotic hadron production.

\appsubsection{The puzzle in hadroproduction}{ssec:AppA:hadroproduction}

From the Tevatron era onward, high‑$q_T$ hadroproduction of vector quarkonia [particularly $\Jpsi$ and $\psi(2S)$] has revealed persistent inconsistencies between theoretical predictions and experimental data. 
In fixed-order calculations restricted to the ${}^3S_1^{(1)}$ color-singlet configuration, the predicted cross sections fall significantly below those observed at the Tevatron and in early LHC runs. These discrepancies motivated the adoption of the NRQCD factorization framework, which includes color-octet contributions from intermediate $c\bar c$ states such as ${}^1S_0^{(8)}$, ${}^3S_1^{(8)}$, and ${}^3P_J^{(8)}$~\cite{Braaten:1993rw,Cho:1995vh,Beneke:1996tk,Butenschoen:2010rq,Gong:2012ug}.

With the advent of the LHC, quarkonium production and polarization measurements have become more precise and systematic. 
The inclusive yields of $\Jpsi$ at $\sqrt{s} = 7$, $8$, and $13$~TeV are well measured by CMS, ATLAS, and the LHCb over a broad range in transverse momentum and rapidity. However, no clear signal of the strong transverse polarization expected from dominant gluon fragmentation into ${}^3S_1^{(8)}$ configurations has emerged. CMS reported nearly unpolarized prompt $\Jpsi$ production up to $|\vec q_T| \sim 70$~GeV~\cite{CMS:2013fpt}, while the LHCb observed similarly small polarization in the forward region~\cite{LHCb:2013izl}. 
These results challenge standard NRQCD expectations and have triggered extensive theoretical efforts to refine fits of LDMEs and incorporate higher-order QCD corrections~\cite{Chao:2012iv,Butenschoen:2012qr}.

Global fits that attempt to reconcile hadroproduction, photoproduction, and $e^+e^-$ data often yield LDME values that vary significantly between processes, casting doubts on their universality. 
In particular, polarization observables remain sensitive to feed-down effects, kinematic cuts, and resummation of soft-gluon emissions---all of which complicate direct comparisons.
Monte Carlo generators such as {\tt PYTHIA8}~\cite{Bierlich:2022pfr} include NRQCD-based production mechanisms and are routinely used to simulate quarkonium observables under various color-octet and singlet assumptions. 
Although useful for phenomenological studies, these tools often rely on effective modeling of the hadronization stage, limiting the precision with which they can constrain LDMEs.
Despite these challenges, NRQCD remains the most systematic framework for analyzing inclusive quarkonium production. 
However, the persistent mismatch between data and theory (especially in polarization) continues to motivate further theoretical development and more differential experimental analyses.

\appsubsection{Color-octet mechanisms in pseudoscalar production}{ssec:AppA:CO_pseudoscalars}

In contrast to vector quarkonia, pseudoscalar mesons such as $\etc$ and $\etb$ offer a phenomenologically cleaner environment to assess the validity of NRQCD factorization and the relevance of color-octet contributions. 
Their quantum numbers, $J^{PC} = 0^{-+}$, forbid a direct coupling to two vector gluons in several production channels at LO. 
As a result, the dominant contribution is expected to arise from the color-singlet ${}^1S_0^{(1)}$ channel, with octet mechanisms entering only at higher orders in the NRQCD velocity expansion and often vanishing at tree level. 
This picture is confirmed by hadroproduction data: for example, measurements of $\etc$ transverse momentum spectra by LHCb~\cite{LHCb:2014oii,LHCb:2019zaj} are well described by singlet-only predictions, with no compelling need for octet terms.

The situation is even more constrained for the $\etb$, where the heavier quark mass implies smaller relative velocities ($v_{\cal Q}^2 \sim 0.1$), and hence stronger suppression of all color-octet channels. 
Thus, any observable $\etb$ cross section is expected to be highly dominated by the singlet contribution, and any discrepancy from theoretical predictions could provide strong evidence for unexpected octet enhancements or for a breakdown of the NRQCD factorization.

However, the situation changes in photon-initiated processes. 
In $\gamma p$ photoproduction, the singlet contribution to $\etc$ production vanishes at both LO and NLO in $\alpha_s$, due to the absence of appropriate final-state configurations. 
This makes $\etc$ photoproduction a clean probe of the ${}^1S_0^{(8)}$ LDME~\cite{Eboli:2003fr,Han:2014jya}, which dominates the cross section in this regime. 
Similar behavior is expected in leptoproduction with virtual photons ($\gamma^*$), though the phenomenology is more involved and remains largely unexplored.

In $e^+e^-$ annihilation and $\gamma\gamma$ collisions, the hierarchy between singlet and octet channels is reversed again. 
For $\etc$, direct color-singlet production is suppressed or forbidden depending on the final state, while color-octet mechanisms can already contribute at LO. 
These channels offer an opportunity to test NRQCD scaling rules and the universality of LDMEs extracted from hadronic collisions.

Overall, pseudoscalar quarkonia provide a sensitive testing ground for the color-octet mechanism across different processes. 
In hadroproduction, the singlet dominance supports the validity of the color-singlet model. 
In contrast, photoproduction and $e^+e^-$ annihilation amplify the octet contribution, offering a complementary perspective on the NRQCD expansion.

\appsubsection{Insights from photoproduction and lepton collisions}{ssec:AppA:photo_leptoproduction}

The relative suppression of color-octet channels observed in pseudoscalar quarkonia hadroproduction motivates the exploration of complementary processes, such as photoproduction and $e^+e^-$ annihilation, where the color-singlet contributions are either suppressed or vanish at leading order. These channels provide an independent handle on color octet--dominated transitions and thus serve as ideal benchmarks to assess the universality of the NRQCD factorization framework.

In $\gamma p$ photoproduction events, extensively studied at HERA and relevant for future EIC programs, the energy distribution of produced vector quarkonia, such as the $\Jpsi$, constitutes a key observable. 
A commonly used variable is
\begin{equation}
\label{z_photoprod}
 z = \frac{E_{\Q}}{E_\gamma} \;,  
\end{equation}
defined in the proton rest frame, where $E_{\Q}$ is the energy of the observed quarkonium and $E_\gamma$ that of the incoming photon. 
Large-$z$ events ($z \gtrsim 0.6$) correspond to nearly elastic configurations, while low-$z$ events ($z \lesssim 0.3$) are associated with highly inelastic processes.

At large $z$, theoretical predictions for $\Jpsi$ photoproduction suffer from end point divergences due to soft-gluon emission in the hadronization of color-octet [$c\bar{c}$ pairs. 
This complicates the extraction of color-octet LDMEs from the tail of the spectrum and requires the inclusion of shape function effects or the resummation of end point logarithms~\cite{Fleming:2006cd,Beneke:1999gq}.
Conversely, in the low-$z$ region, resolved-photon processes---where the photon fluctuates into partons before the hard scattering---become significant. 
These processes are enhanced for color-octet configurations such as ${}^1S_0^{(8)}$ and ${}^3P_J^{(8)}$, making them a crucial probe of subleading NRQCD channels. 
However, experimental precision in this regime remains limited, and more differential measurements would be highly beneficial.

Lepton-lepton collisions offer an even cleaner environment. 
In $e^+e^-$ annihilation, color-singlet contributions such as $e^+e^- \to \Jpsi + gg$ or $c\bar{c}$ are suppressed relative to color-octet channels like $e^+e^- \to \Jpsi + g$ or $q\bar{q}$. 
This makes $e^+e^-$ data a unique laboratory test for color-octet dominance and to extract individual LDMEs through energy-dependent analyses. 
For example, low-energy ($\sqrt{s} < 20$ GeV) data are more sensitive to ${}^3S_1^{(8)}$ transitions, whereas higher energies allow access to linear combinations involving ${}^1S_0^{(8)}$ and ${}^3P_J^{(8)}$~\cite{Lepage:1992tx}.

Finally, semi-inclusive deep inelastic scattering (SIDIS) processes at EIC energies, particularly in $ep$ and $eA$ collisions, open new windows to explore the quarkonium puzzle. Theoretical studies suggest that $\Jpsi$ production in such setups can isolate the interplay between color-octet and singlet channels, and offer sensitivity to nuclear effects, initial-state radiation, and parton saturation dynamics~\cite{Chu:2024fpo}. 
These insights, especially when combined with hadro- and photo-production data, will allow more robust and process-independent determinations of NRQCD matrix elements.

\appsubsection{Velocity scaling of NRQCD LDMEs}{ssec:AppA:velocity_scaling}

The phenomenological relevance of each NRQCD channel depends on its scaling with the relative velocity $v_{\cal Q}$ of the heavy quark--antiquark pair in the quarkonium rest frame. 
Within the NRQCD factorization approach, this scaling governs the relative importance of the LDMEs, and thus determines which color and angular momentum configurations are expected to dominate in a given process.

In the table below, we summarize the expected velocity suppression of singlet and octet contributions for $\etc$, $\etb$, $\Jpsi$, and $\Yps$ states, following the standard power-counting rules~\cite{Bodwin:1994jh}. 
Only the leading terms for each channel are shown. 
Octet contributions suppressed by higher powers of $v_{\cal Q}$ may still be phenomenologically relevant in specific processes, but are generally subdominant.

\begin{table}[h!]
\centering
\begin{tabular}{lcccccccc}
\toprule
State & ${}^1S_0^{(1)}$ & ${}^3S_1^{(1)}$ & ${}^1S_0^{(8)}$ & ${}^3S_1^{(8)}$ & ${}^3P_0^{(8)}$ & ${}^3P_1^{(8)}$ & ${}^3P_2^{(8)}$ \\
\midrule
$\etc$ or $\etb$ & 1 & -- & $v_{\cal Q}^4$ & $v_{\cal Q}^3$ & $v_{\cal Q}^4$ & $v_{\cal Q}^4$ & $v_{\cal Q}^4$ \\
$\Jpsi$ or $\Yps$ & -- & 1 & $v_{\cal Q}^3$ & $v_{\cal Q}^4$ & $v_{\cal Q}^4$ & $v_{\cal Q}^4$ & $v_{\cal Q}^4$ \\
\bottomrule
\end{tabular}
\caption{Velocity scaling of LDMEs for $\Jpsi$ and $\etc$ production in NRQCD~\protect\cite{Bodwin:1994jh}.}
\label{tab:nrqcd_scaling}
\end{table}

As seen in Table~\ref{tab:nrqcd_scaling}, the dominant contribution to $\eta_{c,b}$ production comes from the color-singlet ${}^1S_0^{(1)}$ channel, which is unsuppressed in the NRQCD expansion. 
All octet configurations are power-suppressed, with ${}^3S_1^{(8)}$ contributing at $\mathcal{O}(v_{\cal Q}^3)$ and the others at $\mathcal{O}(v_{\cal Q}^4)$ or beyond. 
This renders the $\etc$ a particularly clean testing ground for singlet-dominated quarkonium production, with reduced sensitivity to color-octet ambiguities. 
Indeed, recent LHCb measurements of $\etc$ production at $\sqrt{s} = 7$ and $13$~TeV are well described by color-singlet-only predictions~\cite{LHCb:2014oii,LHCb:2019zaj}.

In the case of $\Jpsi$ and $\Yps$, the singlet ${}^3S_1^{(1)}$ channel is leading and unsuppressed, but multiple color-octet channels contribute at similar or only mildly suppressed orders, especially the ${}^1S_0^{(8)}$ term at $\mathcal{O}(v_{\cal Q}^3)$. 
This makes the $\Jpsi$ significantly more sensitive to the details of the LDME extraction, and explains the richer phenomenology and associated uncertainties discussed in Appendices~\ref{ssec:AppA:hadroproduction} to~\ref{ssec:AppA:photo_leptoproduction}.

The same power-counting logic applies to the bottomonium sector. 
However, due to the larger quark mass, the characteristic velocity of the ($b\bar{b}$) pair in bound states is significantly smaller: $v_{\cal Q}^2 \sim 0.1$ for bottomonia versus $v_{\cal Q}^2 \sim 0.3$ for charmonia. 
As a consequence, all octet contributions in $\etb$ and $\Upsilon(nS)$ production are more strongly suppressed, and color-singlet dominance becomes even more pronounced. 
This supports the expectation that bottomonium production, especially for $\etb$, is less sensitive to octet mechanisms, though experimental limitations---such as suppressed decay channels---currently hinder detailed comparisons.

\appsubsection{Current status and open directions}{ssec:AppA:conclusions}

The quarkonium production puzzle remains a key test of our understanding of QCD factorization and nonperturbative dynamics. 
Although the inclusion of color-octet channels in NRQCD has enabled significant progress, a fully consistent description across production modes, kinematic regions, and quarkonium species is still lacking. 
The case of $\etc$ and, more generally, of pseudoscalar quarkonia, provides a crucial testing ground due to their reduced sensitivity to octet mechanisms. 
Measurements in complementary channels (photoproduction, $e^+e^-$, and ultraperipheral collisions) and for excited states [\emph{e.g.}, $\eta_{c}(2S)$, $\eta_{c2}$] will be essential to further clarify the picture and validate the universality of LDMEs across processes.

\setcounter{appcnt}{0}
\hypertarget{app:nlo_gluon_SDC}{
\section{NLO correction to the gluon SDC}}
\label{app:nlo_gluon_SDC}
\counterwithout{table}{section}
\renewcommand{\thetable}{\arabic{table}}

In this Appendix, we report the expressions of the regular and distributional functions entering the NLO correction to the [$g \to \etQ$ SDC (see Section~\ref{sssec:NRFF_g}).

The regular part of the SDC is expressed as a piecewise function
\begin{equation}
\label{SDC_SQ_g_NLO_reg}
 \drv_{g, \mathrm{reg}}^{\rm [NLO]} (z, ^{1\!\!}S_0^{(1)}) \;=\; 
 \left\{
  \begin{aligned}
    &\hspace{0.15cm}-\frac{N_c}{2 z} + \sum_{m=0}^{2} \sum_{n=0}^{\infty} \;(\ln z)^m \, (2z)^n \left[n_f \, {\cal A}_{mn}^{(f)} + {N_c \, \cal A}_{mn}^{(1)} + \frac{{\cal A}_{mn}^{(N)}}{N_c} \right] \;, \quad &\mbox{for}& \quad 0<z<\frac{1}{4} \\
    &\hspace{0.15cm}\sum_{n=0}^{\infty} \;(2z-1)^n \left[n_f \, {\cal B}_{n}^{(f)} + N_c \,{\cal B}_{n}^{(1)} + \frac{{\cal B}_{n}^{(N)}}{N_c} \right] \;, \quad &\mbox{for}& \quad \! \frac{1}{4} \le z \le \frac{3}{4} \\
    &\hspace{0.15cm}\sum_{m=0}^{3} \sum_{n=0}^{\infty} \; [\ln (1-z)]^m \, (2-2z)^n \, \left[n_f \, {\cal C}_{mn}^{(f)} + N_c \, {\cal C}_{mn}^{(1)}+ \frac{{\cal C}_{mn}^{(N)}}{N_c} \right] \;, \quad &\mbox{for}& \! \quad \frac{3}{4}<z<1 
  \end{aligned}
 \right. \;,
\end{equation}
with $n_f$ the number of active flavors. 
The numerical values of the first 50 ${\cal A}_{mn}^{(f,1,N)}$, ${\cal B}_{mn}^{(f,1,N)}$, and ${\cal C}_{mn}^{(f,1,N)}$ coefficients are provided in Appendix~E of Ref.~\cite{Zhang:2018mlo}.
These coefficients have been tabulated and implemented in the {\symJethad} interface for direct numerical use.

Then, the distributional part of the SDC reads
\begin{align}
\label{SDC_SQ_g_NLO_dist}
\begin{split}
 \drv_{g, \mathrm{dist}}^{\rm [NLO]} (z, {}^{1\!\!}S_0^{(1)}) 
      \,&=\, 
      N_c \bigg[-2 (z+2) \operatorname{Li}_2(z)
      - 2 (z-1) \ln^2(1-z)
      + 2 (z-1) \ln z \ln(1-z)
      + (z-4) z \ln z
     \\
      \,&\phantom{=}\;\;
      -\, \frac{(2z+1) \left(9 z^2 - 5 z - 6 \right) \ln(1-z)}{6z}
      + \frac{46 z^3 + (8\pi^2 - 3) z^2 + 4(\pi^2 - 9) z + 4}{12z}
      \bigg]
     \\
      \,&-\, 6 n_f \, m_Q^3 \, \drv_{g}^{\rm [LO]} (z, {}^{1\!\!}S_0^{(1)})
      \;,
\end{split}
\end{align}
with $\drv_{g}^{\rm [LO]}$ given in Eq.~\eqref{SDC_SQ_g_LO}.
As in the case of the regular component, the expression above corresponds to the fully regulated form of the distributional part, where all plus-distributions and $\delta(1-z)$ terms have already been integrated against suitable test functions. 
This form is directly usable in numerical codes without requiring additional distributional handling.

\end{appendices}

\setlength{\bibsep}{0.6em}
\bibliographystyle{apsrev}
\bibliography{references}

\end{document}